\documentclass[conference]{IEEEtran}
\IEEEoverridecommandlockouts
\usepackage{cite}
\usepackage{amsmath,amssymb,amsfonts}
\usepackage{algorithmic}
\usepackage{graphicx}
\usepackage{textcomp}
\usepackage{xcolor}
\usepackage{subfigure}
\usepackage{multirow}
\usepackage{amsthm}
\usepackage[explicit]{titlesec}
\usepackage[noend, linesnumbered, ruled, lined]{algorithm2e}
\usepackage{hyperref}
\usepackage{verbatim}
\usepackage{adjustbox}
\usepackage{diagbox}
\usepackage{multirow}
\usepackage[normalem]{ulem}
\usepackage{ulem}
\usepackage{balance}
\usepackage{amsthm}
\usepackage{pifont}
\usepackage{tabularx}
\hypersetup{
colorlinks=true,
linkcolor=black,
citecolor=black
}
\begin{document}

% \title{HyGNN: Accelerating Graph Neural Networks Training with Hybrid GPU Cores}
\title{HC-SpMM: Accelerating Sparse Matrix-Matrix Multiplication for Graphs with Hybrid GPU Cores}
% \title{HySpMM: Accelerating SpMM for Graphs \\with Hybrid GPU Cores}
% {\texorpdfstring{HyGNN: Accelerating Graph Neural Networks Training \\with Hybrid GPU Cores}{HyGNN: Accelerating Graph Neural Networks Training with Hybrid GPU Cores}}

\author{
\IEEEauthorblockN{Zhonggen Li\IEEEauthorrefmark{1}, Xiangyu Ke\IEEEauthorrefmark{1}, Yifan Zhu\IEEEauthorrefmark{1}, Yunjun Gao\IEEEauthorrefmark{1}, Yaofeng Tu\IEEEauthorrefmark{2}}
\IEEEauthorblockA{\IEEEauthorrefmark{1}Zhejiang University, Hangzhou, China \; \IEEEauthorrefmark{2}ZTE Corporation, Nanjing, China \\
\IEEEauthorrefmark{1}\{zgli, xiangyu.ke, xtf\_z, gaoyj\}@zju.edu.cn, \IEEEauthorrefmark{2}tu.yaofeng@zte.com.cn
}}

\maketitle

\begin{abstract}
  Sparse Matrix-Matrix Multiplication (SpMM) is a fundamental operation in graph computing and analytics. However, the irregularity of real-world graphs poses significant challenges to achieving efficient SpMM operation for graph data on GPUs. Recently, significant advancements in GPU computing power and the introduction of new efficient computing cores within GPUs offer new opportunities for acceleration. 

  In this paper, we present \textsf{HC-SpMM}, a pioneering algorithm that leverages \underline{h}ybrid GPU \underline{c}ores (Tensor cores and CUDA cores) to accelerate \underline{SpMM} for graphs. 
  To adapt to the computing characteristics of different GPU cores, we investigate the impact of sparse graph features on the performance of different cores, develop a data partitioning technique for the graph adjacency matrix, and devise a novel strategy for intelligently selecting the most efficient cores for processing each submatrix. 
  Additionally, we optimize it by considering memory access and thread utilization, to utilize the computational resources to their fullest potential. 
  To support complex graph computing workloads, we integrate \textsf{HC-SpMM} into the GNN training pipeline. Furthermore, we propose a kernel fusion strategy to enhance data reuse, as well as a cost-effective graph layout reorganization method to mitigate the irregular and sparse issues of real-world graphs, better fitting the computational models of hybrid GPU cores.  
  % Additionally, we propose a cost-effective graph layout reorganization method to mitigate the irregular and sparse issues of real-world graphs, better fitting the computational models of hybrid GPU cores. 
  % Furthermore, we optimize it by considering memory access, thread utilization, etc., to utilize the computational resources to their fullest potential. 
  % To verify the effectiveness of HySpMM, we systematically integrate it into \textsf{PyTorch} to accelerate GNN training. 
  Extensive experiments on 14 real-world graph datasets demonstrate that \textsf{HC-SpMM} achieves an average speedup of 1.33$\times$ and 1.23$\times$ over state-of-the-art SpMM kernels and GNN frameworks. 
\end{abstract}


\section{Introduction}
\label{sec:introduction}
Sparse Matrix-Matrix Multiplication (SpMM) is a fundamental operation that multiplies a sparse matrix and a dense matrix, which has been widely used in various graph computing tasks such as Graph Neural Networks (GNNs) training \& inference~\cite{zhang2023ducati, li2024daha, gunduz2022scalable}, PageRank~\cite{hou2021massively, luo2019efficient, shi2019realtime} and graph clustering~\cite{zhang2022effective, chang2017mathsf, xu2012model}. Recent studies employ matrix multiplication operations on GPUs to accelerate commonly used graph algorithms such as triangle counting and shortest path computing, with SpMM emerging as one of the core operations and becoming the efficiency bottleneck~\cite{sundaram2015graphmat, yang2022graphblast}. SpMM is also a fundamental operation in GNNs, accounting for more than 80\% of the GNN training time~\cite{wang2023tc}. 
% However, the irregularity of real-world graphs results in SpMM being characterized by irregular memory access patterns and low computational intensity, posing significant challenges for GPU acceleration~\cite{wang2023tc}. 
However, characterized by {\em irregular memory access patterns} and {\em low computational intensity}, the real-world graphs pose significant challenges for achieving efficient SpMM on GPU~\cite{wang2023tc}. 
Especially for GNNs, as the scale of graph data proceeds to grow and the complexity of GNN architectures escalates with an increasing number of layers, the efficiency issue becomes more serious, constraining the large-scale applicability\footnote{Training a GNN-based recommendation system over a dataset comprising 7.5 billion items consumes approximately 3 days on a 16-GPU cluster \cite{duan2022comprehensive}.}, even though it achieves state-of-the-art performance on a myriad of problems such as node classification \cite{song2022towards, zhang2022mul, wen2021meta}, link prediction \cite{gu2022hybridgnn, zou2023embedx, zhang2023disconnected, ran2023differentially}, recommendation \cite{zhang2021group, xia2021multi, wu2022graphnew} and beyond \cite{fangshu, sun2023graph, li2023graph}. 
% There are few SpMM kernels specifically designed for graph computing workloads. 
% During both forward and backward propagations in GNNs, two fundamental operations are performed: \textit{Aggregation} and \textit{Update}. The \textit{Update} operation can be efficiently computed using {\sf cuBLAS} \cite{refcublas}. 
% % , developed by Nvidia and optimized with Tensor cores within GPUs. 
% In contrast, \textit{Aggregation} involves sparse matrix-matrix multiplication (SpMM), which is characterized by {\em irregular memory access patterns} and {\em low computational intensity}, especially on real-world graphs, posing significant challenges for GPU acceleration~\cite{wang2023tc}. A recent study indicates that \textit{Aggregation} accounts for more than 80\% of the execution time \cite{wang2023tc}, emerging as the primary bottleneck of GNNs. 
% In this paper, we aim to devise a graph-friendly SpMM kernel using hybrid CUDA and Tensor cores to support high-performance graph computing workload. As a popular application, we also apply the SpMM kernel to GNN training. 

To enhance the efficiency of SpMM, some efforts have been made to employ CUDA cores in GPU for acceleration~\cite{refcusparse,hong2019adaptive,gale2020sparse,huang2020ge}. As the performance ceil is constrained by the hardware structure of CUDA cores, recent attempts have turned to explore harnessing the Tensor cores, which are specialized for efficient matrix multiplication, to accelerate SpMM~\cite{zachariadis2020accelerating,chen2021efficient,li2022efficient}. 
% CUDA cores and Tensor cores represent distinct computing units in GPUs. CUDA cores are versatile units capable of general computing tasks, while Tensor cores are specialized for efficient matrix multiplication operations~\cite{zachariadis2020accelerating}.
% Tensor cores exhibit a remarkable ability to perform fixed-size (e.g., 4 $\times$ 4) matrix multiplications in a single clock cycle~\cite{nvidia_v100}, which is more efficient than CUDA cores. 
However, despite significantly improving the efficiency of dense matrix multiplication~\cite{li2022efficient}, 
% Tensor cores require fixed-size input (e.g. two 4$\times$4 matrices), 
the irregularity and sparsity of real-world graphs impeding Tensor cores from achieving satisfactory performance ~\cite{zachariadis2020accelerating, wang2023tc, fan2024dtc}. Various preprocessing techniques have been investigated to enhance the density of sparse matrices, yet significant sparse portions persist, hindering performance improvement\footnote{For instance, the average sparsity of the matrices output by the preprocessing method proposed in \textsf{TC-GNN}~\cite{wang2023tc} is still 90.9\% on 10 tested datasets. }.

\begin{figure}[tbp]
    \centering    
    
    \hspace{-2mm}
    \subfigure[Varying sparsity.]{
    \includegraphics[width=0.23\textwidth,height=0.17\textwidth]{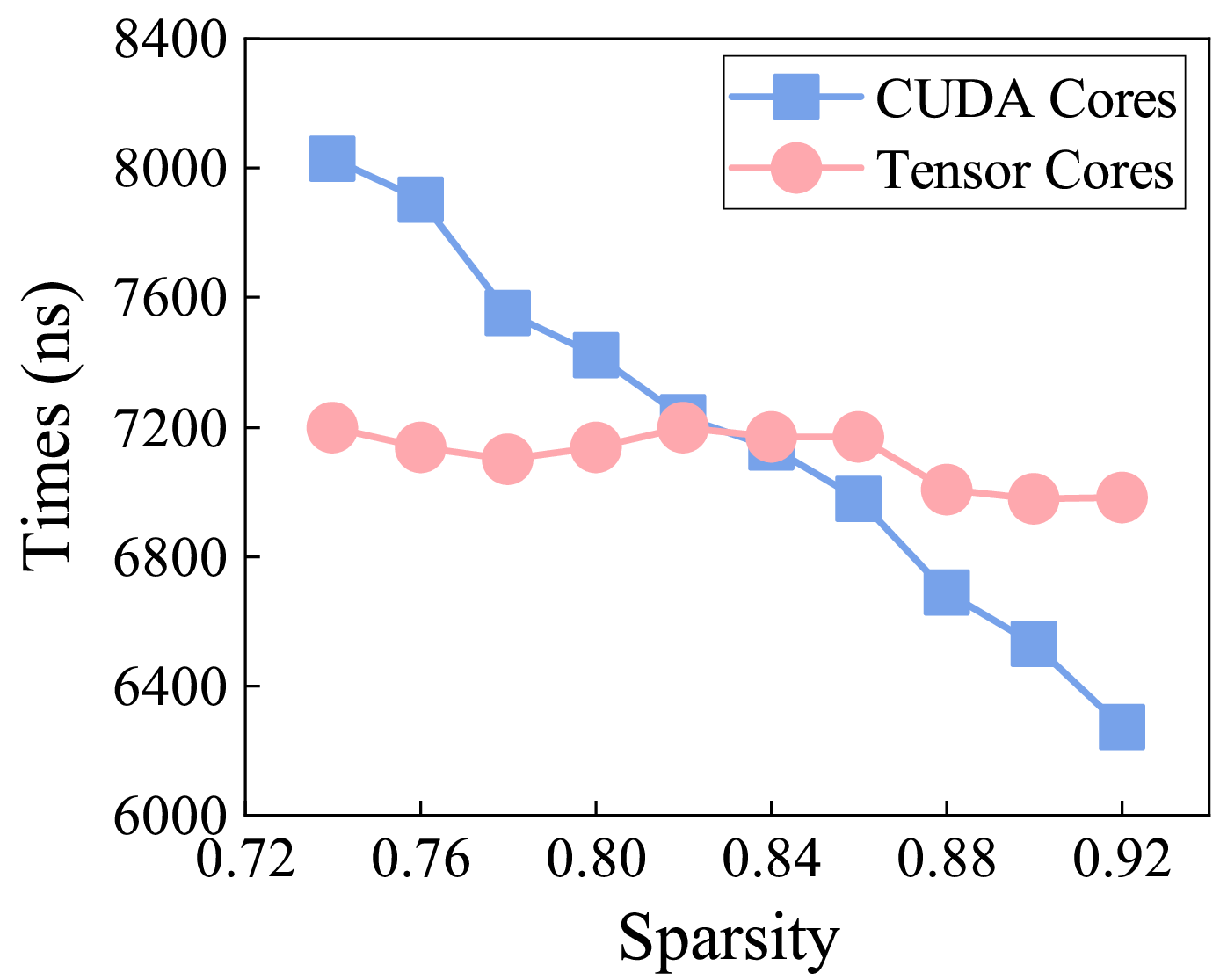}}
    \hspace{-1mm}
    \subfigure[Varying non-zero columns.]{
    \includegraphics[width=0.23\textwidth,height=0.17\textwidth]{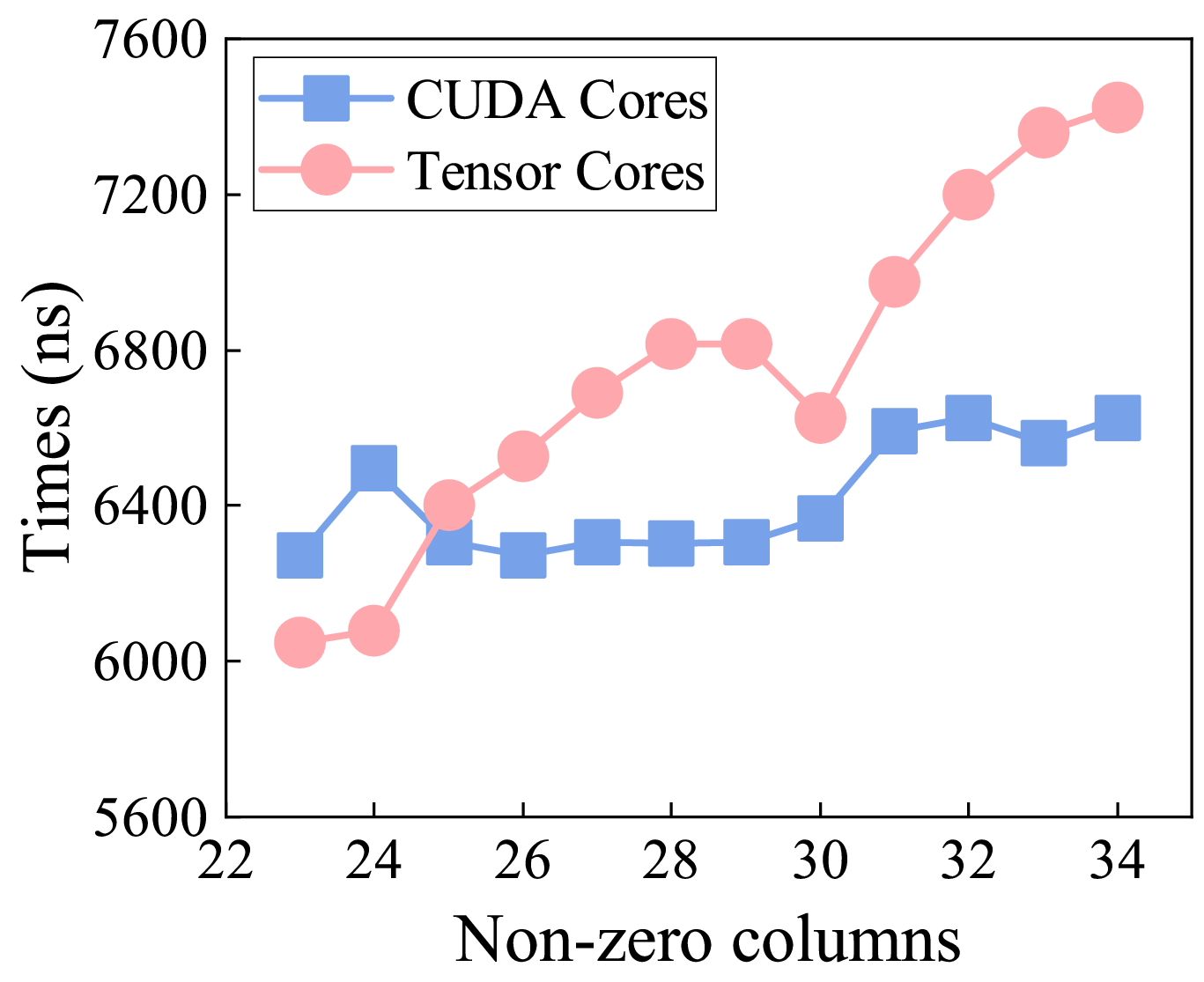}}
    \vspace{-2mm}
    \caption{SpMM execution time with varying sparsity and non-zero columns.}
    \label{executiontime}
\vspace{-6mm}
\end{figure}

%Although extensive efforts have been dedicated to harnessing either CUDA cores or Tensor cores to expedite SpMM for GNN training, the irregularity of real graphs, characterized by a mixture of dense and sparse regions within the corresponding adjacency matrices, renders none of them holding absolute superiority. 
Figure~\ref{executiontime} illustrates the time overhead incurred by CUDA cores and Tensor cores across matrices with varying sparsity and the number of non-zero columns, which highlights their distinct applicabilities\footnote{An in-depth analyzation is provided in \S~\ref{adaptivecoreselection}.}. 
CUDA cores demonstrate superior performance when handling sparser matrices with a higher number of non-zero columns, whereas Tensor cores exhibit greater efficacy in denser matrices characterized by fewer non-zero columns. 
Thereafter, {\em relying solely on CUDA cores or Tensor cores for SpMM acceleration fails to fully exploit their respective computational strengths}, primarily due to the irregularity of real-world graphs, which typically comprise {a mixture of dense and sparse regions~\cite{yan2014blogel}}. 
%The requisite of fixed-size input precludes Tensor cores from dynamically circumventing the computation of zero elements within sparse matrices, thereby limiting their applicability to dense matrix multiplication. Conversely, CUDA cores, while less proficient in matrix multiplication, offer greater adaptability within sparse computational contexts. 
% Consequently, there arises a need for a hybrid strategy tailored to the varying computational demands of different graph regions, enabling more effective utilization of the capabilities of different GPU cores. 
% Consequently, we and devise a hybrid strategy tailored to the varying computational demands of different graph regions, enabling more effective utilization of the capabilities of GPU cores.

In this paper, we aim to devise a graph-friendly SpMM kernel using hybrid CUDA cores and Tensor cores, tailored to the varying computational demands of different graph regions, to support high-performance graph computing workloads. Furthermore, as a DB technique for AI acceleration, we integrate the hybrid SpMM kernel into the GNN training pipeline to enhance the training efficiency and demonstrate its effectiveness in complex graph computing scenarios.

However, crafting a hybrid SpMM kernel using CUDA cores and Tensor cores based on the characteristics of graph data and systematically integrating it into the GNN training pipeline encounters non-trivial challenges.  
\noindent \textbf{Challenges of designing hybrid SpMM kernel.} {\bf (1)} Effectively partitioning the adjacency matrix into sub-regions with distinct sparsity characteristics to leverage the different cores for collaborative computation presents a considerable challenge. {\bf (2)} The computational characteristics of CUDA and Tensor cores vary significantly, which is also influenced by the graph features, making the selection of the optimal core for matrices with different sparsity levels crucial for enhancing efficiency. {\bf (3)} Accurately modeling the computational performance of CUDA and Tensor cores for SpMM to facilitate precise core selection is another major difficulty. {\bf (4)} Last but not least, inefficiencies inherent in the SpMM kernel, such as suboptimal memory access patterns and underutilized threads, limit the overall performance. 
To address these issues, we {\bf firstly} propose a fine-grained partition strategy, which divides the adjacency matrix into equal-sized submatrices along the horizontal axis, allocating each submatrix to the appropriate GPU cores for efficient computation (\S~\ref{combinationstrategy}). This allows CUDA and Tensor cores to perform calculations independently, eliminating the need to merge results between cores. 
% and avoiding establishing separate data structures for each core, thus preserving the sequential memory access of edges. 
{\bf Secondly}, comprehensive quantitative experiments reveal that CUDA cores are \textit{memory-efficient} while Tensor cores are \textit{computing-efficient} (\S~\ref{computingchar}). These experiments identify two pivotal factors for submatrix characterization: sparsity and the number of non-zero columns, which dominate the most expensive parts for CUDA and Tensor cores, \textit{computation} and \textit{memory access}, respectively. 
{\bf Thirdly}, leveraging these factors, we develop an algorithm tailored for the selection of appropriate GPU cores for submatrices, optimizing computational capability (\S~\ref{computingchar}). 
{\bf Finally}, we conduct in-depth optimizations of the SpMM kernel, considering thread collaboration mode (\S~\ref{bottleneckofccu}) and memory access patterns (\S~\ref{bootleneckTCU}).

\noindent \textbf{Challenges of integrating the SpMM kernel into GNN.} When integrating our proposed hybrid SpMM kernel into the GNN training pipeline, there arise new challenges. {\bf (1)} The isolation among GPU kernels within a GNN layer in prevalent GNN training frameworks~\cite{wang2019deep, wang2023tc} impedes data reuse, leads to additional memory access overhead, and introduces significant kernel launch overhead. {\bf (2)} Tensor cores incur significantly higher throughput than CUDA cores, potentially offering substantial efficiency gain. However, real-world graph layouts inherently exhibit irregularity and sparsity, resulting in a majority of segments partitioned from the adjacency matrix being sparse and less amenable to processing via Tensor cores. Consequently, the performance gains achievable with Tensor cores are often negligible~\cite{wang2023tc}. 
% However, more parts suitable for CUDA cores diminishes the incremental efficiency gains afforded by the introduction of Tensor cores.  
To tackle the {\bf first} problem, we discover opportunities to reuse data in the shared memory of GPU and present a kernel fusion strategy to mitigate kernel launch costs and global memory access (\S~\ref{subsec:kernelfusion}). 
To address the {\bf second} problem, we first introduce a metric termed \textit{computating intensity} to estimate the calculation workload of the submatrices multiplication (\S~\ref{subsec:layout}), which is calculated by the quotient of the number of non-zero elements and the number of non-zero columns. 
Higher computational intensity is achieved with more non-zero elements and fewer non-zero columns.
Subsequently, we propose an efficient algorithm to reconstruct submatrices, adjusting the graph layouts to obtain more dense segments suitable for processing by Tensor cores, gaining significant efficiency with Tensor cores (\S~\ref{subsec:layout}). 
This adjustment has a relatively small cost compared with GNN training but renders the graph data more compatible with hybrid GPU cores, unlocking the full computational potential of Tensor cores and thereby significantly enhancing the efficiency of GNN training (\S~\ref{eva:layout}).

\noindent\textbf{Contributions.} In this work, we propose \textsf{HC-SpMM}, a novel approach for accelerating SpMM using hybrid GPU cores. Our key contributions are summarized as follows: 
% \vspace{-1mm}
\begin{itemize}
    \item We quantify the difference between CUDA and Tensor cores in SpMM, and propose a hybrid SpMM kernel, which partitions the graph adjacency matrix and intelligently selects appropriate GPU cores for the computation of each submatrix based on its characteristics. We further optimize the SpMM kernel considering thread collaboration mode and memory access patterns (\S~\ref{sec:hybridspmm}). 
    \item We propose a kernel fusion strategy for integrating {\sf HC-SpMM} into the GNN training pipeline, eliminating kernel launch time and enabling efficient data reuse, and present a lightweight graph layout optimization algorithm to enhance irregular graph layouts and better align with the characteristics of both GPU cores (\S~\ref{sec:spmmintegration}). 
    \item We conduct comprehensive evaluations demonstrating that \textsf{HC-SpMM} outperforms existing methods, achieving 1.33$\times$ speedup in SpMM and 1.23$\times$ speedup in GNN training on average (\S~\ref{sec:evaluation}).  
\end{itemize}
% \vspace{-9mm}
% \textcolor{red}{1-2 sentences about other sections not mentioned above}
% \vspace{-1mm}
%To the best of our knowledge, this is the first attempt to combine CUDA cores and Tensor cores for algorithm optimization, which will offer novel and valuable insights for future research. 

% \begin{figure*}[htbp]
%     \centering
%     \includegraphics{figures/overview.eps}
%     \caption{Overview of \textsf{HyGNN}.}
%     \label{fig:overview}
% \end{figure*}

% \textbf{Roadmap}. 
% To summarize, we explore and utilize the distribution characteristics of graph adjacency matrices, and reformat the graphs to fully release the computing potential of hardware and further optimize the efficiency of GNN training systems. 

The rest of this paper is organized as follows. 
Section \ref{relatedworks} reviews the related work. 
Section \ref{sec:preliminaries} gives the preliminaries. 
Section \ref{sec:hybridspmm} presents the design of the hybrid SpMM kernel. 
Section \ref{sec:spmmintegration} describes the optimization of integrating the SpMM kernel into GNNs. 
Section \ref{sec:evaluation} exhibits the experimental results. 
We conclude the paper in Section \ref{sec:conclusion}. 
% \balance
\section{Related Work}
\label{relatedworks}

\noindent\textbf{SpMM Using CUDA Cores.} Optimization of SpMM has been a subject of extensive study \cite{buluc2008challenges, zhang2020sparch, hong2019adaptive, yang2018design, xia2023flash}. In recent years, Yang et al.~\cite{yang2018design} leveraged merge-based load balancing and row-major coalesced memory access strategies to accelerate SpMM. Hong et al.~\cite{hong2019adaptive} designed an adaptive tiling strategy to enhance the performance of SpMM. 
\textsf{cuSPARSE} \cite{refcusparse} offers high-performance SpMM kernels and has been integrated into numerous GNN training frameworks such as \textsf{DGL} \cite{wang2019deep}. 
Gale et al. point out that \textsf{cuSPARSE} is efficient only for matrices with sparsity exceeding 98\%. Consequently, they proposed \textsf{Sputnik} \cite{gale2020sparse} to accelerate the unstructured SpMM in deep neural networks and achieve state-of-the-art performance. Dai et al.~\cite{dai2022heuristic} introduced an approach capable of heuristically selecting suitable kernels based on input matrices. However, it demonstrates superior performance only when the matrix dimension is less than 32. Furthermore, Fan et al.~\cite{fan2023fast} proposed an SpMM kernel employing a unified hybrid parallel strategy of mixing nodes and edges, achieving efficient performance. To the best of our knowledge, there are few SpMM kernels specifically designed for graph computing workloads. Huang et al.~\cite{huang2020ge} introduced a GNN-specified SpMM kernel, utilizing coalesced row caching and coarse-grained warp merging to optimize memory access. However, it did not fully consider the characteristics of the graph adjacency matrix. 
Despite highly optimized algorithms, the computational capabilities of CUDA cores are inherently restricted by the hardware structures and are less efficient than Tensor cores, hindering the efficiency improvement in dense matrix multiplication. 

\noindent\textbf{SpMM Using Tensor Cores.} Numerous recent works have been shifted to accelerate SpMM with the assistance of Tensor cores \cite{chen2021efficient, li2022efficient, wang2022qgtc, wang2023tc, fan2024dtc}. However, most of them require structured input matrices, which is often impractical for adjacency matrices of real-world graphs. 
Alternatively, other works focus on unstructured SpMM and employ preprocessing methods to reduce the number of tiles needing traversal and to avoid unnecessary computation~\cite{zachariadis2020accelerating, wang2023tc, fan2024dtc}. For example, 
% \textsf{tSparse} \cite{zachariadis2020accelerating} skips the tiles with all zero elements and 
\textsf{TC-GNN}~\cite{wang2023tc} compresses all zero columns within a row window and \textsf{DTC-SpMM}~\cite{fan2024dtc} transforms the matrix into memory-efficient format named ME-TCF. Although \textsf{TC-GNN} implements a CUDA and Tensor cores collaboration design, it just employs CUDA cores for data loading, not for computing. Different from it, our proposed \textsf{HC-SpMM} employs both cores for computing according to their characteristics. Besides, Xue et al.~\cite{xue2023releasing} propose an unstructured SpMM kernel using Tensor cores, introducing a format named Tile-CSR to reduce the zero elements in submatrices traversed by Tensor cores. However, this kernel only supports half precision. As mentioned, Tensor cores are not natively suitable for unstructured SpMM. While preprocessing can densify submatrices, they still involve some computational waste and lead to suboptimal performance.

\noindent\textbf{GNN Training Frameworks.} Motivated by the remarkable performance but inefficient training of GNN, plenty of GNN training frameworks have been proposed, focusing on the graph locality~\cite{wang2021gnnadvisor, zhang2023ducati}, the GPU memory utilization~\cite{li2024daha}, etc. 
% Deep Graph Library (\textsf{DGL})~\cite{wang2019deep} and PyTorch Geometric (\textsf{PyG})~\cite{fey2019fast} stand out as comprehensive GNN training frameworks. 
% Additionally, \textsf{fuseGNN} \cite{chen2020fusegnn} employs diverse programming abstractions and kernel fusion to optimize the \textit{Aggragation} operation. 
% Additionally, \textsf{GNNAdvisor}~\cite{wang2021gnnadvisor} identifies the graph locality and leverages 2D workload management to accelerate the training. 
% \textsf{DUCATI}~\cite{zhang2023ducati} designs a novel dual-cache strategy to better utilize the locality of the adjacency matrix and the GPU memory. 
% \textsf{GNNLab}~\cite{yang2022gnnlab} presents an efficient graph sampling method for GNN training.
% \textsf{DAHA}~\cite{li2024daha} presents a GNN training framework with data and hardware-aware execution planning. 
Besides, various frameworks for distributed GNN training have emerged recently~\cite{wang2022neutronstar, peng2022sancus, zhang2023lotan, wan2023scalable, gunduz2022scalable, tang2024xgnn, min2021large, zhu2019aligraph}, which represent an orthogonal research direction. 
Our work focuses on optimizing SpMM and employing hybrid GPU cores for accelerating basic matrix multiplication, which can be seamlessly integrated into any aforementioned frameworks. This integration allows the \textit{Aggregation} phase to directly call the optimized SpMM kernel, significantly reducing the execution time of \textit{Aggregation}. 
Works addressing similar problems include \textsf{GE-SpMM}~\cite{huang2020ge} and \textsf{TC-GNN}~\cite{wang2023tc} mentioned above, both of which have integrated the kernels into GNN training frameworks. However, they solely employ CUDA or Tensor cores to perform SpMM operations in GNNs, failing to fully exploit the respective computational strengths of GPU cores. 

\noindent\textbf{Heterogeneous processing in data computing.} Some efforts are dedicated to employing heterogeneous systems for queries in databases~\cite{bogh2017template,chrysogelos2019hetexchange,rosenfeld2022query}, general computing tasks~\cite{hsu2023simultaneous} and deep learning~\cite{kwon2021heterogeneous,zheng2020flextensor}. Other research employs heterogeneous processing elements for SpMM~\cite{gerogiannis2024hottiles} and employs CUDA \& Tensor cores for general matrix multiplication~\cite{ho2022improving}. Our work focuses on accelerating SpMM by heterogeneous GPU cores, a topic that has barely been explored.
\section{Preliminaries}
\label{sec:preliminaries}

In this section, we provide a concise description of GPU architecture, including the characteristics of CUDA and Tensor cores, followed by an overview of GNNs. 

\subsection{GPU Architecture}
\label{subsec:gpus}
GPU is a highly parallel hardware comprising tens of Streaming Multiprocessors (SMs). Each SM features dedicated local memory, registers, and processing cores. These SMs individually schedule the execution of threads in warps, each consisting of 32 threads. Threads in a warp run simultaneously in a Single Instruction Multiple Threads (SIMT) fashion. A block consisting of multiple warps is allocated to an SM.  
% The Compute Unified Device Architecture (CUDA) provides an abstraction of the GPU's architecture and serves as a programming model. In CUDA, 

The GPU's memory hierarchy includes global memory, shared memory, and registers. Global memory, shared by all SMs, offers the largest capacity but lower I/O bandwidth. Each SM contains fast but limited shared memory, typically ranging from 16 KB to 64 KB. Additionally, each SM includes registers, which serve as the fastest storage structure.

Access to global memory within a warp is granular at 128 bytes when the L1 cache is enabled or 32 bytes otherwise. Consequently, when 32 threads within a warp request consecutive data within 128 bytes, only one memory access transaction is necessary, resulting in coalesced memory access. 
Shared memory is partitioned into multiple independent storage areas known as banks. Each bank can independently serve a thread in a single clock cycle. Concurrent access by multiple threads to the same bank results in a conflict, diminishing memory access efficiency. In shared memory, data is allocated across 32 consecutive banks, with each bank having a granularity of 4 bytes. For instance, when storing 64 numbers of float type in shared memory, the first 32 numbers and the last 32 numbers will be mapped to the 32 banks respectively. The 1st and 33rd numbers are assigned to the same bank, with the remaining numbers following a similar pattern. 

\subsection{Computing Cores in GPU}
\label{subsec:gpucores}

\begin{figure}[tbp]
    \centering
    \includegraphics[width=0.46\textwidth]{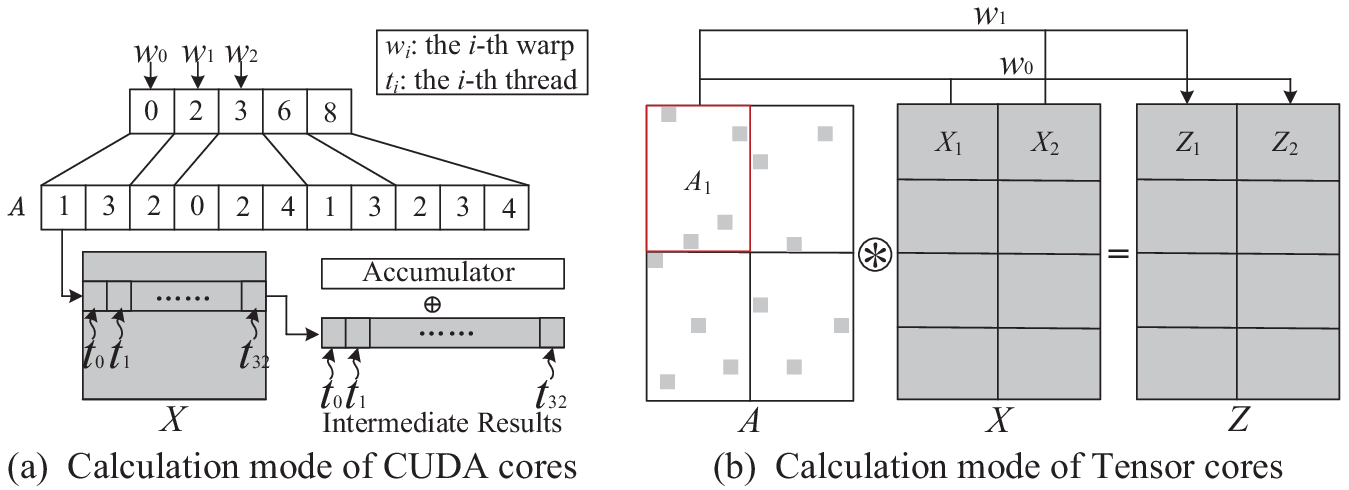}
    \vspace{-3mm}
    \caption{SpMM computing procedure of CUDA cores and Tensor cores.}
    \label{fig:tcuccu}
    \vspace{-6mm}
\end{figure}

\noindent\textbf{CUDA Cores.} CUDA cores are the primary computing cores used in GPUs. Each thread is assigned to a CUDA core for performing various calculations. Furthermore, CUDA cores can execute only one operation per clock cycle. Multiplying two 4×4 matrices involves 64 multiplications and 48 additions. 

\noindent\textbf{Tensor Cores.} Tensor cores, specifically designed for deep learning computation, have been integrated into advanced GPUs since 2017. Unlike CUDA cores, Tensor cores operate at the warp level, where threads within a warp collaboratively multiply two fixed-size matrices (e.g. 4$\times$ 4) in a single clock cycle. In addition to utilizing the matrix multiplication APIs provided by \textsf{cuBLAS}~\cite{refcublas}, Tensor cores can also be leveraged through the Warp Matrix Multiply-Accumulate ({\sf WMMA}) API for more flexible programming. The {\sf WMMA} API requires a fixed-size input matrix, with the required size varying depending on the data type. In this paper, we use TF-32 as the input data type of Tensor cores, following previous work \cite{wang2023tc}, which requires 16 $\times$ 8 $\times$ 16 as the input size. Although more efficient than CUDA cores in matrix multiplication, the fix-sized input requirement restricts the flexibility to avoid computing numerous zero elements in sparse adjacency matrices of graphs, limiting its efficiency in performing SpMM. 

Figure \ref{fig:tcuccu} illustrates the disparity between CUDA and Tensor cores during SpMM. $w_i$ represents the $i$-th warp and $t_j$ represents the $j$-th thread. As depicted in Figure \ref{fig:tcuccu}(a), each thread computes a single element in the result matrix, i.e., $Z$ in Equation \ref{aggregation}. CUDA cores possess the flexibility to efficiently skip zero elements in the sparse matrix and perform multiplication operations based on the CSR format. Conversely, Tensor cores conduct computations at the warp level as shown in Figure \ref{fig:tcuccu}(b). Each warp retrieves a submatrix from both the sparse matrix $A$ and the dense matrix $X$, subsequently computing the multiplication products. Despite the presence of numerous zeros in the sparse matrix, Tensor cores are unable to skip them, resulting in the waste of computational resources. 

Based on the characteristics outlined above, we can observe that while Tensor cores offer high efficiency in matrix multiplication, they lack flexibility compared to CUDA cores, particularly in handling sparse matrices.  
%For example, in a SpMM operation involving a matrix comprising 90\% zeros, CUDA cores only calculate the remaining 10\% non-zero elements, whereas 
Tensor cores process all elements indiscriminately due to their strict input criteria. 
Therefore, determining which cores are more efficient in the context of SpMM is challenging. This paper aims to devise a strategy for dynamically selecting the appropriate cores for submatrices within a sparse matrix for enhanced performance. 

% \vspace{-1mm}
% \subsection{Solution Overview}
% The overview of \textsf{HyGNN} is depicted in Figure \ref{fig:overview}. In detail, \S~\ref{sec:hybridspmm} presents the design of hybrid SpMM kernels, which leverages a logistic regression model to intelligently categorize submatrices and allocate them to the appropriate cores. In \S~\ref{sec:layoutoptimization}, we propose a graph layout reformat algorithm to improve the irregular and sparse graph layouts, which incurs a relatively small cost but renders the graph data more conducive to hybrid GPU cores. Furthermore, considering the entire training pipeline from a holistic perspective, we propose a kernel fusion method and refine the SpMM kernel in \S~\ref{overalconsideration} to bolster overall performance.

% \begin{figure*}[htbp]
%     \centering
%     \includegraphics{figures/overview.eps}
%     \vspace{-4mm}
%     \caption{Overview of \textsf{HyGNN}.}
%     \label{fig:overview}
% \end{figure*}

\subsection{Graph Neural Networks}
\label{subsec:gnns}

\begin{figure}[tbp]
    \centering
    \includegraphics[width=0.47\textwidth,height=0.22\textwidth]{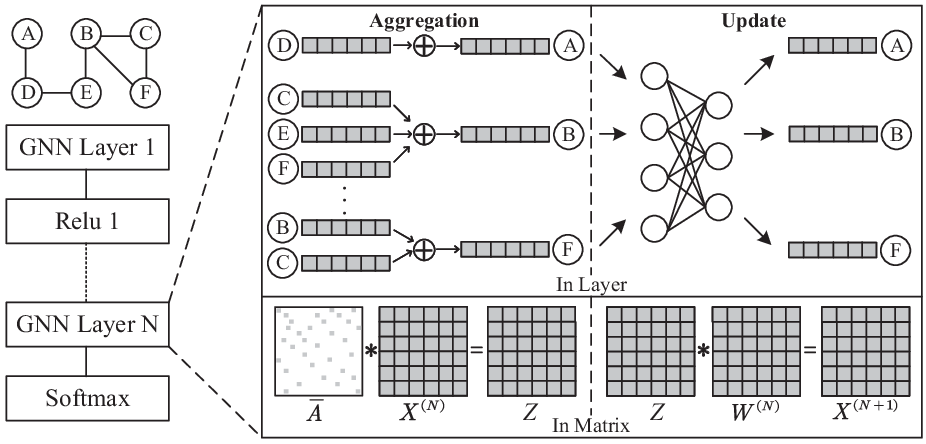}
    \vspace{-3mm}
    \caption{An example of GNN.}
    \label{fig:gnnexample}
    \vspace{-6mm}
\end{figure}

% Consider a graph $G=(V,E)$, where $V$ and $E$ are the sets of vertices and edges, respectively. Suppose the embedding matrix of vertices is $X_{|V|\times D}$ and the adjacency matrix of $G$ is $A_{|V|\times|V|}$, where $D$ is the embedding dimension. There are two main phases in a GNN layer, namely \textit{Aggregation} and \textit{Update}. An example of GNN is shown in Figure \ref{fig:gnnexample}. 
% The \textit{Aggregation} phase is about graph processing. In this phase, each node retrieves the feature vectors of its neighbors from $X$ and then aggregate them into a new vector. This process at layer $k$ can be formalized as a matrix multiplication operation, as shown in Equation \ref{aggregation}, where $\Bar{A}$ is calculated from $A$, and $Z$ is the intermediate result. The matrix $\Bar{A}$ is always sparse matrix so that the \textit{Aggregation} phase can be seen as an SpMM-like operation. 

In a graph $G=(V,E)$, where $V$ and $E$ represent the node set and edge set, 
respectively, the adjacency matrix of $G$ is denoted by $A_{|V|\times|V|}$. Each node has an embedding, which is a continuous vector representation with dimension $D$. The embedding encodes various properties of a node according to the downstream tasks. In GNNs, node embeddings are represented as matrix $X_{|V|\times D}$. 
Typically, a GNN layer consists of two fundamental operations: \textit{Aggregation} and \textit{Update}. 
An example of GNN is depicted in Figure \ref{fig:gnnexample}. 

In the \textit{Aggregation} operation, each node aggregates its feature vector and those of its neighbors from $X$ to form the new one. 
This process at layer $k$ can be formalized as a matrix multiplication operation, as depicted in Equation \ref{aggregation}, where $\Bar{A}$ is 
% the normalized adjacency matrix with self-loop, 
calculated from $A$, 
and $Z$ is the aggregated result. The matrix $\Bar{A}$ is always sparse, making the \textit{Aggregation} operation analogous to an SpMM-like operation. 
\vspace{-1mm}
\begin{equation}
\small
\vspace{-1mm}
\label{aggregation}
    Z = \Bar{A}X^{(k)}
\end{equation}

% The \textit{Update} phase is about neural network processing. This phase usually employs a neural network such as MLP to transform the intermediate vector of each node. 
% Suppose the network parameters is $W^{(k)}$ at layer $k$ and the output updated feature matrix is $X^{(k+1)}$. The \textit{Update} phase is formalized as Equation \ref{update}. This operation can be efficiently done by \textit{gemm} kernel in cuBLAS library. 

% The \textit{Update} operation involves neural processing, typically employing a neural network like Multilayer Perceptron which usually contains a simple matrix multiplication operation, to transform the aggregated vector of each node to an updated vector, serving as the new embedding vector for each node. 
Equation \ref{update} formalizes the \textit{Update} operation, where $W^{(k)}$ denotes the network parameters at layer $k$ and $X^{(k+1)}$ is the updated feature matrix. 
This operation can be efficiently executed using the \textit{gemm} kernel in {\sf cuBLAS}.
\vspace{-1mm}
\begin{equation}
\small
\vspace{-1mm}
\label{update}
    X^{(k+1)} = ZW^{(k)}
\end{equation}

Suppose the gradient is $X^{\prime (k+1)}$. Backward propagation also involves \textit{Aggregation} and \textit{Update}. The \textit{Update} phase in layer $k$ is formalized as Equation \ref{backward_update}, containing two \textit{gemm} operations.
\vspace{-2mm}
\begin{equation}
\vspace{-1mm}
\small
\label{backward_update}
    W^{\prime (k)} = Z^{T}X^{\prime (k+1)}; 
    Z^{\prime (k)} = X^{\prime (k+1)}W^{(k) T}
\end{equation}

The \textit{Aggregation} operation can be abstracted as an SpMM operation as Equation \ref{backward_aggre}. 
\vspace{-1mm}
\begin{equation}
% \vspace{-2mm}
\small
\label{backward_aggre}
    X^{\prime (k)} = \Bar{A}Z^{\prime (k)}
\end{equation}
%\vspace{-2mm}
\section{Hybrid SpMM Kernel}
\label{sec:hybridspmm}
% \balance
%In this section, we introduce our proposed SpMM kernel using hybrid GPU cores, which is the key component of \textsf{HyGNN}.
In this section, we justify our selection of row window as the fundamental hybrid unit, contrasting it with the straightforward choice of submatrices. Additionally, we delve into the essential characteristics that influence the efficiency of GPU cores, serving as the driving force behind our subsequent designs.

\vspace{-2mm}
\subsection{Combination Strategy}
\label{combinationstrategy}
\begin{figure}[tbp]
    \centering
    \includegraphics[width=0.47\textwidth]{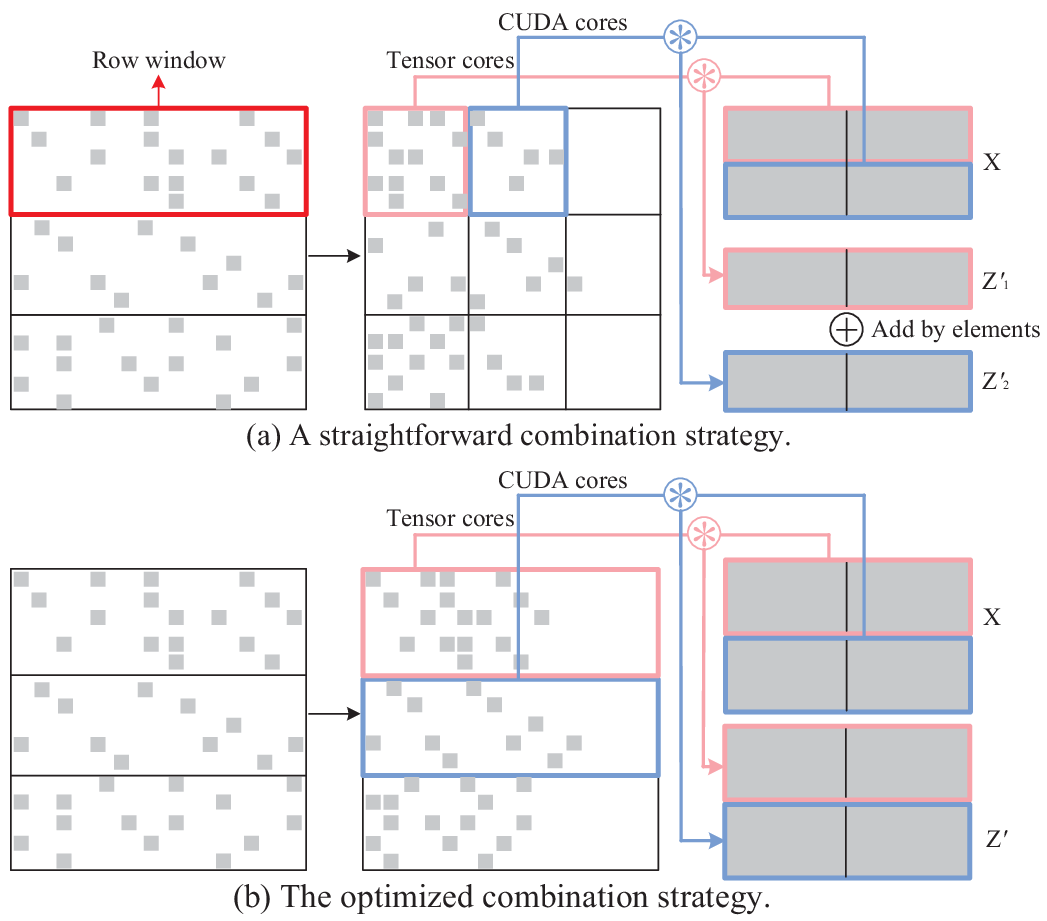}
    \vspace{-3mm}
    \caption{Combination strategies for SpMM on CUDA and Tensor cores.}
    \label{fig:combstra}
    \vspace{-6mm}
\end{figure}

Initially, we need to consider how to combine the two GPU cores in a sparse matrix computation, i.e., the partitioning of the sparse matrix into submatrices and the processing strategy of each submatrix using different GPU cores. A straightforward strategy is illustrated in Figure \ref{fig:combstra}(a). Due to the fixed-size input requirement of the {\sf WMMA} API, it divides the input matrix into submatrices sized $16 \times \#nodes$, named \textit{row windows}. Within each row window, columns are rearranged based on the count of non-zero elements. Intuitively, the first few columns within a row window tend to be denser, while subsequent ones are sparser. It further splits each row window into $16\times 8$ submatrices and traverses the matrix in units of $16\times8$ which is the minimum granularity required by {\sf WMMA} API for input. Each $16\times8$ submatrix is assigned to appropriate GPU cores based on our identified characteristics (\S~\ref{adaptivecoreselection}). 
% In a row window, the initial units which intend to be denser are designated for calculation by Tensor cores, while the subsequent sparser units are assigned to CUDA cores. 

While the straightforward strategy offers fine-grained traversal, upon closer examination, it reveals several limitations. 
{\bf (1)} Merging the results from CUDA and Tensor cores introduces extra overhead. 
Results computed by Tensor cores are initially buffered in registers before being transferred to shared or global memory, while those computed by CUDA cores are directly stored in shared or global memory. Within each row window, the merging of results from Tensor and CUDA cores necessitates additional I/O and adds to the overhead of addition operations\footnote{The overhead is up to 31\%, which is unacceptable.}.
{\bf (2)} Hybrid computation within a row window necessitates the separate storage of edges for Tensor cores and CUDA cores, resulting in increased preprocessing overhead and poor access locality.  
{\bf (3)} The execution time in the granularity of $16\times 8\times16$ matrix multiplication complicates accurate measurement\footnote{Execution time remains in microseconds in this granularity, whose tendency regarding sparsity is not easily visible.}. Furthermore, sparsity is the only characteristic that can be leveraged for core selection in $16\times 8$ submatrices. 
These constraints hinder a comprehensive analysis of the relationship between GPU cores and matrix characteristics.

To tackle the aforementioned issues while maintaining fine granularity, instead of using a $16\times8$ submatrix, we adopt the row window as the minimum hybrid unit. 
Following a strategy similar to \textsf{TC-GNN} \cite{wang2023tc}, we position non-zero columns at the forefront of row windows, thereby density the submatrices, i.e., the row windows.  
As depicted in Figure \ref{fig:combstra}(b), within each row window, Tensor cores execute across all 16$\times$8 submatrices, while CUDA cores compute directly using the CSR format. 
This approach eliminates the need for merging the results from CUDA and Tensor cores, reducing the extra computation overhead. 
Edges within each row window can be stored consecutively in this granularity, which ensures the access locality of edges. 
The sufficient duration of execution in the scale of row windows aids in the analysis of appropriate GPU core selection. Another pivotal characteristic in this granularity, the number of non-zero columns, stands out as a reference for better core allocation (\S~\ref{adaptivecoreselection}). 
%This combination strategy also simplifies the parallel computation of the matrix. 

\subsection{Computing Characteristics of GPU Cores}
\label{computingchar}
In this subsection, we identify key features of row windows that impact the GPU cores' performance and quantify the difference in computing characteristics between different cores. 

\setlength{\textfloatsep}{0pt}
\begin{algorithm}[tbp]
    \caption{SpMM on CUDA cores}
    \label{spmm_ccu}
    \LinesNumbered
    \KwIn{rowPtr, colInd, val and dense matrix $X$.}
    \KwOut{The result of SpMM $Z$.}
    \For{$i=0$ to $m-1$ in parallel}{
        \For{$j=0$ to $n-1$ in parallel}{
            $res \leftarrow 0$;\\
            \For{$k=rowPtr[i]$ to $rowPtr[i+1]$}{
                $res += val[k]*X[colInd[k],j]$;\\
            }
            $Z[i,j]=res$;\\
        }
    }
\end{algorithm}

\makeatletter
% Remove right hand margin in algorithm
\patchcmd{\@algocf@start}% <cmd>
  {-1.5em}% <search>
  {0pt}% <replace>
  {}{}% <success><failure>
\makeatother
\begin{algorithm}[tbp]
    \caption{SpMM on Tensor cores}
    \label{spmm_tcu}
    \LinesNumbered
    \KwIn{rowPtr, colInd, val and dense matrix $X$.}
    \KwOut{The result of SpMM $Z$.}
    $\_\_shared\_\_$ $ASh[16*8], XSh[WarpSize][8*embDim]$;\\
    \ForEach{\textit{row window} in parallel}{
        \ForEach{$16\times8$ block in \textit{row window}}{
            Convert CSR into $ASh$ in matrix format;\\
            Load submatrices of $X$ into $XSh$;\\
            Each warp calculates $ASh \times XSh$ using Tensor cores;\\
        }
        Store the result into $Z$;
    }
\end{algorithm}
\setlength{\textfloatsep}{.55\baselineskip plus  0.2\baselineskip minus  0.4\baselineskip}

%We begin by providing an overview of how CUDA cores and Tensor cores handle computations to clarify the relationship. 
The algorithms presented in Algorithm \ref{spmm_ccu} and \ref{spmm_tcu} highlight the distinct computational strategies employed by CUDA and Tensor cores in SpMM. 
CUDA cores operate in parallel to compute elements in the result matrix $Z$, utilizing the CSR format to efficiently skip zeros in $A$. On the other hand, the process for Tensor cores, adhering to the specifications of the {\sf WMMA} API, necessitates retrieving a $16\times8$ submatrix from $A$ and an $8\times16$ submatrix from $X$, followed by caching these submatrices in shared memory for subsequent loading and computation using the {\sf WMMA} API.
The computational cost of SpMM on CUDA cores is primarily influenced by the \textit{number of non-zero elements} in the sparse matrix. In contrast, for Tensor cores, the computational cost is tied to the \textit{quantity of $16\times8$ blocks}.
To validate these observations, we conducted a series of experiments to evaluate the impact of these two metrics on the performance of GPU cores.

\noindent\textbf{Sparsity of the Input Matrix.} The sparsity of a sparse matrix is reflected in the number of zero elements it contains. To explore the relationship between the performance of GPU cores and sparsity, we conducted a series of evaluations following the strategy outlined in \S~\ref{combinationstrategy}. 
We specify the size of a row window as $16\times32$ and the dense matrix dimension as $32$. Sparse matrices with varying degrees of sparsity were generated for evaluation. 
The execution times of different GPU cores are visualized in Figure \ref{executiontime}(a). 
Interestingly, the execution times of Tensor cores remained stable as sparsity increased. This can be attributed to the fixed number of $16\times8$ submatrices, resulting in relatively consistent execution times for lines 4-6 in Algorithm \ref{spmm_tcu}.
Conversely, the execution times of CUDA cores decrease with higher sparsity, surpassing Tensor cores when sparsity exceeds 83\%. This observation suggests a positive correlation between the computational costs of CUDA cores and the number of non-zero elements in sparse matrices.

\begin{table}[tbp]
    \centering
    \caption{Computing and memory access costs of GPU cores in SpMM\protect\footnotemark[6].}
    \vspace{-2mm}
    \resizebox{\linewidth}{!}{
    \begin{tabular}{|c||c|c|c|c|c|c|}
        \hline
        \textbf{Datasets} & \textbf{C-m} & \textbf{C-c} & \textbf{m/c(C)} & \textbf{T-m} & \textbf{T-c} & \textbf{m/c(T)} \\ \hline
        \hline
        \textit{DD} & 5.04 & 7.13 & \textbf{0.71} & 15.15 & 11.13 & \textbf{1.36} \\
        \hline
        \textit{YS} & 25.09 & 31.77 & \textbf{0.79} & 48.41 & 21.14 & \textbf{2.29} \\
        \hline
        \textit{RD} & 71.16 & 82.84 & \textbf{0.86} & 130.44 & 55.10 & \textbf{2.37} \\
        \hline
        % \multicolumn{7}{l}{\small C: CUDA cores, T: Tensor cores, m: memory access cost, c: computing cost, m/c: the quotient of memory access cost and computing cost, Units: $10^{-2}$ ms.}\\
    \end{tabular}}
    \label{tab:computing_memory_costs}
    \vspace{-2mm}
\end{table} 
\footnotetext[6]{C: CUDA core, T: Tensor core, m: memory access cost, c: computing cost, m/c: the quotient of memory access and computing cost, Units: $10^{-2}$ ms. }
\noindent\textbf{Non-zero Columns of the Input Matrix.} 
% The number of $16\times8$ blocks is a coarse-grained metric which is unable to accurately depict the relationship with the performance of GPU cores. To refine this metric, we adopt the number of non-zero columns as a more granular substitute. 
The number of non-zero columns is also a pivotal characteristic of the performance of both GPU cores. 
To evaluate this characteristic, we maintain a constant level of sparsity while varying the number of non-zero columns in a row window. Figure \ref{executiontime}(b) depicts the execution times of different GPU cores under these conditions. 
We observe distinct behaviors: CUDA cores demonstrate relatively consistent computational costs as the number of non-zero columns rises, while the computational costs of Tensor cores increase.
This discrepancy arises because, for CUDA cores, the bottleneck lies in computation. With the constant sparsity, the increase in the number of non-zero columns will not bring about a large increase in calculations, leading to relatively consistent computational costs of CUDA cores. For Tensor cores, they are unable to skip zero elements in adjacent matrices due to the input requirements, thereby triggering more data loading as the number of non-zero columns increases. However, the bottleneck of Tensor cores exactly lies in the loading of $X$. Experimental results indicate that the loading time for $X$ is about 2$\times$ longer than the time spent on multiplication, constituting more than 60\% of the total execution time. As a result, the significant increase in Tensor cores' computational costs is attributed to the memory access overhead. As the number of non-zero columns increases, more memory access is required to load $X$ for Tensor cores, resulting in low efficiency.
% Different from the SpMM kernel on CUDA cores, the input requirements of the WMMA API restricts the ability to achieve global memory coalesced access and eliminates the shared memory bank conflict simultaneously. Therefore, the memory access cost of Tensor cores exceeds that of CUDA cores. As a result, as the number of non-zero columns increases, more memory access is required to load $X$, resulting in a sharp increase in the execution time of Tensor cores.

Other factors such as the distribution of non-zero elements within sparse matrices may also influence the performance of GPU cores. However, their impact is considerably insignificant\footnote[7]{The variation of the execution time is less than 10\%.} compared to the aforementioned two characteristics, therefore we choose to disregard them.
To further discover the difference between CUDA cores and Tensor cores, we evaluate the computing and memory access costs of both types of GPU cores in SpMM operation, which is reported in Table \ref{tab:computing_memory_costs}. The data loading of CUDA cores is faster than the calculation, while it is exactly the opposite for Tensor cores. As a result, CUDA cores are optimal for tasks that are \textit{\textbf{memory access-intensive}}, while Tensor cores excel in \textit{\textbf{computation-intensive}} tasks. 
Our identified key characteristics, sparsity and the number of non-zero columns, exactly govern \textit{computation} and \textit{memory access}, respectively. These two characteristics offer sufficient information to select the appropriate cores, resulting in a high accuracy reported in the next paragraph. 
The row windows with low sparsity and a small number of non-zero columns require more computation overhead and less memory access overhead, which are suitable for Tensor cores. Otherwise, we opt for CUDA cores.  
%From the preceding discussion, we can infer that the row windows with high sparsity and a large number of non-zero columns require less computation overhead and more memory access overhead, which are suitable for CUDA cores. Conversely, row windows with low sparsity and small number of non-zero columns require more computation overhead and less memory access overhead, which are suitable for Tensor cores. 
% CUDA cores are better suited for processing row windows with high sparsity and a large number of non-zero columns, whereas Tensor cores are more suitable for row windows with low sparsity and a small number of non-zero columns. 
%In summary, CUDA cores are optimal for computations that are \textit{\textbf{memory access-intensive}}, while Tensor cores excel in \textit{\textbf{computation-intensive}} tasks.

\subsection{Adaptive Core Selection}
\label{adaptivecoreselection}
We train a logistic regression model to determine the appropriate GPU cores for matrix multiplication of a row window, based on the performance of both GPU cores on synthetic sparse matrices with diverse sparsity and non-zero column counts. This model takes the sparsity and the number of non-zero columns as inputs and predicts the appropriate GPU cores. 
%Once trained, the model can be universally applied to any dataset without alteration, provided the GPU remains unchanged. 
%The adoption of a logistic regression model to determine the appropriate GPU cores offers numerous benefits. Firstly, only two main characteristics are the model inputs, making a concise classification process. Secondly, 
The logistic regression model is lightweight and efficient, 
%, comprising only three parameters, 
allowing for rapid computation. 
%Taking into account these factors, we opt for the logistic regression model over alternative models. 
It can complete the selection of GPU cores for a row window in a few nanoseconds, with an accuracy greater than 90\%. 
% More details of the logistic regression model are provided in our technical report \cite{technicalreport}.

The training pipeline of the model consists of 4 procedures: (1) sparse matrices generating, (2) execution results collecting, (3) model training, and (4) model encoding. 

\noindent\textbf{Sparse matrices generating.} We begin by generating a set of matrices for evaluating GPU cores. The logistics regression model is used to identify the appropriate GPU cores for each row window with 16 rows. Therefore, the number of rows in the generated matrices is set to 16. The number of columns ranges from 1 to 130\footnote[8]{The maximum number of non-zero columns is set to 130 because the average degree of most graphs is less than 8. If all nodes within a row window have no common neighbors, the number of non-zero columns is $16\times8=128$. Setting the maximum at 130 accommodates most cases.}. Each column contains at least one non-zero element. The matrix sparsity ranges from $\frac{1}{16}$ to $\frac{15}{16}$\footnote[9]{The maximum sparsity is set to $\frac{15}{16}$ to ensure each column has at least one non-zero element. The minimum sparsity is $\frac{1}{16}$ because the observed sparsity of most row windows is greater than 0.06. Lower minimum sparsity is optional for higher accuracy according to user needs.}, with the corresponding number of non-zero elements ranging from $\#columns$ to $\#columns\times 15$. To ensure each column contains at least one non-zero element, we first generate an element per column, with the row position determined by a uniform random number function. Subsequently, the row and column positions of the remaining elements are generated randomly using the same function. 

\noindent\textbf{Execution results collecting.} Using the matrices generated in the previous step, we execute the SpMM kernel on Tensor cores and CUDA cores, respectively. For each matrix, we perform 100 executions to evaluate the average runtime. It's important to note that the kernels used are identical to the deployed SpMM kernel, with the same parameter settings.

\noindent\textbf{Model training.} We use the logistics regression model API from Sklearn. Each matrix in the previous step serves as a training sample characterized by two features: sparsity and the number of non-zero columns. 
% These two features dominate the computing and memory access respectively, which represent the primary difference between CUDA and Tensor cores. 
For each sample, we have measured the execution times on Tensor and CUDA cores, respectively. If the execution time of CUDA cores is shorter than that of Tensor cores, the sample is labeled as 1; otherwise, it's labeled as 0. Then we train the model until it converges. 

\noindent\textbf{Model encoding.} Finally, we extract the coefficients from the logistics regression model and hard-code them into the cores selection function. The classification is performed during the preprocessing step, with the classification results for each row window stored in a boolean array. In this array, 0 represents the selection of CUDA cores and 1 denotes Tensor cores. During SpMM computation, the kernel refers to the boolean array to assign the appropriate GPU cores for calculation. 

The model is trained offline using synthetic data. Once trained, it can be reused for inference without retraining, provided the GPU architecture and precision remain unchanged. For a new dataset (sparse matrix), the classification overhead is minimal, as it only involves computing $w_1*x_1+w_2*x_2+b$, which can be completed within a few nanoseconds. 

\subsection{Kernel Optimizations}
In this subsection, we further optimize the SpMM kernel considering the thread collaboration mode and memory access pattern to achieve higher efficiency.

\subsubsection{Optimizations on CUDA Cores}
\label{bottleneckofccu}

\begin{figure}[tb]
    \centering
    \includegraphics[width=0.48\textwidth]{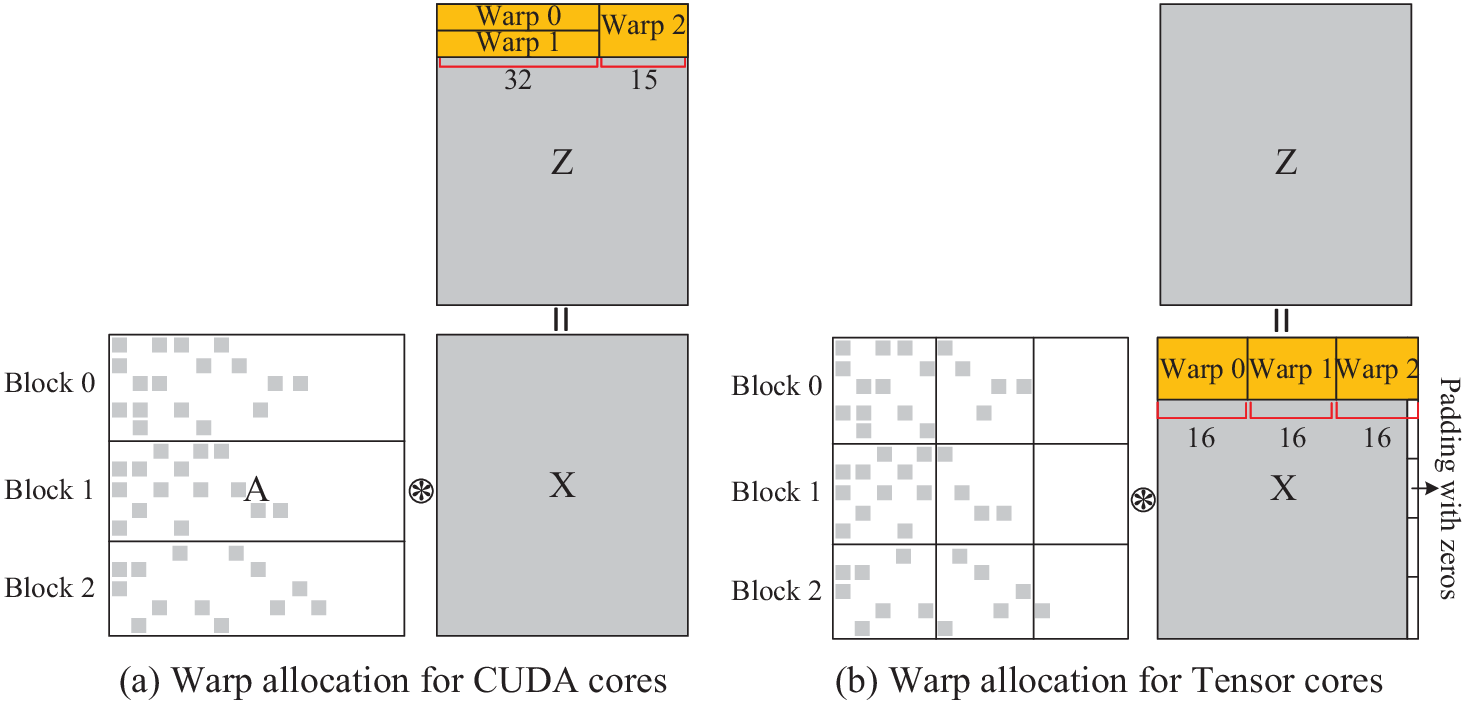}
    \vspace{-3mm}
    \caption{Warp allocation strategies on different GPU cores.}
    \label{fig:parallelstrategy}
    \vspace{-2mm}
\end{figure}

\begin{figure*}[htb]
    \centering
    \includegraphics[width=0.95\textwidth]{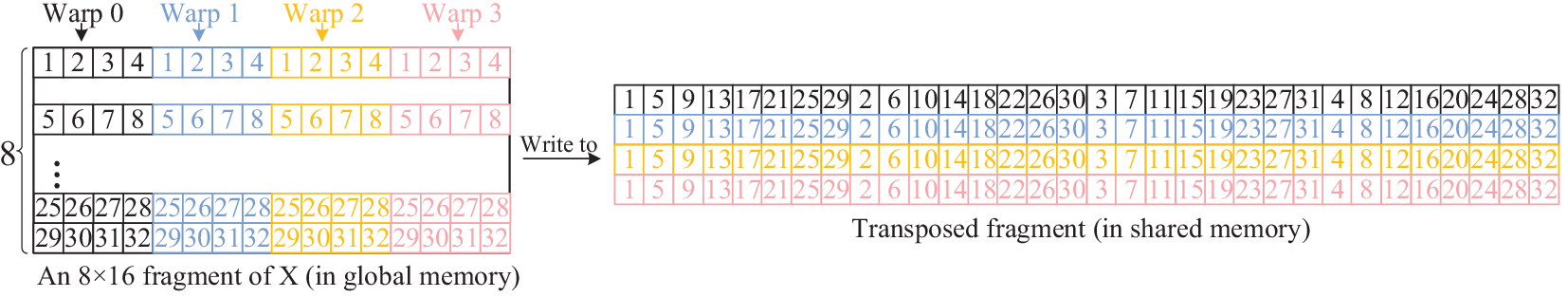}
    \vspace{-4mm}
    \caption{Data loading strategy for the dense matrix $X$. The $8\times 16$ submatrix is transposed during loaded into shared memory.}
    \label{fig:memoryaccess}
    \vspace{-4mm}
\end{figure*}

In the SpMM kernel designed for CUDA cores, we utilize multiple warps to concurrently compute the results of a row window. For simplicity, let's assume a dense matrix dimension of 32, which we will generalize in the subsequent discussion. Each warp within a thread block is allocated a row from the sparse matrix. Within each warp, each thread is responsible for computing one element in the result matrix $Z$. This approach guarantees coalesced memory access when retrieving elements from the dense matrix stored in global memory.

\noindent\textbf{Generalization.} In scenarios where the dense matrix dimension is not a multiple of 32, using a warp to compute one row may lead to inefficient thread utilization. For instance, with a dimension of 47, a warp requires two iterations over a row, leaving 17 idle threads during the second iteration. 
To address this inefficiency, we develop an adaptive kernel that can handle varying dimensions. This kernel utilizes different numbers of threads to calculate a row, such as 16 or 8. When employing 16 threads, for example, a warp is utilized to compute 2 rows of the result matrix, as depicted in Figure \ref{fig:parallelstrategy}(a).

\noindent\textbf{Memory Management.} Optimizing memory access in the SpMM kernel on CUDA cores commonly involves storing the dense submatrix in shared memory due to its frequent usage. However, our observations indicate that loading either all data or only the frequently accessed (``hot'') data into shared memory is not as efficient as directly accessing it from global memory for GNN-tailored SpMM\footnote[10]{Loading all the data including the sparse and dense matrices will result in a performance loss of up to 9\%, while only loading hot data has the similar performance to directly accessing it from the global memory.}, even with vectorized data loading techniques. This inefficiency stems from modern GPUs' efficient cache management mechanisms, which can effectively handle such scenarios. Therefore, we opt to load only the edge indices in CSR into shared memory. Threads within a warp accessing identical addresses when calculating results can trigger a broadcast mechanism in global memory, leading to time-consuming operations. By loading this data into shared memory, we circumvent this overhead.

\setlength{\textfloatsep}{0pt}
\begin{algorithm}[tbp]
    \caption{Optimized SpMM on CUDA cores}
    \label{opt_spmm_ccu}
    \LinesNumbered
    \KwIn{rowPtr, colInd, val and dense matrix $X$.}
    \KwOut{The result of SpMM $Z$.}
    \tcp{Memory Management}
    \_\_shared\_\_ TmpEdges[MaxSize],TmpVals[MaxSize];\\
    \For{$i=0$ to $m-1$ in parallel}{
        \For{$j=rowPtr[i]$ to $rowPtr[i+1]$ in parallel}{
            $TmpEdges[j]\leftarrow colIdx[j]$;\\
            $TmpVals[j]\leftarrow val[j]$;\\
        }
    }
    \tcp{Generalization}
    $AlignN\leftarrow dim/32$;\\
    \For{$r=0$ to $AlignN$}{
        \For{$i=0$ to $m-1$ in parallel}{
            \For{$j=0$ to $n-1$ in parallel}{
                $res \leftarrow 0$;\\
                \For{$k=rowPtr[i]$ to $rowPtr[i+1]$}{
                    $res += TmpVals[k]*X[TmpEdges[k],j]$;\\
                }
                $Z[i,j]=res$;\\
            }
        }
    }
    \tcp{Process the unaligned dimension of $X$}
    \For{$(i=0; i < m-1; i+=2)$ in parallel}{
        \For{$j=AlignN*32$ to $dim$ in parallel}{
            $res \leftarrow 0$;\\
            \For{$k=rowPtr[i]$ to $rowPtr[i+1]$}{
                $res += TmpVals[k]*X[TmpEdges[k],j]$;\\
            }
            $Z[i,j]=res$;\\
        }
    }
    \Return{Z};
\end{algorithm}
\setlength{\textfloatsep}{.55\baselineskip plus  0.2\baselineskip minus  0.4\baselineskip}

In Algorithm \ref{opt_spmm_ccu}, we provide the implementation details of the optimized SpMM kernel on CUDA cores. The memory management strategy is implemented in lines 1-5. We employ all threads to load the $colIdx$ and $val$ to shared memory in parallel. Lines 6-19 implement the strategy to handle general dense matrix dimensions besides 32. $dim$ in line 6 denotes the dimension of the dense matrix $X$. In lines 7-13, we first calculate the parts that can align with 32 like the processing parts of warp 0 and warp 1 in Figure \ref{fig:parallelstrategy}. The remaining parts are calculated in lines 14-19, with each warp processing 2 rows like warp 2 in Figure \ref{fig:parallelstrategy}. 

\subsubsection{Optimizations on Tensor Cores}  
\label{bootleneckTCU}
To enhance data loading efficiency for SpMM on Tensor cores, we address the bottleneck of loading data from the dense matrix by employing all warps within a block to cooperatively load data into shared memory. This strategy allows data from the dense matrix to be loaded in units of $8\times16$ matrices, aligning with the input constraints of the {\sf WMMA} API.

Specifically, as illustrated in Figure \ref{fig:memoryaccess}, rather than retrieving 2 or 1 row, a warp is utilized to fetch 8 rows of an $8\times 16$ matrix\footnote[11]{For Tensor cores, we split the matrix into $8\times 16$ submatrices according to the input requirement of {\sf WMMA} API.}, with each row containing 4 elements for a warp. This approach ensures that threads within the warp can write these retrieved elements into distinct banks of shared memory, thereby preventing bank conflicts and eliminating extra overhead. In shared memory, the data is organized in a one-dimension format. However, for visualization purposes and space limitations, we display the data in a $4\times 32$ layout. 

Additionally, we adopt the parallel strategy for Tensor cores presented in \cite{wang2023tc}, which is illustrated in Figure \ref{fig:parallelstrategy}(b). Following collaboratively loading the required data from $X$ by warps, each warp is assigned an $8\times16$ submatrix to independently compute the results using Tensor cores concurrently. With each block responsible for calculating a row window, this method effectively harnesses the parallel processing capabilities of GPU and enhances computational efficiency.

\setlength{\textfloatsep}{0pt}
\makeatletter
% Remove right hand margin in algorithm
\patchcmd{\@algocf@start}% <cmd>
  {-1.5em}% <search>
  {0pt}% <replace>
  {}{}% <success><failure>
\makeatother
\begin{algorithm}[tbp]
    \caption{Optimized SpMM on Tensor cores}
    \label{opt_spmm_tcu}
    \LinesNumbered
    \KwIn{rowPtr, colIdx, rowIdx, val and dense matrix $X$.}
    \KwOut{The result of SpMM $Z$.}
    $\_\_shared\_\_$ $ASh[16*8], XSh[WarpSize][8*embDim]$;\\
    \ForEach{\textit{row window} in parallel}{
        \ForEach{$16\times8$ block in \textit{row window}}{
            \For{$i = rowPtr[block*16]$ to $rowPtr[(block+1)*16]$ in parallel}{
                $ASh[colIdx[i]\%8+(rowIdx[i]\%16)*8]\leftarrow val[i]$;\\
            }
            \For{$idx = warpId; idx < dim/4; idx += warpNum$}{
                $denseRowId\leftarrow$calculate the row idx;\\
                $srcIdx\leftarrow denseRowId*dim+threadIdx\%4+idx*4$;\\
                $dstIdx\leftarrow warpId*32+(threadIdx\%4)*8 + (threadIdx/4)$;\\
                $XSh[dstIdx]\leftarrow X[srcIdx]$;\\
            }
            Each warp calculates $ASh \times XSh$ using Tensor cores;\\
        }
        Store the result into $Z$;
    }
    \Return{Z};
\end{algorithm}
\setlength{\textfloatsep}{.55\baselineskip plus  0.2\baselineskip minus  0.4\baselineskip}

In lines 6-10 of Algorithm \ref{opt_spmm_tcu}, we present the implementation of data loading strategy in Figure \ref{fig:memoryaccess}. It's worth noting that the parallel data loading strategy is designed for the retrieval of the dense matrix. For loading CSR entries, we follow the strategy in {\sf TC-GNN}. Line 6 adopts the parallel strategy depicted in Figure \ref{fig:memoryaccess}, with each warp performing the data loading of 4 elements within each row. It calculates the index of data to load in $X$, denoted as $srcIdx$ and the index to write, denoted by $dstIdx$ in lines 8-9. Data is loaded to $XSh$ according to the calculated indices in line 10. More details can be found in our source code. 

\textbf{Discussion.} The distinction between {\sf HC-SpMM} and existing SpMM methods is twofold. First, existing techniques solely rely on CUDA or Tensor cores for calculation, which leads to suboptimal performance. In contrast, {\sf HC-SpMM} is an SpMM method specifically designed for the irregular distribution of graphs, adaptively selecting the proper cores according to the characteristics of submatrices, which achieves better performance. Second, for both CUDA and Tensor cores, data loading is one of the bottlenecks. {\sf HC-SpMM} optimizes the data loading for CUDA and Tensor cores by reasonably utilizing shared memory and multi-thread, which are not considered in existing methods for Tensor cores. 
\section{SpMM Integration into GNN Training}
\label{sec:spmmintegration}
% \balance
\begin{figure*}[htb]
    \centering
    \includegraphics[width=0.96\textwidth]{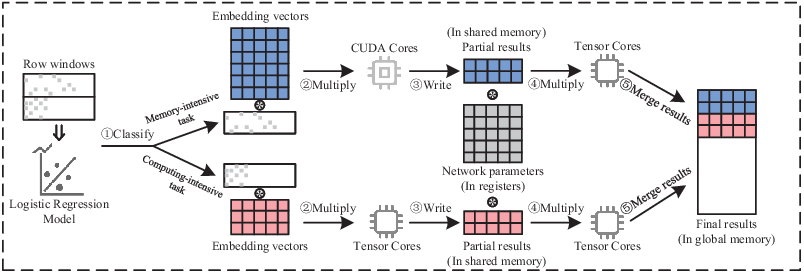}
    \vspace{-3mm}
    \caption{Process in a GNN layer with kernel fusion.}
    \label{fig:kernelfusion}
    \vspace{-6mm}
\end{figure*}

In this section, we integrate {\sf HC-SpMM} into the GNN training pipeline and propose a series of optimization techniques from system and graph data management perspectives to enhance the GNN training efficiency.  

\subsection{Kernel Fusion}
\label{subsec:kernelfusion}
% \balance
To integrate {\sf HC-SpMM} into the GNN training pipeline and achieve higher efficiency, we rethink the training pipeline and propose a novel kernel fusion strategy. Existing implementations, such as \textsf{DGL} \cite{wang2019deep} and \textsf{TC-GNN} \cite{wang2023tc}, typically treat the \textit{Aggregation} and \textit{Update} phases independently, leading to increased kernel launch overhead and additional memory access overhead. 
During the \textit{Aggregation} phase, results are written to global memory; then, during the \textit{Update} phase, they are read from the same address of global memory. Moreover, the separate launching of the \textit{Update} and \textit{Aggregation} kernels contributes to significant time overhead\footnote[12]{The launch time of a single matrix multiplication kernel is measured to be around 0.03ms. }. 

Upon analyzing the allocation of row windows to thread blocks, we recognize an opportunity to optimize performance by fusing the \textit{Update} and \textit{Aggregation} kernels. 
Kernel fusion proves particularly effective when the \textit{Update} phase directly follows the \textit{Aggregation} phase, as seen in the backward propagation of GCN \cite{kipf2016semi} and the forward propagation of GIN \cite{xu2018powerful}. 
In such scenarios, the data required for the \textit{Aggregation} phase is readily available and consistent, facilitating efficient fusion. Otherwise, implementing kernel fusion becomes complex due to the varying and partially overlapping embedding vectors needed for each row window, making it challenging to determine and calculate the precise embedding vectors required for the \textit{Aggregation} phase during the \textit{Update} phase. 
Fortunately, during the forward and backward propagation stages, there is always one stage where the \textit{Update} phase follows the \textit{Aggregation} phase, enabling effective kernel fusion.

% As depicted in Figure \ref{fig:kernelfusion}, in forward propagation with kernel fusion, the computed results for row windows are stored in shared memory rather than global memory, serving as the input for the \textit{Update} phase. Following this, the data, along with the network parameters, is loaded into registers. Tensor cores are then used to compute the partial results of matrix $Z$, which are stored in the global memory. 
As depicted in Figure \ref{fig:kernelfusion}, in forward propagation with kernel fusion, the row windows within the matrix are \textcircled{1}classified by the logistic regression model and distributed to the corresponding GPU cores. The GPU cores \textcircled{2}multiply the row windows with the corresponding embedding vectors and \textcircled{3}write the intermediate results into shared memory. Following this, Tensor cores \textcircled{4}multiply the results stored in shared memory with the network parameters used during \textit{Update} phase, and \textcircled{5}store the final results in the global memory. 
During backward propagation, the process is similar, with the addition of computing $W^{\prime (k)}$ in Equation \ref{backward_update}. To calculate $W^{\prime (k)}=Z^TX^{\prime (k+1)}$, the submatrix result in shared memory is transposed first, as $X^{\prime (k+1)}$ is the right-hand side of the equation. Subsequently, Tensor cores compute the final results. The kernel fusion approach consolidates the number of kernels from two to one, thereby reducing the kernel launch overhead. Additionally, storing results in shared memory rather than global memory significantly alleviates memory access overhead. 

\subsection{Layout Optimization}
\label{subsec:layout}

\begin{figure}[tbp]
    \centering
    \includegraphics[width=0.5\textwidth]{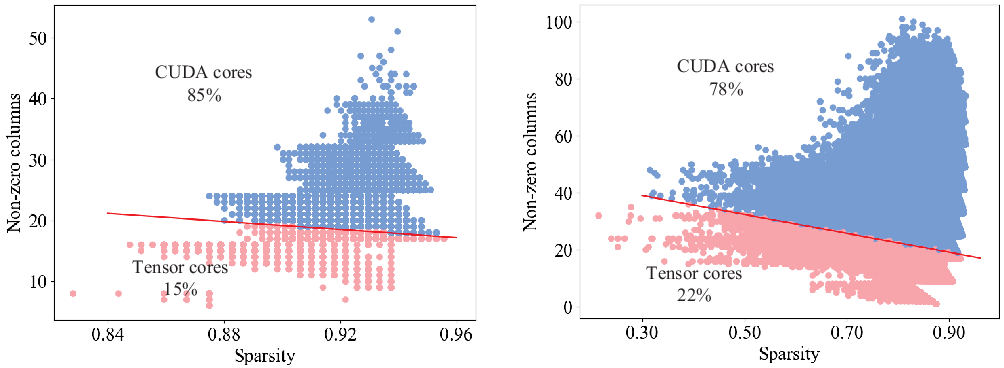}
    \vspace{-8mm}
    \caption{Sparsity of row windows in representative graphs.}
    %\vspace{-6mm}
    \label{fig:layoutdis}
    \vspace{-2mm}
\end{figure}

As discussed in \S~\ref{adaptivecoreselection}, CUDA cores are ideally suited for memory access-intensive row windows, while Tensor cores excel in computing-intensive row ones. 
Real-world graphs typically exhibit row windows with numerous non-zero columns but few non-zero elements in each of them. To quantify this, we compute the sparsity and number of non-zero columns for row windows within two representative datasets, as depicted in Figure \ref{fig:layoutdis}. 
The red line denotes the classification boundary learned by the logistic regression model. Only 15\% and 22\% of row windows are deemed appropriate for Tensor cores in the two datasets, respectively. Given that Tensor cores are introduced to efficiently handle computations, which CUDA cores may struggle with, a low proportion of suitable row windows restricts the potential for acceleration.
%Recall that the motivation behind introducing Tensor cores is to perform calculations efficiently, which CUDA cores cannot achieve. 
%If the proportion of row windows suitable for Tensor cores is minimal, the potential for acceleration will be limited. 
In response, we propose a novel algorithm named \textsf{LOA} in this section to reformat the graph layout, increasing density and obtaining more row windows conducive to Tensor cores. 

Our objective is to reconstruct each row window to enhance its computing intensity. For each row window, we define the \textit{computing intensity} in Equation \ref{comintensity}. 
% \vspace{-1mm}
% \setlength{\textfloatsep}{0pt}
\begin{equation}
% \vspace{-1mm}
\small
\label{comintensity}
    computing\; intensity = \frac{\#non zero\; elements}{\#non zero\; columns}
\end{equation}

A higher \textit{computing intensity} indicates a denser layout of row windows, making them more suitable for Tensor cores, as illustrated in Figure \ref{executiontime} and \S~\ref{adaptivecoreselection}. Based on this objective, we devise an efficient greedy strategy as below. 

Our basic algorithm is to iteratively select an initial vertex for each row window and incrementally expand it by adding the vertex that maximizes the \textit{computing intensity} in conjunction with the existing vertices in the row window. 
To mitigate the computational overhead of identifying the vertex with the maximum \textit{computing intensity} among all vertices, we introduce a vertices window $VW$ to confine the search scope.
As outlined in Algorithm \ref{reorder}, we sort all vertices based on their neighbors' smallest index (line 1). $N(v)$ in line 1 represents the set of neighbors for vertex $v$. For each row window, we select the first unvisited vertex from the sorted list as the initial vertex and add it to the set row window $RW$ (line 4). Subsequently, we compute the \textit{computing intensity} when considering adding each vertex to the row window according to Equation \ref{comintensity}. The loop in line 5 is executed 15 times because the row window is fixed to 16. Then, we add the vertex that maximizes the \textit{computing intensity} to $RW$ and proceed to the next iteration. Multiple vertices may yield the highest \textit{computing intensity}. To prioritize low sparsity, we select the vertex with the highest degree to add to $RW$ (lines 7-8).

\setlength{\textfloatsep}{0pt}
\makeatletter
% Remove right hand margin in algorithm
\patchcmd{\@algocf@start}% <cmd>
  {-1.5em}% <search>
  {0pt}% <replace>
  {}{}% <success><failure>
\makeatother
\begin{algorithm}[tbp]
    \caption{Layout reformat}
    \label{reorder}
    \LinesNumbered
    \KwIn{Graph $G=(V,E)$, vertices window $VW$.}
    \KwOut{Graph $G'$ after layout optimization.}
    $NRW\leftarrow \emptyset$;\\
    $soList\leftarrow$ sort $v\in V$ by $min(N(v))$;\\
    \ForEach{row window}{
        $RW\leftarrow$ the first vertex in $soList$ hasn't been visited;\\
        \For{$i=1$ to $15$}{
            $CanVtxs\leftarrow$ vertices with the highest \textit{computing intensity} within $VW$;\\
            $v\leftarrow$ vertex with the highest degree in $CanVtxs$;\\
            $RW \leftarrow RW\cup\{v\}$;\\
        }
        $NRW\leftarrow NRW.append(RW)$;\\
    }
    Reorder $G$ using $NRW$;\\
\end{algorithm}
\setlength{\textfloatsep}{.55\baselineskip plus  0.2\baselineskip minus  0.4\baselineskip}

\begin{figure}[tbp]
    \centering
\vspace{-4mm}
    \includegraphics[width=0.4\textwidth]{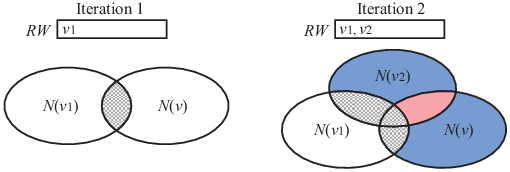}
    \vspace{-4mm}
    \caption{An example of avoiding redundant computing.}
    \label{fig:unionexample}
    \vspace{-2mm}
\end{figure}

\noindent \textbf{Efficiency Optimization.} To address the computational bottleneck in line 6, where $VW$ vertices are added to the current row window and their corresponding \textit{computing intensity} is calculated, we need to optimize the computation of $\#nonzero\ columns$. The most time-consuming aspect of this process is the frequent set union operations. 
Suppose $RW$ denotes the vertices in a row window. The number of non-zero columns within this row window can be computed as $\#nonzero\ columns = |\cup_{v\in RW}N(v)|$. However, this direct method results in significant redundancy, especially in the 15 iterations in lines 5-8 of Algorithm \ref{reorder}. 
For instance, as depicted in Figure \ref{fig:unionexample}, the union of $N(v_1)$ and $N(v)$ in the first iteration has been calculated, where $v_1\in RW$ and $v$ is a vertex from the range window. In the second iteration, we proceed to compute $N(v_1)\cup N(v_2)\cup N(v)$. However, we have already computed $N(v_1)\cup N(v)$ and $N(v_2)\cup N(v)$, resulting in redundant computation if calculated using a brute-force approach. Instead, we only need to calculate the union of the blue parts, namely $(N(v_2) - N(v_1)) \cup (N(v) - N(v_1))$.  Similarly, we can also avoid redundant computations in subsequent iterations. 
In general, in the $i$-th iteration, assuming the appended vertex is $v_i$ from the previous iteration, the calculation can be formulated as $N(v)\cup N(v_i) - \cup_{u\in RW\backslash\{v_i\}}{N(u)}$. 

In addition, we calculate the intersection instead of the union result. This enables efficient calculation by traversing the vertices in $N(N(v_i))$. Algorithm \ref{alg:LOA} illustrates the optimized layout reformatting algorithm. In each iteration, \textsf{LOA} calculates the intersection of each vertex with the vertices in the current row window in lines 7-9, i.e., $|N(v)\cap(\cup_{u\in RW}N(u))|$. In lines 10-14, \textsf{LOA} traverses the vertices in the range window and calculates the \textit{computation intensity} by Equation \ref{equ:cominten}. 

\vspace{-2mm}
\begin{equation}
% \vspace{-1mm}
\small
\label{equ:cominten}
    \frac{|N(v)|+\Sigma_{u\in RW}|N(u)|}{|N(v)|+|\cup_{u\in RW}N(u))|-|N(v)\cap(\cup_{u\in RW}N(u))|}
\end{equation}

Subsequently, the vertex with the highest \textit{computation intensity} is appended to row window (line 15). At the end of each iteration, the intermediate results are updated in lines 16-19. 

\setlength{\textfloatsep}{0pt}
\makeatletter
% Remove right hand margin in algorithm
\patchcmd{\@algocf@start}% <cmd>
  {-1.5em}% <search>
  {0pt}% <replace>
  {}{}% <success><failure>
\makeatother
\begin{algorithm}[tbp]
    \caption{\textsf{LOA}}
    \label{alg:LOA}
    \LinesNumbered
    \KwIn{Graph $G=(V,E)$, vertices window $VW$.}
    \KwOut{Reconstructed row windows $NRW$.}
    $NRW\leftarrow \emptyset$;\\
    $soList\leftarrow$ sort $v\in V$ by $min(N(v))$;\\
    \ForEach{row window}{
        $RW\leftarrow$ the first vertex $v_0$ in $soList$ hasn't been visited;\\
        $Resi\leftarrow N(v_0), allCols\leftarrow N(v_0)$;\\
        \For{$i=1$ to $15$}{
            \ForEach{$u\in Resi$}{
                \ForEach{$w\in N(u)$}{
                    $w.cns$++;\\
                }
            }
            \For{$j=v_0.id$ to $(v_0.id + VW)$}{
                $v\leftarrow soList[j]$;\\
                $P=\frac{curEles+|N(v)|}{curCols + |N(v)| - v.cns}$;\\
                \If{$P>maxP$}{
                    $maxP\leftarrow P, v_{max}\leftarrow v$;\\
                }
            }
            $RW \leftarrow RW\cup\{v_{max}\}$;\\
            $Resi\leftarrow N(v_{max}) - allCols$;\\
            $allCols\leftarrow allCols\cup N(v_{max})$;\\
            $curEles\leftarrow curEles + |N(v_{max})|$;\\
            $curCols\leftarrow |allCols|$;\\
        }
        $NRW\leftarrow NRW.append(RW)$;\\
    }
    \Return{$NRW$}
\end{algorithm}
\setlength{\textfloatsep}{.55\baselineskip plus  0.2\baselineskip minus  0.4\baselineskip}

\section{Experiments}
\label{sec:evaluation}
In this section, we evaluate \textsf{HC-SpMM} and conduct a comparative evaluation with state-of-the-art SpMM kernels as well as GNN training frameworks. More information on the experiments is provided in Appendix \ref{supexp}. 

\subsection{Experimental Setup}
\noindent\textbf{Datasets.} In order to fully evaluate the proposed SpMM kernel, we run experiments on 14 datasets, some of which have been used in previous related work \cite{fey2019fast,wang2019deep,wang2023tc,fan2024dtc}. All the datasets are accessible at SNAP\footnote[13]{snap.stanford.edu/data/index.html}, TUDataset\footnote[14]{chrsmrrs.github.io/datasets/docs/datasets} and KONECT\footnote[15]{konect.cc/networks}. Table \ref{datasetdetail} summarizes the properties of the 13 datasets. Most datasets do not have information on the number of classes, so we uniformly use 22. 

\noindent\textbf{Baselines.} We compare \textsf{HC-SpMM} with existing methods in two aspects: SpMM and GNN forward/backward propagation. 

$1)$ SpMM kernels: \textsf{Sputnik} \cite{gale2020sparse} is a library 
% of sparse linear algebra kernels and utilities for deep learning 
developed by Google, which provides the state-of-the-art unstructured SpMM kernel for full precision on CUDA cores. \textsf{GE-SpMM} \cite{huang2020ge} is another efficient SpMM kernel on CUDA cores specifically designed for SpMM operation in GNN. \textsf{cuSPARSE} \cite{refcusparse} is a CUDA sparse matrix library developed by Nvidia, which contains a set of high-performance SpMM kernels. In our experiments, we employ the SpMM kernel taking the CSR format as its input. \textsf{TC-GNN} \cite{wang2023tc} is a method to use Tensor cores to accelerate the full-precision unstructured SpMM operation in GNNs. \textsf{DTC-SpMM} \cite{fan2024dtc} is the state-of-the-art SpMM kernel using Tensor cores. Notably, {using CUDA cores or Tensor cores alone for SpMM cannot achieve comparable efficiency to the methods in the experiment}. 
% Notably, solely using a certain type of cores in \textsf{HySpMM} will degrade it to the aforementioned methods.
% These are SOTA methods on CUDA cores and Tensor cores in our problem scenario. 

$2)$ GNN training: We conduct end-to-end experiments and compare the frameworks targeting the optimization of SpMM to demonstrate the ability of {\sf HC-SpMM} to handle complex graph computing tasks. Due to the GNN algorithm remaining unchanged, the training results of these frameworks are identical. Other frameworks, such as \cite{zhang2023ducati,wang2021gnnadvisor}, are not considered in experiments, {which are orthogonal to our work}. \textsf{GE-SpMM} and \textsf{TC-GNN} integrate the optimized SpMM kernels into \textsf{PyTorch}, which are the state-of-the-art works for accelerating GNN by optimizing SpMM on CUDA cores and Tensor cores, respectively. As the efficiency of \textsf{GE-SpMM} and \textsf{TC-GNN} is much more significant\footnote[16]{\textsf{TC-GNN} achieves 1.7$\times$ speedup on average over \textsf{DGL}, while \textsf{GE-SpMM} brings up to 3.7$\times$ speedup over \textsf{PyG}.} than \textsf{PyG} and \textsf{DGL}, we do not evaluate these two GNN training frameworks in experiments. 

\noindent\textbf{Platforms.} All experiments are conducted on a CentOS 7 server, featuring an Intel Core i9-10900K CPU and an Nvidia RTX 3090 GPU. The GPU has 82 SMs, 10,496 CUDA cores, and 328 Tensor cores. We implement \textsf{HC-SpMM}\footnote[17]{{https://github.com/ZJU-DAILY/HC-
SpMM}} in C++ under Nvidia CUDA 12.2 and integrate it into \textsf{PyTorch} 1.8. 

\begin{table}[tbp]
    \vspace{-2mm}
    % \begin{center}   
        \caption{Details of datasets.}  
        \vspace{-3mm}
        \label{datasetdetail} 
        \begin{tabularx}{\linewidth}{|X||X|X|l|}   
        \hline   \textbf{Datasets} & \textbf{\#Vertex} & \textbf{\#Edges} & \textbf{Dim} \\ \hline 
        \hline   Citeseer (\textit{CS}) & 3,327 & 9,464 & 3,703 \\ 
        \hline   Cora (\textit{CR}) & 2,708 & 10,858 & 1,433 \\
        \hline   Pubmed (\textit{PM}) & 19,717 & 88,676 & 500 \\
        \hline   PROTEINS (\textit{PT}) & 43,471 & 162,088 & 29 \\
        \hline   DD (\textit{DD}) & 334,925 & 1,686,092 & 89 \\
        \hline   Amazon (\textit{AZ}) & 410,236 & 3,356,824 & 96 \\
        \hline   Yeast (\textit{YS}) & 1,710,902 & 3,636,546 & 74 \\
        \hline   OVCAR (\textit{OC}) & 1,889,542 & 3,946,402 & 66 \\
        \hline   Github (\textit{GH}) & 1,448,038 & 5,971,562 & 64 \\
        \hline   YeastH (\textit{YH}) & 3,138,114 & 6,487,230 & 75 \\
        \hline   Reddit (\textit{RD}) & 4,859,280 & 10,149,830 & 96 \\
        \hline   Twitch (\textit{TT}) & 3,771,081 & 22,011,034 & 96 \\
        \hline   CitPatents (\textit{CP}) & 3,774,768 & 16,518,948 & 96 \\
        \hline   Depedia (\textit{DP}) & 18,268,981 & 172,183,984 & 96 \\
        \hline   
        \end{tabularx}
    % \end{center}
    \vspace{-2mm}
\end{table}

\subsection{Evaluations on SpMM kernel}
\label{eva:spmm}
In this subsection, we compare {\sf HC-SpMM} with 4 state-of-the-art SpMM kernels, and conduct comprehensive experiments to evaluate our proposed optimization techniques. 

\subsubsection{Comparison with SpMM kernels}
\begin{figure*}[tbp]
    \centering    
    \hspace{-2mm}
    \includegraphics[width=0.7\textwidth]{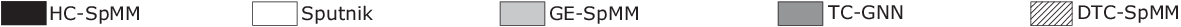}\\
    % \vspace{-1mm}
    \hspace{-2mm}
    \includegraphics[width=0.47\textwidth]{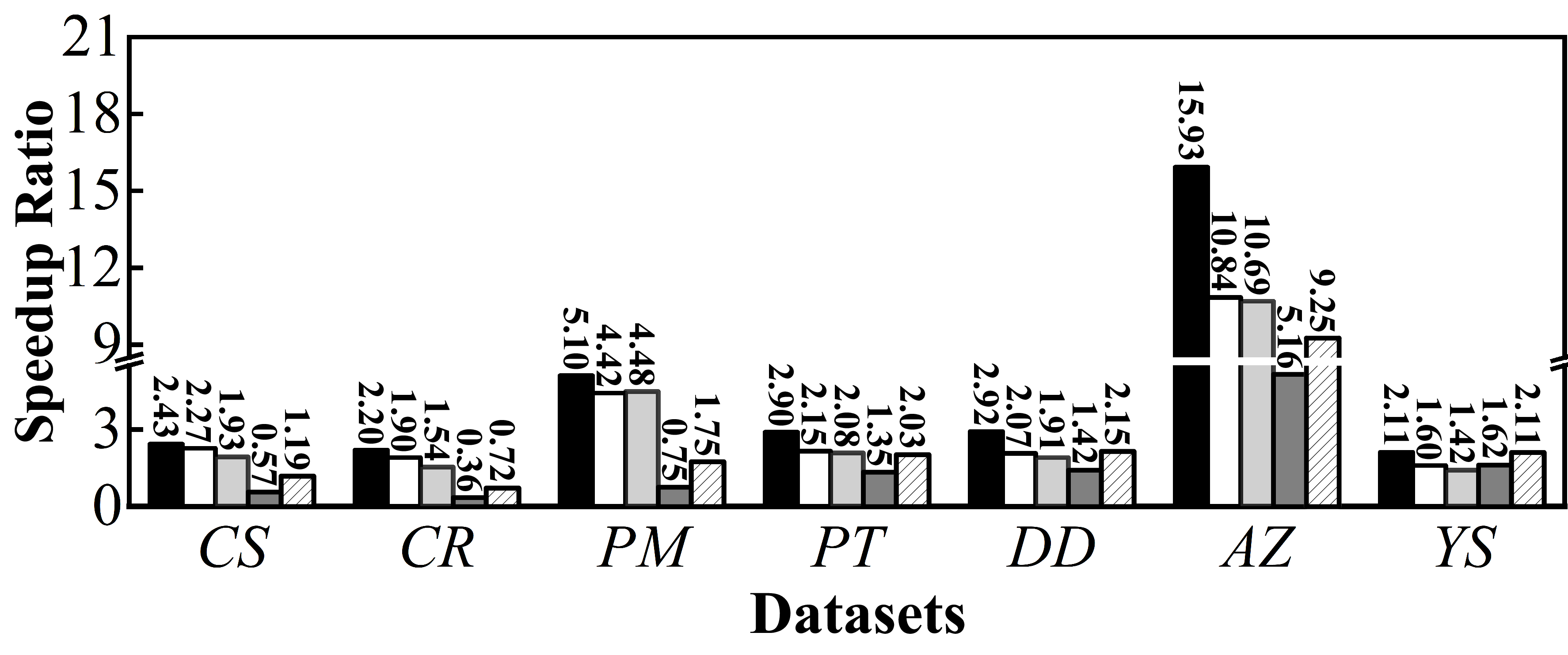}
    \hspace{6mm}
    \includegraphics[width=0.47\textwidth]{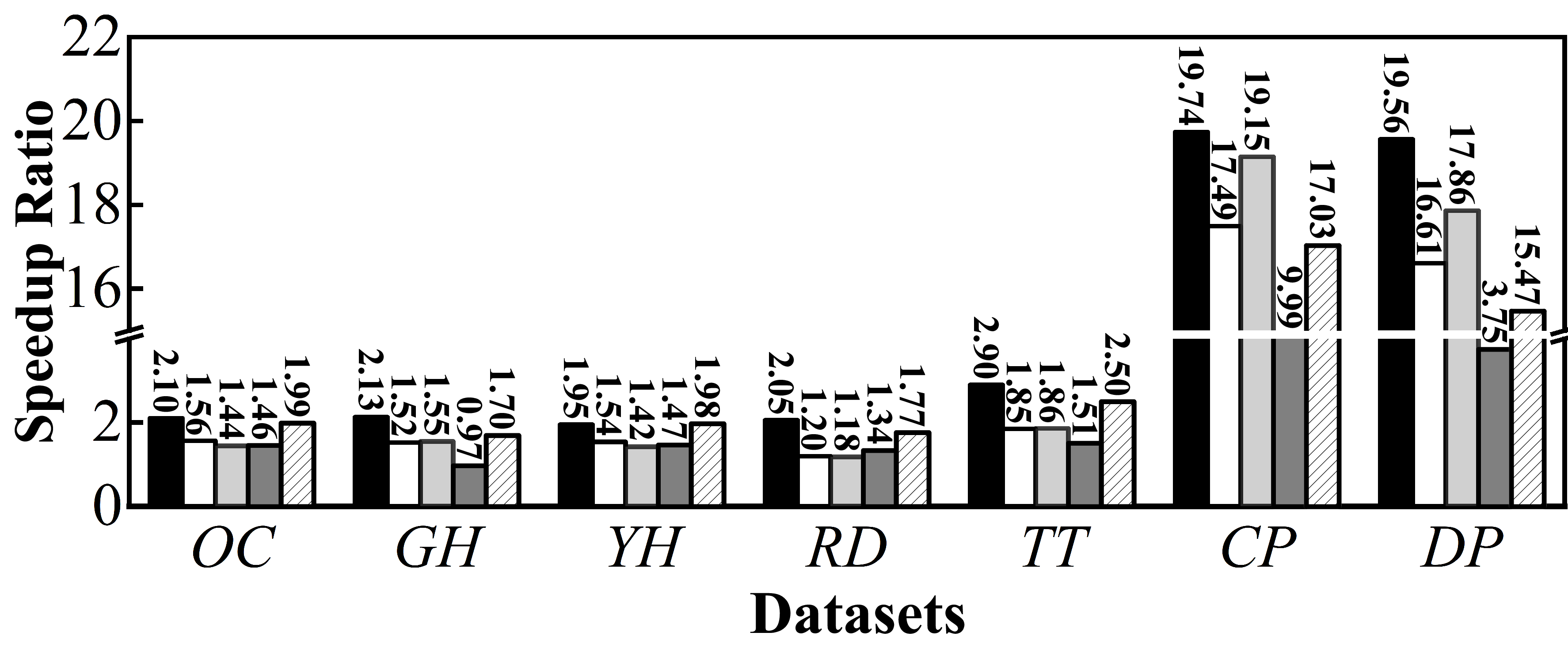}
    
    \vspace{-4mm}
    \caption{Overall performance of SpMM kernels.}
    \label{spmm}
\vspace{-4mm}
\end{figure*}

% To substantiate the efficacy of our methodology, we engage in a comprehensive comparative analysis of our proposed SpMM kernel integrated within \textsf{HC-SpMM} against SOTA kernels. 
Figure \ref{spmm} presents the overall performance of various SpMM kernels across 13 datasets, using \textsf{cuSPARSE} as the baseline. Execution durations are measured employing \textsf{nvprof}. We report the SpMM kernel time and exclude the preprocessing and PCIe transfers overhead. The data transferred through PCIe includes only the dense matrix and the CSR format of the sparse matrix. The preprocessing step prepares the data used for Tensor cores and classifies row windows to appropriate cores using GPU. It is 13.0$\times$ of a single SpMM execution in {\sf HC-SpMM} on average, which is negligible in many real-world scenarios that require thousands of SpMM operations such as GNN. Further discussions about the preprocessing are detailed in Appendix \ref{exp:preprocess}. The caches are cold when performing SpMM kernels. The absolute numbers of the execution time are detailed in Appendix \ref{absoluteSpMM}. 
%所有实验都是用同一个logistic模型
Our results show that \textsf{HC-SpMM} consistently outperforms all compared methods across all datasets. \textsf{HC-SpMM} achieves speed improvements of 1.85-18.89$\times$ over \textsf{cuSPARSE}, 1.07-1.57$\times$ over \textsf{Sputnik}, 1.05-1.57$\times$ over \textsf{GE-SpMM}, 1.30-6.76$\times$ over \textsf{TC-GNN} and 0.99-3.03$\times$ over \textsf{DTC-SpMM}. Notably, in the best cases, \textsf{HC-SpMM} achieves a remarkable speedup of 19.56$\times$ over \textsf{cuSPARSE} on \textit{DP}, 1.57$\times$ speedup over \textsf{Sputnik} on \textit{TT}, 1.57$\times$ speedup over \textsf{GE-SpMM} on \textit{RD}, 6.76$\times$ and 3.00$\times$ speedup over \textsf{TC-GNN} and \textsf{DTC-SpMM} on \textit{PM}. We also compare \textsf{HC-SpMM} with the SpMM kernel in \textsf{PyTorch} executed on CPU, which achieves an average speedup of 183.77$\times$. These results underscore the effectiveness of leveraging hybrid cores to perform SpMM. \textsf{HC-SpMM} demonstrates its capability to intelligently identify submatrices suitable for CUDA and Tensor cores to fully utilize the characteristics of different hardware structures and maximize the performance of the GPU. Please note that \textsf{HC-SpMM} don't need to merge the results because of the combination strategy proposed in \S~\ref{combinationstrategy} even if we employ two cores for simultaneous calculation. 

The speedup ratios of \textsf{HC-SpMM} vary across different datasets when contrasted against SOTA methods. Compared with \textsf{Sputnik} and \textsf{GE-SpMM} which employ CUDA cores for computation, the improvements of \textsf{HC-SpMM} on \textit{CS}, \textit{CR} and \textit{PM} are less pronounced than on other datasets. As detailed in Table \ref{datasetdetail}, this discrepancy can be attributed to the density of the adjacency matrices. Datasets such as \textit{CS}, \textit{CR} and \textit{PM}, characterized by relatively few vertices and edges, possess limited dense parts in the adjacency matrices. Consequently, the efficiency of Tensor cores, which excel at dense matrix multiplication, is not fully leveraged. This conclusion is further corroborated by the performance of \textsf{TC-GNN} and \textsf{DTC-SpMM}, which use Tensor cores for SpMM on \textit{CS}, \textit{CR}, and \textit{PM}. On these three datasets, the speedup ratio of \textsf{TC-GNN} and \textsf{DTC-SpMM} is considerably behind \textsf{Sputnik} and \textsf{GE-SpMM}, which is because the potential of Tensor cores in dense parts cannot be fully realized on small-scale matrices. 
% 进一步证明了hybrid策略的有效性，可以根据具体条件判断到底适合什么cores执行
Conversely, \textsf{HC-SpMM}, \textsf{Sputnik} and \textsf{GE-SpMM} demonstrate remarkable speedups on datasets like \textit{AZ} and \textit{DP}. Analysis of these datasets reveals that the adjacency lists of vertices exhibit poor locality, with neighbor IDs scattered rather than compactly distributed as in other datasets. This scattered distribution leads to inefficient memory access for \textsf{cuSPARSE}.  

We also evaluate the performance of the five kernels in various sparsity to further demonstrate the necessity and effectiveness of using hybrid GPU cores for SpMM. The results are reported in Appendix \ref{additionexp}. We also evaluate the adaptability of the regression model, the sensitivity of performance to specific parameters, and the utilization metrics of GPU cores, which are discussed in Appendix \ref{absoluteSpMM}\&\ref{evaFPtype}, \ref{exp:sensity}, and \ref{exp:utilization}, respectively. 

\subsubsection{Effectiveness of the optimization techniques}
Next, we evaluate the three optimization techniques respectively to demonstrate their effectiveness. 

\noindent\textbf{Optimizations on CUDA cores}. \textbf{(i)} \textit{Generalization.} We evaluate the execution time of kernels using the generalization technique as well as without using this optimization. The datasets used for this evaluation are those featuring an unaligned embedding dimension (not a multiple of 32) in Table \ref{datasetdetail}. Table \ref{tab:generazation} illustrates the results. On average, the generalization technique achieves 18.8\% time savings across tested datasets, which demonstrates that the generalization of SpMM on CUDA cores saves threads in each warp when the dimension is not a multiple of 32. {\textbf{(ii)} \textit{Memory Management.}} In \S~\ref{bottleneckofccu}, shared memory is used to store the column indices of CSR. We evaluate the effectiveness of this technique and report the results in Table \ref{tab:sharedmemory}. The usage of shared memory achieves an average speedup of 2.85\%. On \textit{YS}, this strategy has the most significant improvement. This is because \textit{YS} has more row windows suitable for CUDA cores before performing \textsf{LOA}. In addition, \textit{YS} has a low average degree which leads to less shared memory usage, thus increasing the number of warps that can be concurrently scheduled by GPU. 

\begin{table}[tp]
    \centering
    \vspace{-2mm}
    \caption{Effectiveness of generalization.}
    \vspace{-2mm}
    \resizebox{\linewidth}{!}{
    \begin{tabular}{|c||c|c|c|c|}
        \hline
        \textbf{Datasets} & \textbf{Generalization} & \textbf{No optimization} & \textbf{Speedup} \\ \hline
        \hline
        \textit{DD} & 0.398ms & 0.498ms & 25.1\% \\
        \hline
        \textit{YS} & 1.456ms & 1.593ms & 9.4\% \\
        \hline
        \textit{OC} & 1.576ms & 1.869ms & 18.6\% \\
        \hline
        \textit{YH} & 3.205ms & 3.912ms & 22.1\% \\
        \hline
    \end{tabular}
    }
    \label{tab:generazation}
    \vspace{-2mm}
\end{table}

\begin{table}[t]
    \centering
    \vspace{-2mm}
    \caption{Effectiveness of shared memory using strategy.}
    \vspace{-2mm}
    \resizebox{\linewidth}{!}{
    \begin{tabular}{|c||c|c|c|c|}
        \hline
        \textbf{Datasets} & \textbf{Shared memory} & \textbf{No optimization} & \textbf{Speedup} \\ \hline
        \hline
        \textit{YS} & 0.581ms & 0.603ms & 3.79\% \\
        \hline
        \textit{OC} & 0.625ms & 0.639ms & 2.24\% \\
        \hline
        \textit{YH} & 1.046ms & 1.072ms & 2.49\% \\
        \hline
        \textit{RD} & 1.575ms & 1.614ms & 2.48\% \\
        \hline
        \textit{TT} & 1.384ms & 1.429ms & 3.25\% \\
        \hline
    \end{tabular}
    }
    \label{tab:sharedmemory}
    \vspace{-2mm}
\end{table}

\noindent\textbf{{Optimizations on Tensor Cores.}} Finally, we conduct experiments to validate the effectiveness of data loading. In this evaluation, we only record the calculation time of Tensor cores. Table \ref{tab:dataloading} presents the results. Our proposed data loading strategy leads to an average speedup of 17.50\%. This improvement is attributed to the increased participation of thread warps in data loading and reduced bank conflicts in shared memory. It is worth noting that although our data loading strategy enhances performance, data loading remains the bottleneck of SpMM on Tensor cores and requires further exploration. 

\subsection{Evaluations on GNN training}
In this subsection, with the support of {\sf HC-SpMM}, we compare the GNN training efficiency with 2 methods, which also focus on accelerating the SpMM operation in GNN training. Subsequently, we evaluate the performance of kernel fusion strategy and {\sf LOA} algorithm. 

\subsubsection{Comparison with GNN training frameworks}
Although our proposed SpMM kernel accelerated by hybrid GPU cores can be used for general sparse matrix-matrix multiplication, our primary objective is to utilize this kernel to enhance the efficiency of graph computing. To demonstrate the efficiency, we integrate our SpMM kernel into \textsf{PyTorch} and compare its performance with two efficient GNN training frameworks. Specifically, We implement \textsf{GCN} \cite{kipf2016semi} and \textsf{GIN} \cite{xu2018powerful} as benchmarking models. It's worth noting that the backward propagation of \textsf{GCN} and forward propagation of \textsf{GIN} employs a unique kernel fusion technique distinct from the forward(backward) propagation. Consequently, we separately report the performance of these two stages within end-to-end evaluations. {Due to the large memory requirements of the dataset \textit{DP}, an out-of-memory error occurs when training GNN with all three frameworks. So we omit to report the results of \textit{DP}. Additional experiments in Appendix \ref{exp:memory} suggest that the memory usage of {\sf HC-SpMM} is only up to 2\% and 6\% more than {\sf GE-SpMM} and {\sf TC-GNN}, respectively.} All the reported time below is the average execution time of one epoch unless noted otherwise. 

\begin{table}[tbp]
    \centering
    \vspace{-2mm}
    \caption{Effectiveness of data loading strategy.}
    \vspace{-2mm}
    \resizebox{\linewidth}{!}{
    \begin{tabular}{|c||c|c|c|c|}
        \hline
        \textbf{Datasets} & \textbf{Opt. data loading} & \textbf{No optimization} & \textbf{Speedup} \\ \hline
        \hline
        \textit{YS} & 0.387ms & 0.456ms & 17.83\% \\
        \hline
        \textit{OC} & 0.330ms & 0.386ms & 16.97\% \\
        \hline
        \textit{YH} & 0.552ms & 0.663ms & 20.11\% \\
        \hline
        \textit{RD} & 0.880ms & 1.006ms & 14.32\% \\
        \hline
        \textit{TT} & 0.678ms & 0.802ms & 18.29\% \\
        \hline
    \end{tabular}
    }
    \label{tab:dataloading}
    \vspace{-3mm}
\end{table}

\begin{figure}[tbp]
    \centering
    \hspace{-2mm}
    % \vspace{-1mm}
    \includegraphics[width=0.3\textwidth]{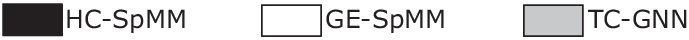}\\
    \hspace{-2mm}
    \includegraphics[width=0.23\textwidth]{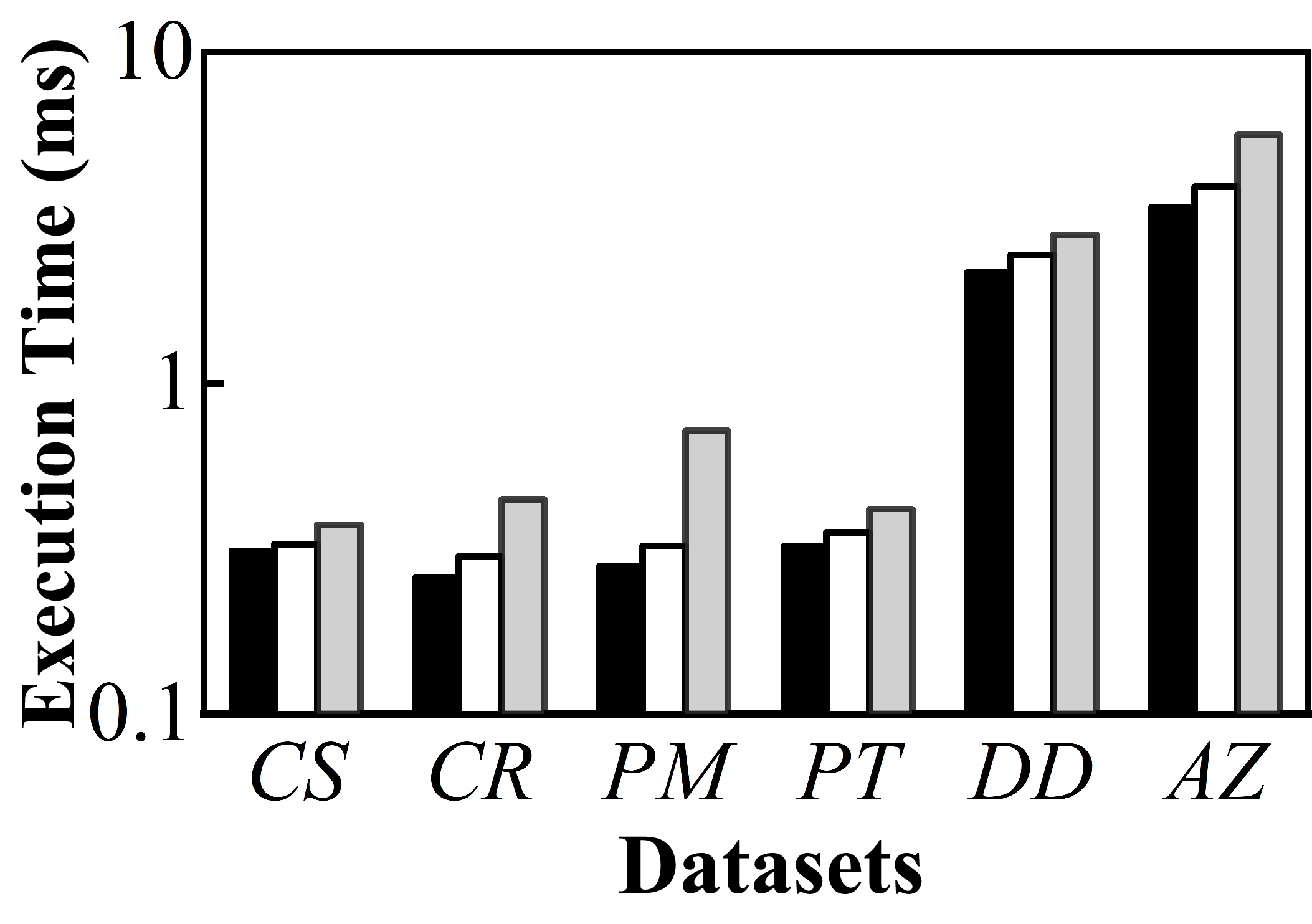}
    \hspace{-1mm}
    \includegraphics[width=0.23\textwidth]{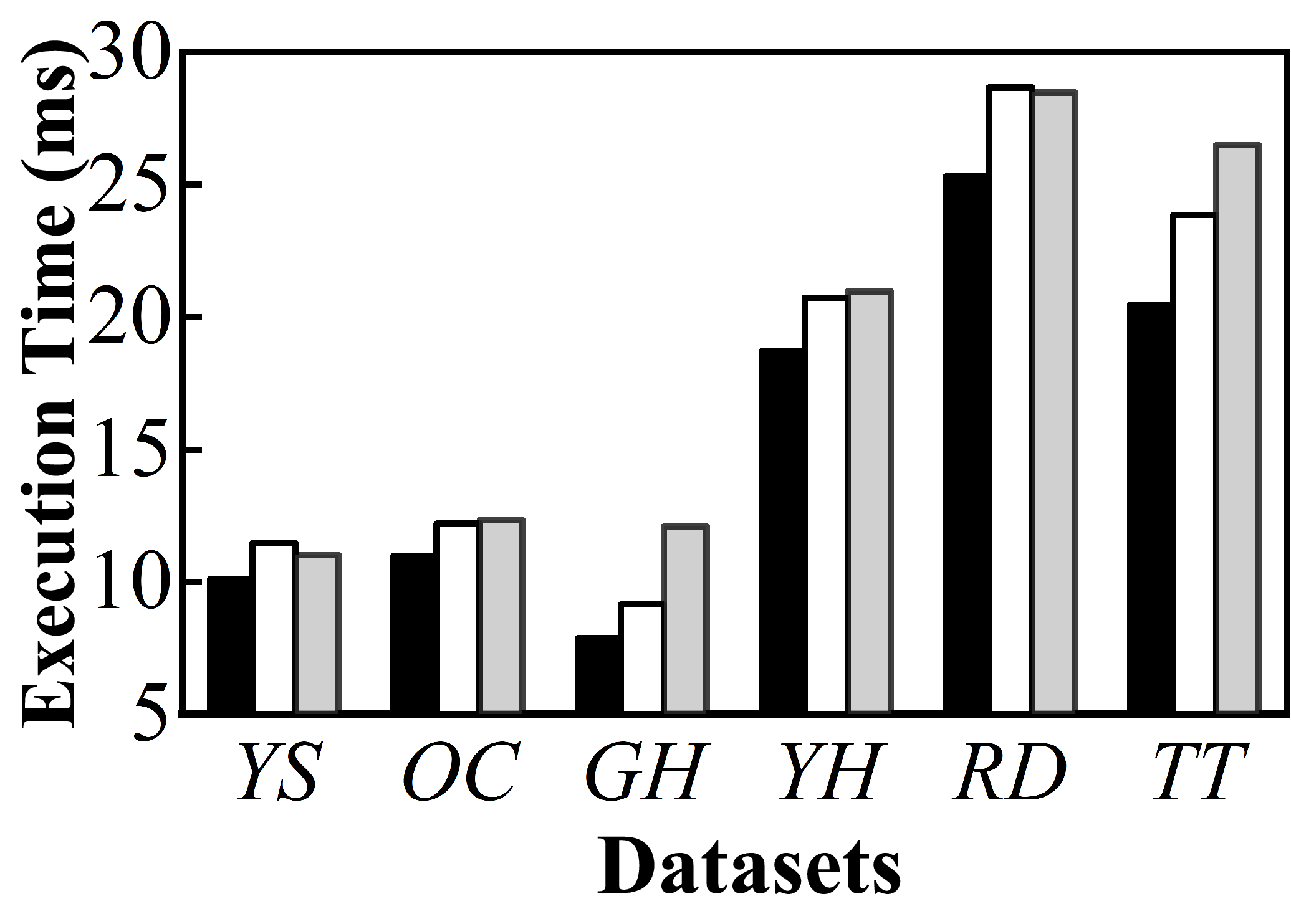}
    \vspace{-3mm}
    \caption{Comparison of \textsf{GCN} forward propagation.}
    \label{forward}
\vspace{-2mm}
\end{figure}

\begin{figure}[tbp]
    \centering
    \hspace{-2mm}
    % \vspace{-1mm}
    \includegraphics[width=0.3\textwidth]{figures/GNN_legend.eps}\\
    \hspace{-2mm}
    \includegraphics[width=0.23\textwidth]{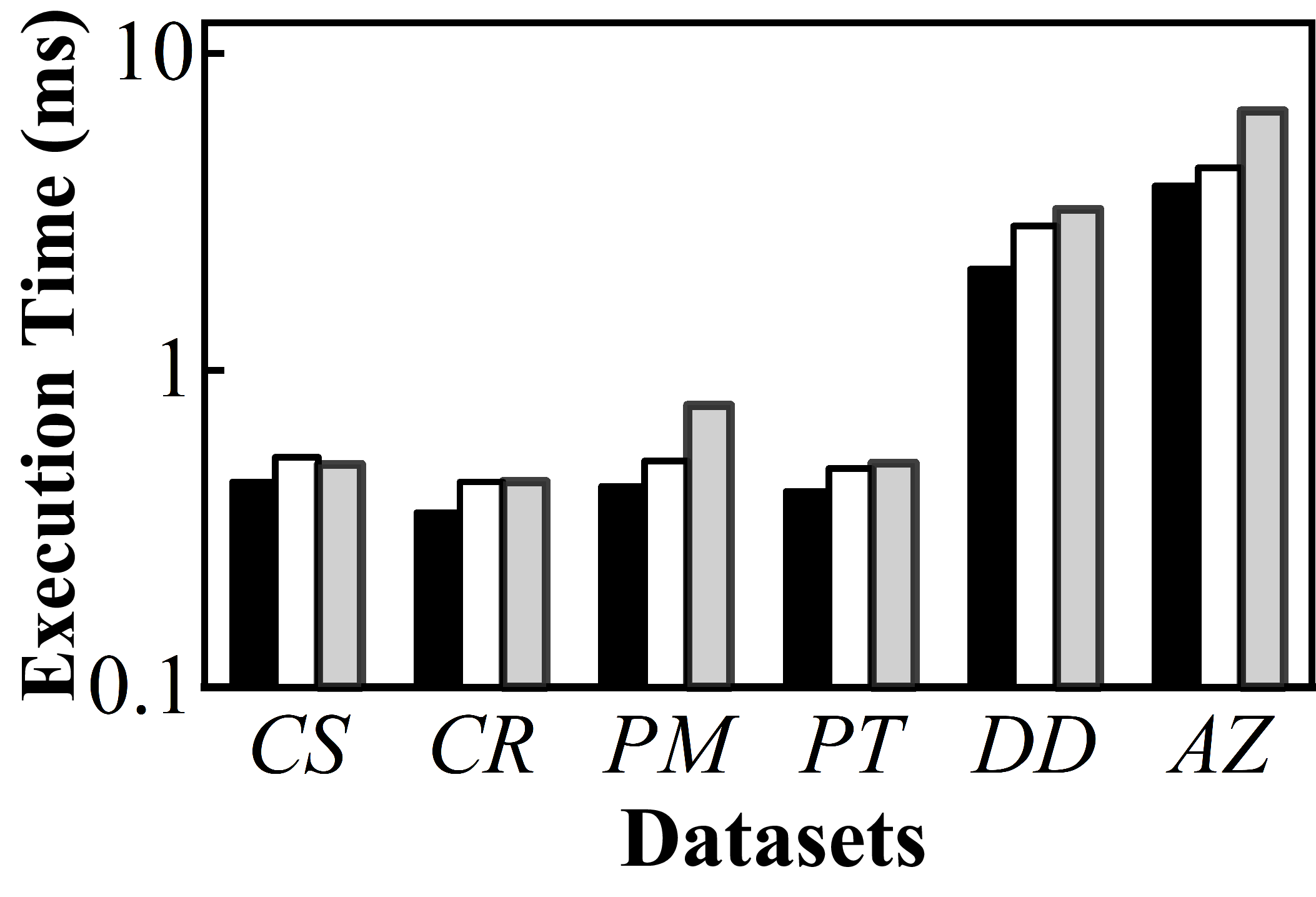}
    \hspace{-1mm}
    \includegraphics[width=0.23\textwidth]{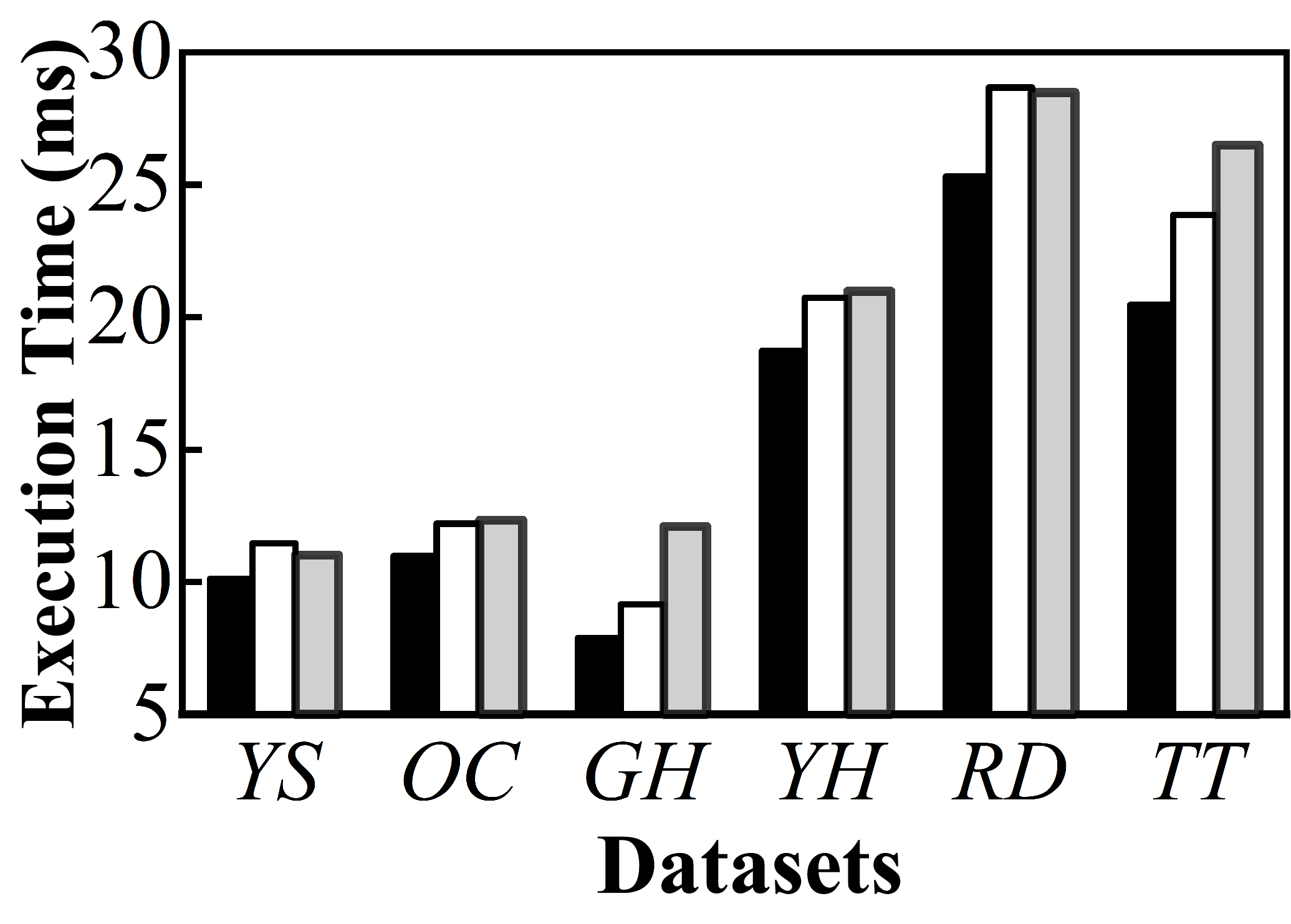}
    \vspace{-2mm}
    \caption{Comparison of \textsf{GCN} backward propagation.}
    \label{backward}
\vspace{-2mm}
\end{figure}

Figure \ref{forward} illustrates the performance of GNN frameworks in forward propagation of \textsf{GCN}. Overall, \textsf{HC-SpMM} outperforms \textsf{GE-SpMM} and \textsf{TC-GNN} across all datasets. Specifically, \textsf{HC-SpMM} achieves an average speedup of 1.42$\times$ over \textsf{TC-GNN} and 1.12$\times$ over \textsf{GE-SpMM}. Forward propagation comprises two operations: sparse matrix-matrix multiplication and dense matrix-matrix multiplication, corresponding to the \textit{Aggregation} and \textit{Update} phases respectively. Since our optimization primarily targets the \textit{Aggregation} phase, the performance of \textsf{HC-SpMM} in forward propagation is similar to that of SpMM in \S~\ref{eva:spmm}. The acceleration of \textsf{HC-SpMM} becomes more pronounced for large graphs, such as \textit{GH}, \textit{RD} and \textit{TT}. As previously discussed, there are more dense parts in large graphs and Tensor cores are efficient on the dense parts. 

% Figure \ref{backward} illustrates the performance of backward propagation of \textsf{GCN} among the three methods. It is evident that \textsf{HySpMM} demonstrates the best performance in all instances. In general, \textsf{HySpMM} achieves a 1.48$\times$ speedup over \textsf{TC-GNN} and a 1.33$\times$ speedup over \textsf{GE-SpMM} on average. Notably, \textsf{HySpMM} exhibits a higher speedup ratio during backward propagation compared with forward propagation. The enhanced acceleration is attributed to both the hybrid SpMM kernel and the sophisticated kernel fusion technique. 
% \textsf{HySpMM} showcases superior performance on large datasets such as \textit{YS}, \textit{OC}, \textit{GH}, \textit{YH}, \textit{RD} and \textit{TT}. On these 6 datasets, \textsf{HySpMM} consistently surpasses \textsf{GE-SpMM} by at least 1.4$\times$. 
% In addition, the most significant speedup cases of \textsf{HySpMM} compared to \textsf{TC-GNN} occur on \textit{AZ} and \textit{GH}, where the speedup ratio reaches at least 1.7$\times$. The reason is that \textit{AZ} and \textit{GH} possess less dense row windows. The utilization of Tensor cores alone for SpMM calculation leads to suboptimal performance in such scenarios. These results further demonstrate the necessity and effectiveness of our hybrid strategy. 

Figure \ref{backward} illustrates the performance of backward propagation of \textsf{GCN} among the three methods. It is evident that \textsf{HC-SpMM} demonstrates the best performance in all instances. In general, \textsf{HC-SpMM} achieves a 1.48$\times$ speedup over \textsf{TC-GNN} and a 1.33$\times$ speedup over \textsf{GE-SpMM} on average. Notably, \textsf{HC-SpMM} exhibits a higher speedup ratio during backward propagation compared with forward propagation. The enhanced acceleration is attributed to both the hybrid SpMM kernel and the sophisticated kernel fusion technique. \textsf{HC-SpMM} showcases superior performance on large datasets such as \textit{YS}, \textit{OC}, \textit{GH}, \textit{YH}, \textit{RD} and \textit{TT}. On these 6 datasets, \textsf{HC-SpMM} consistently surpasses \textsf{GE-SpMM} by at least 1.4$\times$. In addition, the most significant speedup cases of \textsf{HC-SpMM} compared to \textsf{TC-GNN} occur on \textit{AZ} and \textit{GH}, where the speedup ratio reaches at least 1.7$\times$. The reason is that \textit{AZ} and \textit{GH} possess less dense row windows. The utilization of Tensor cores alone for SpMM calculation leads to suboptimal performance in such scenarios. These results further demonstrate the necessity and effectiveness of our hybrid strategy. 

\begin{figure}[tbp]
    \centering    
    \hspace{-4mm}
    \vspace{-1mm}
    \includegraphics[width=0.3\textwidth]{figures/GNN_legend.eps}\\
    \hspace{-2mm}
    % \vspace{-4mm}
    \subfigcapskip=-5pt
    \subfigure[Forward propagation of \textsf{GIN}]{
    \includegraphics[width=0.23\textwidth]{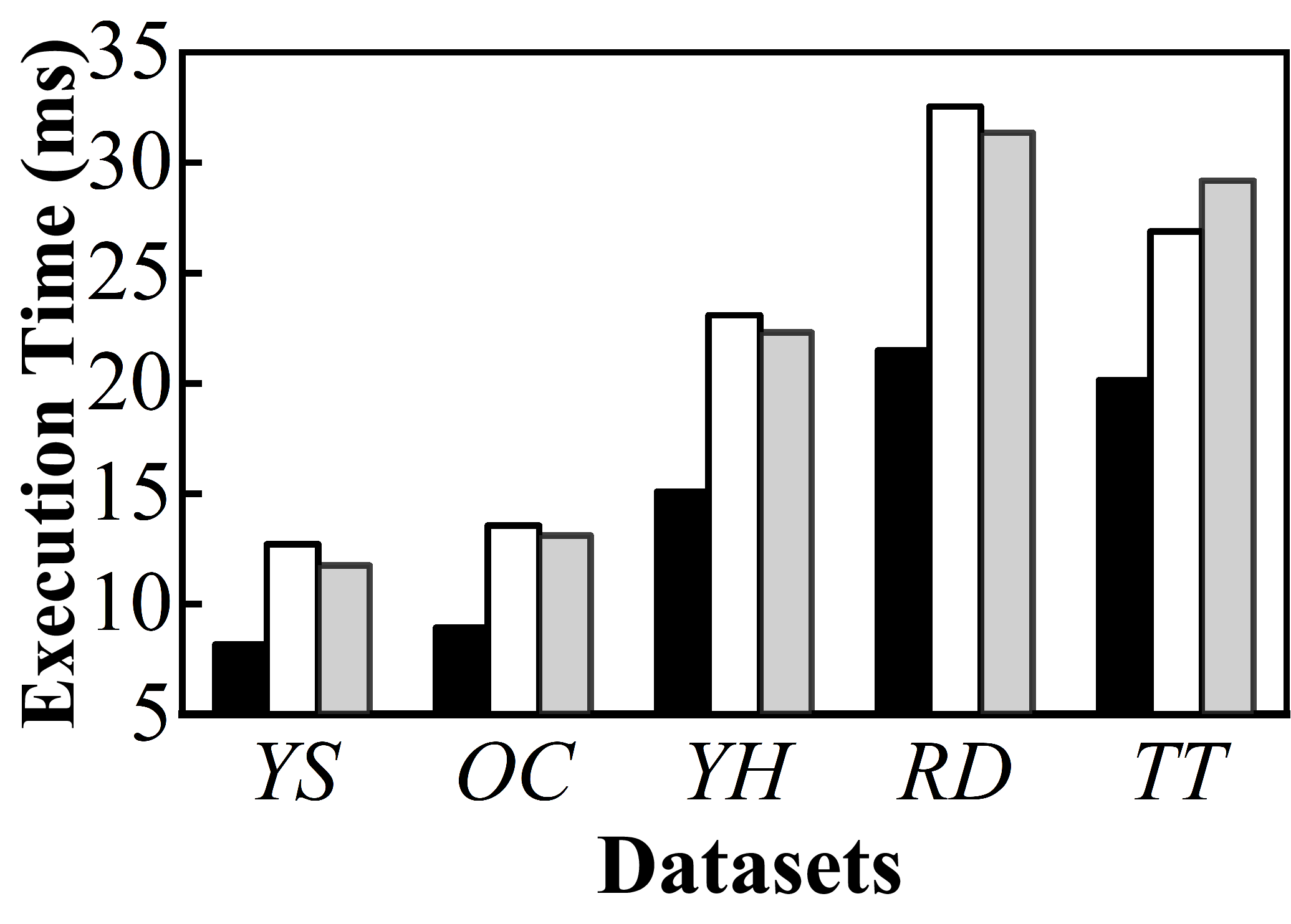}}
    \hspace{-1mm}
    \subfigure[Backward propagation of \textsf{GIN}]{
    \includegraphics[width=0.23\textwidth]{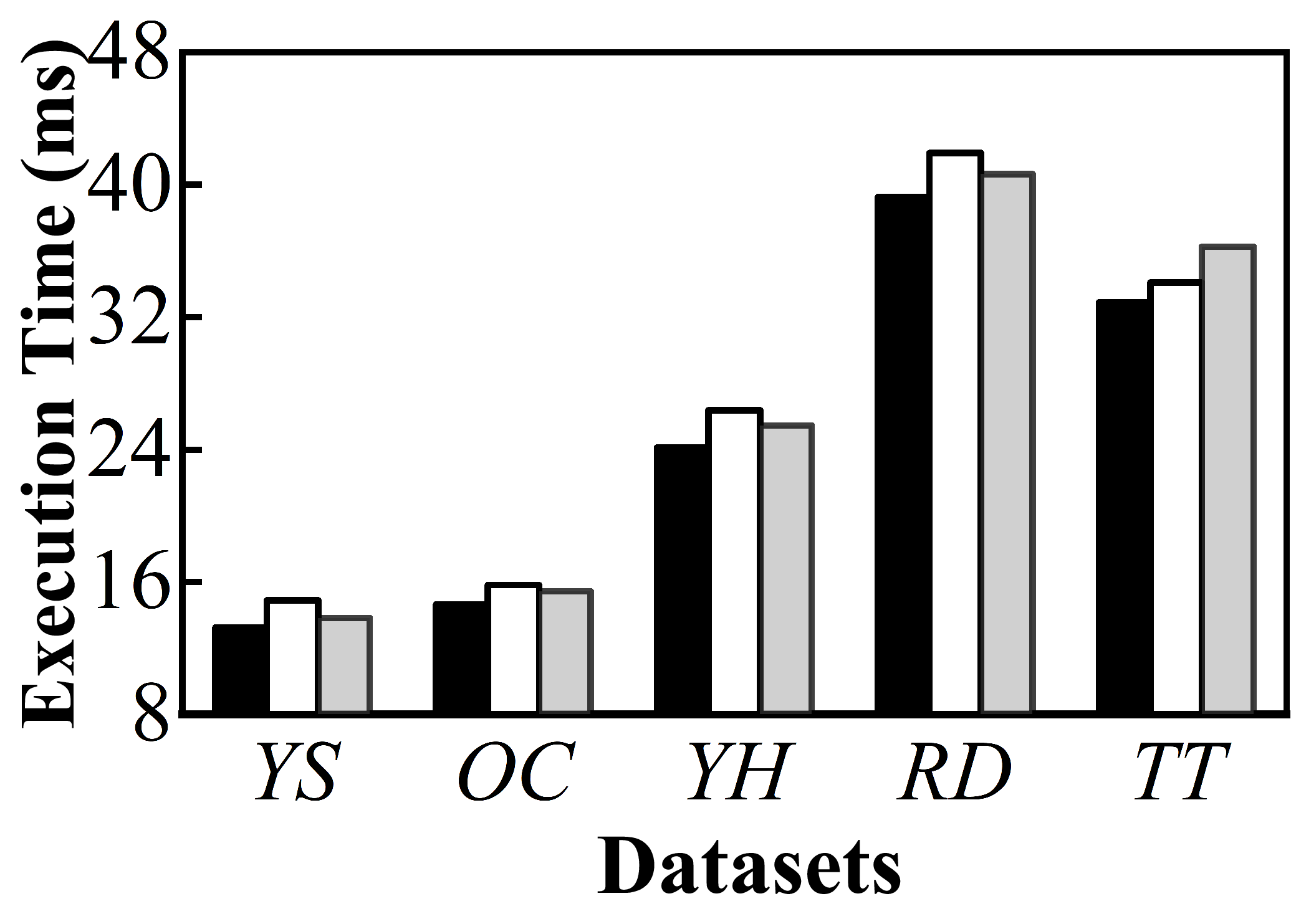}}
    \vspace{-3mm}
    \caption{Comparison of \textsf{GIN} propagation.}
    \label{gineva}
\vspace{-3mm}
\end{figure}

Regarding \textsf{GIN},  Figure \ref{gineva} depicts the results. On average, \textsf{HC-SpMM} achieves a speedup of 1.49$\times$ and 1.08$\times$ over \textsf{GE-SpMM} in forward and backward propagation, respectively. Compared with \textsf{TC-GNN}, it achieves 1.46$\times$ and 1.06$\times$ speedup in the forward and backward propagation on average. The performance is similar to \textsf{GCN}, hence warranting no further in-depth elucidation. 

The state-of-the-art performance of {\sf HC-SpMM} demonstrates its ability to handle complex graph computing tasks.    

% \begin{figure}[tbp]
%     \centering    
%     \hspace{-2mm}
%     \vspace{-1mm}
%     \includegraphics[width=0.3\textwidth]{figures/GNN_legend.eps}\\
%     \hspace{-2mm}
%     % \subfigcapskip=-5pt
%     \subfigure[Forward propagation of \textsf{GCN}]{
%     \includegraphics[width=0.23\textwidth]{figures/forward_select.eps}}
%     \hspace{-1mm}
%     \subfigure[Backward propagation of \textsf{GCN}]{
%     \includegraphics[width=0.23\textwidth]{figures/backward_select.eps}}
%     \vspace{-2mm}
%     \caption{Comparison of \textsf{GCN} propagation.}
%     \label{backward}
% \vspace{-2mm}
% \end{figure}

\begin{table}[tbp]
\vspace{-2mm}
    \centering
    \caption{Effectiveness of kernel fusion method.}
    \vspace{-2mm}
    \resizebox{\linewidth}{!}{
    \begin{tabular}{|c||c|c|c|c|}
        \hline
        \textbf{Datasets} & \textbf{Fusing kernel} & \textbf{No optimization} & \textbf{Speedup} \\ \hline
        \hline
        \textit{YS} & 9.24ms & 12.20ms & 32.03\% \\
        \hline
        \textit{OC} & 10.12ms & 13.36ms & 32.02\% \\
        \hline
        \textit{YH} & 16.82ms & 22.05ms & 31.09\% \\
        \hline
        \textit{RD} & 26.46ms & 34.76ms & 31.37\% \\
        \hline
        \textit{TT} & 21.94ms & 27.74ms & 26.44\% \\
        \hline
    \end{tabular}
    }
    \label{tab:kernelfusion}
    \vspace{-2mm}
\end{table}

\subsubsection{Evaluation of kernel fusion strategy}
% Reducing the cost of data loading would significantly enhance the efficiency of Tensor cores in sparse matrix-matrix multiplication, especially when coupled with our layout optimization algorithm \textsf{LOA}. 
We assess the execution time of a single GNN layer during backward propagation with and without kernel fusion. The results are illustrated in Table \ref{tab:kernelfusion}. The kernel fusion technique achieves an average speedup of 30.6\% over the 5 representative datasets. The performance of kernel fusion is stable on different datasets, which demonstrates the efficacy of the technique in mitigating the overhead associated with memory access and kernel launch. As mentioned in \S~\ref{bootleneckTCU}, the bottleneck in matrix multiplication using Tensor cores is global memory access. 
% The overhead of global memory access is several times as large as the calculation time of Tensor cores. 
By directly storing the results of \textit{Aggregation} phase into shared memory and fetching them in \textit{Update} phase, the memory access overhead can be significantly reduced. 

\subsubsection{Layout optimization}
\label{eva:layout}
\begin{figure}[tbp]
    \centering
    \hspace{4mm}
    % \vspace{-4mm}
    \includegraphics[width=0.48\textwidth]{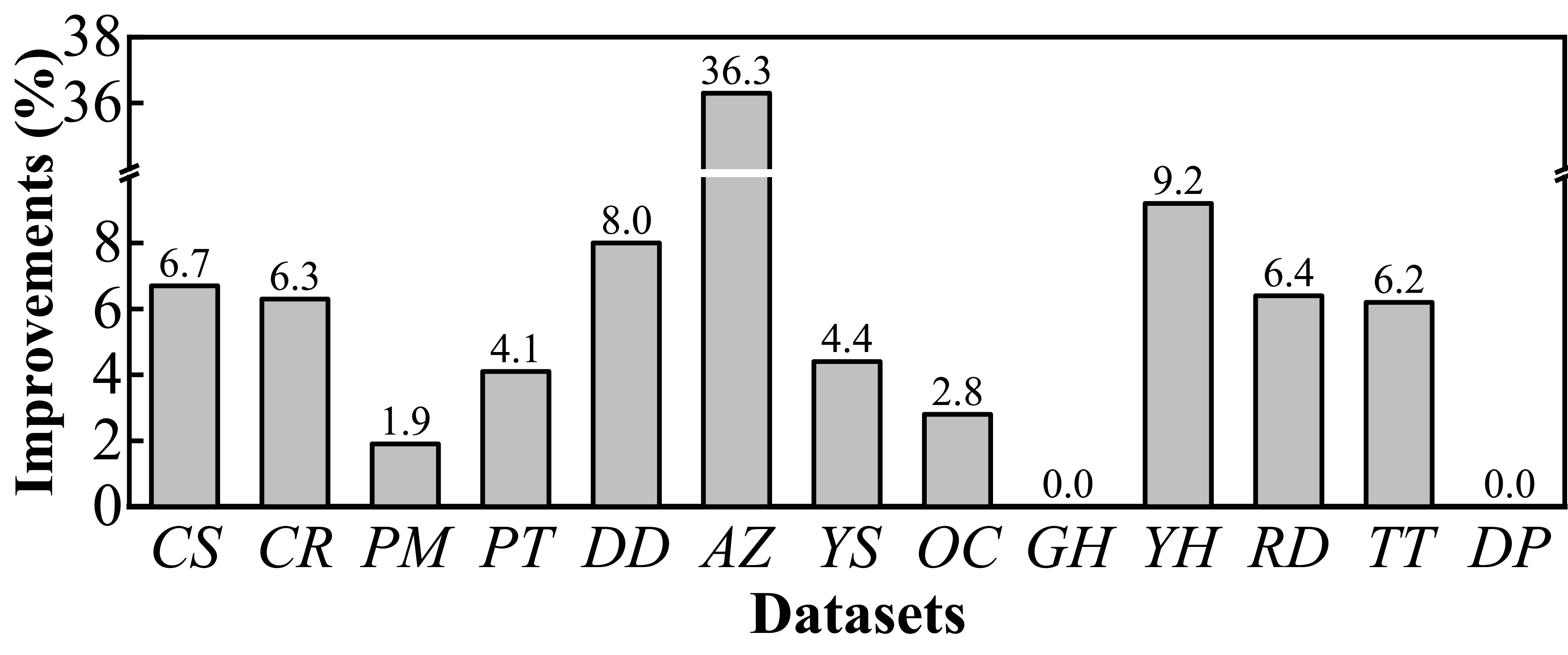}
    \vspace{-4mm}
    \caption{Improvements of layout optimization.}
    \label{ggg}
    \vspace{-4mm}
\end{figure}

% In this section, we evaluate the effectiveness of our proposed layout optimization algorithm \textsf{LOA}. 
The improvements brought by \textsf{LOA} are depicted in Figure \ref{ggg}. \textsf{LOA} achieves an average 8.40\% performance improvement across the 11 datasets, with the exception of \textit{GH} and \textit{DP}. \textsf{LOA} can achieve a maximum performance improvement of 36.3\%. The remarkable improvement on \textit{AZ} is attributed to two factors. Firstly, the original layout of \textit{AZ} is suboptimal, with vertex IDs in the adjacency list being scattered, leading to poor data locality. Secondly, each row window in the original \textit{AZ} is relatively sparse, with only one row window deemed suitable for Tensor cores according to the logistic regression model. \textsf{LOA} adjusts the original layout heuristically, densifying it to allow for the efficient computation of more row windows by Tensor cores. In contrast, the impact of \textsf{LOA} is sometimes not pronounced on some datasets, such as \textit{GH} and \textit{DP}, as these datasets already have favorable original layouts. 

\begin{figure}[tbp]
    \centering
    % \vspace{-2mm}
    \subfigcapskip=-5pt
    \subfigure[Before performing \textsf{LOA}]{
    \includegraphics[width=0.23\textwidth]{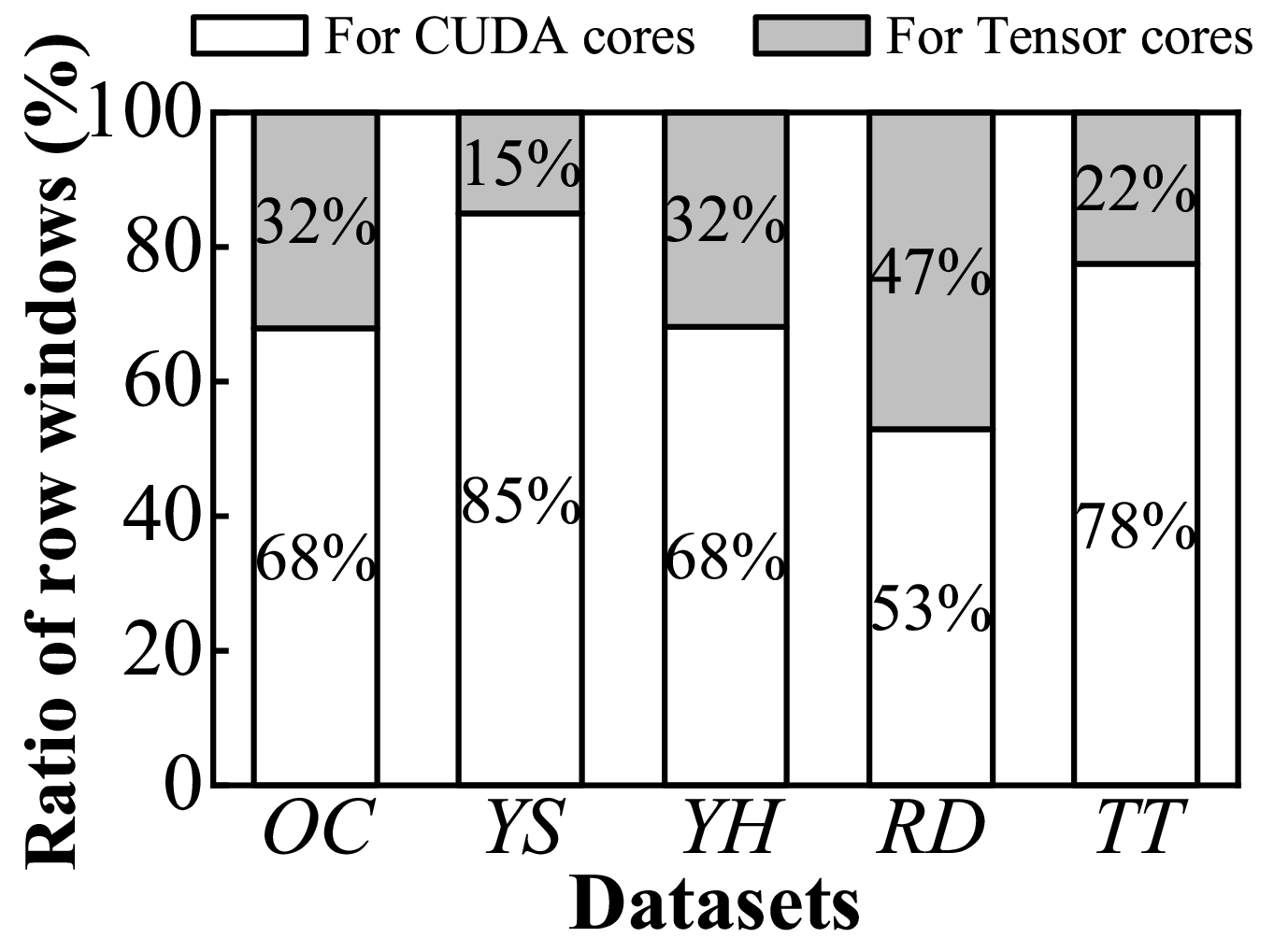}}
    \subfigure[After performing \textsf{LOA}]{
    \includegraphics[width=0.23\textwidth]{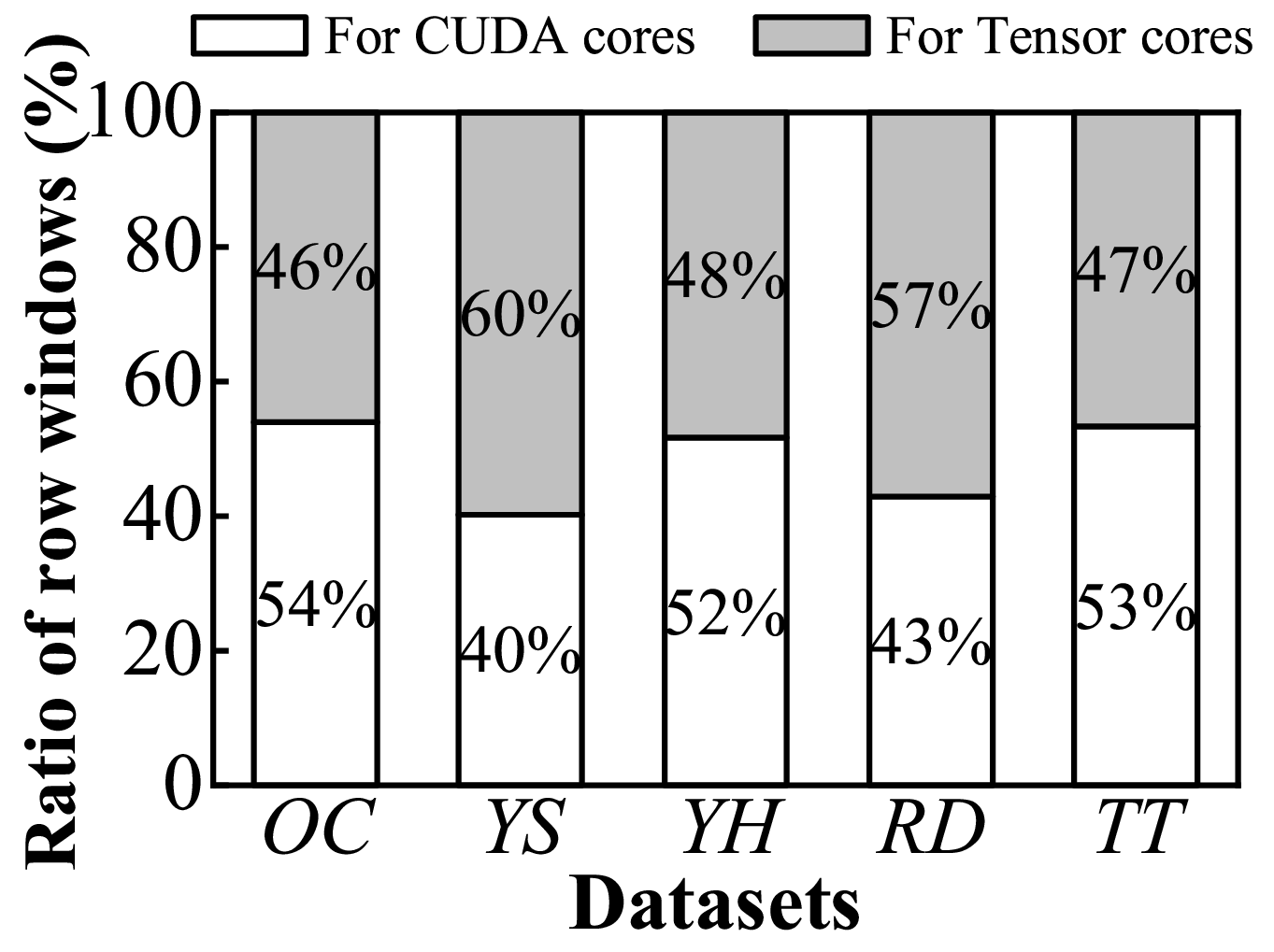}}
    \vspace{-3mm}
    \caption{Effectiveness of \textsf{LOA}.}
    \label{fig:effect}
\vspace{-2mm}
\end{figure}

We then conduct a thorough analysis of the effectiveness of \textsf{LOA}. We quantify the number of row windows calculated by Tensor cores and CUDA cores before and after employing \textsf{LOA}. An increase in the count of row windows suitable for Tensor cores suggests that the calculation time using Tensor cores is shorter than using CUDA cores for more row windows, potentially resulting in more improvements relative to solely using CUDA cores. The results are presented in Figure \ref{fig:effect}. It is evident that across most datasets before using \textsf{LOA}, there are significantly fewer row windows suitable for Tensor cores compared to CUDA cores. This phenomenon arises from the inherently sparse nature of the original graph layout. 
% However, CUDA cores are less efficient than Tensor cores. 
We can improve the graph layout with little overhead by leveraging \textsf{LOA}. As shown in Figure \ref{fig:effect}(b), \textsf{LOA} increases the number of row windows suitable for Tensor cores, leading to more efficient calculations. 

\begin{figure}[tp]
    \centering
    \includegraphics[width=0.3\textwidth]{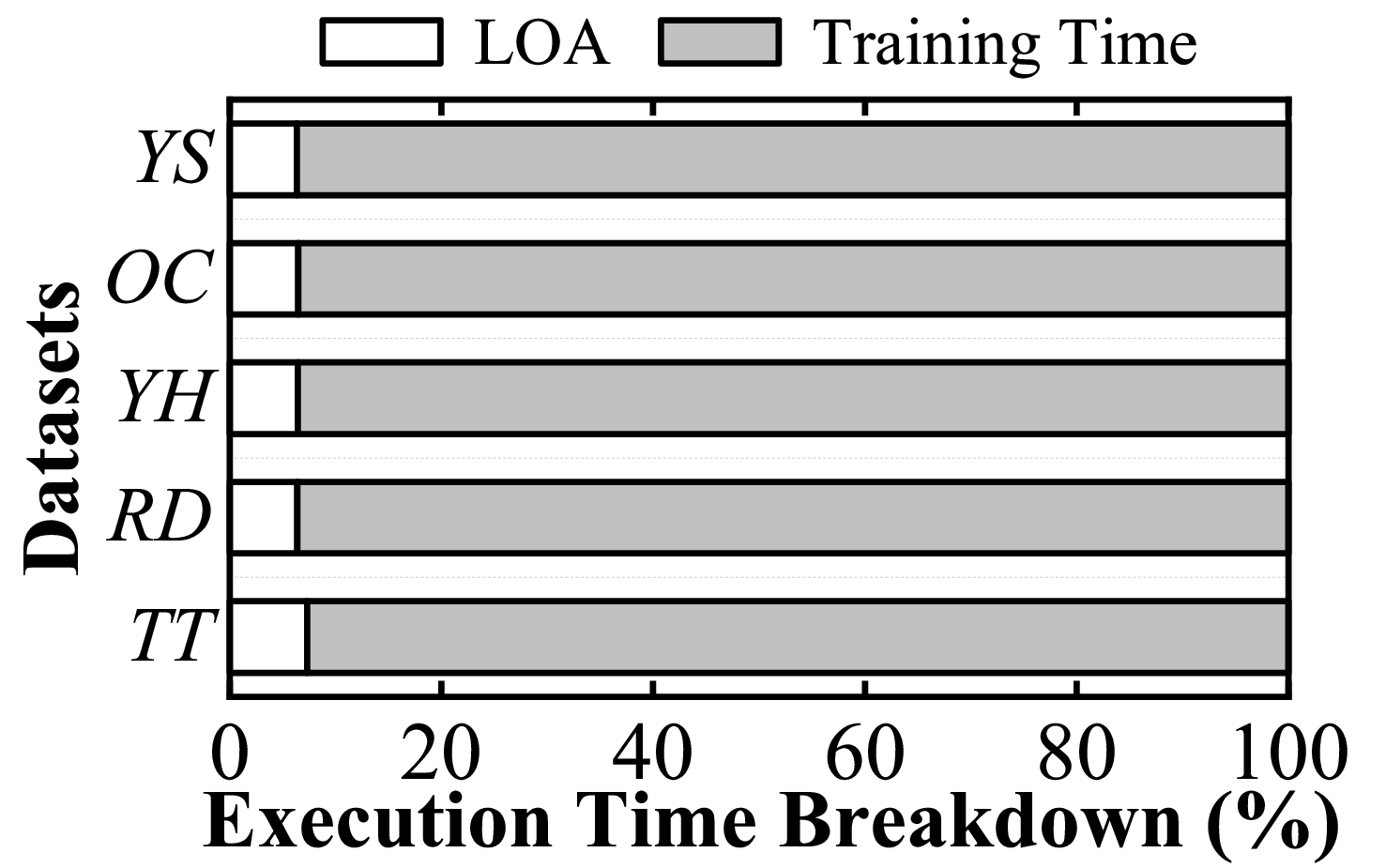}
    \vspace{-3mm}
    \caption{Overhead of \textsf{LOA} in the entire training.}
    \label{fig:prop}
    \vspace{-2mm}
\end{figure}

As a preprocessing algorithm, \textsf{LOA} necessitates that its time cost be maintained within a tolerable threshold. 
Figure \ref{fig:prop} compares the overhead of \textsf{LOA} with the training time (200 epochs). 
% We compare the overhead of \textsf{LOA} with the training time (200 epochs). 
The overhead consistently remains low, accounting for only 6.58\% of the training time on average, which is already lower than the benefit of \textsf{LOA} (8.40\%) in 200 epochs. The execution of \textsf{LOA} is offline, which has constant overhead regardless of the number of epochs and layers. Conversely, the benefit of \textsf{LOA} will accumulate as the epochs increase. With an increase in the number of epochs, particularly for larger datasets and deeper models that require more epochs to converge, this preprocessing overhead becomes more negligible\footnote[18]{For example, the overhead of \textsf{LOA} accounts for only about 3\% of the GNN training time in the case of 400 epochs, while the speedup ratio of \textsf{LOA} remains unchanged, which is 8.4\% on average. }. In such scenarios, we advocate for utilizing \textsf{LOA}.

% \begin{figure}[tbp]
%     \centering
%     \includegraphics[width=0.3\textwidth]{figures/reorderbreakdown.eps}
%     \vspace{-3mm}
%     \caption{Overhead of \textsf{LOA} in the entire training.}
%     \label{fig:prop}
%     \vspace{-4mm}
% \end{figure}

\section{Conclusion}
\label{sec:conclusion}
We present \textsf{HC-SpMM}, a novel algorithm for accelerating SpMM on graphs. Being the first GPU accelerator to utilize hybrid GPU cores, i.e., CUDA and Tensor cores, \textsf{HC-SpMM} offers pioneering ideas that lay the foundation for future research. It intelligently selects hardware structures suitable for submatrices, fully leveraging the computational characteristics of different GPU cores. Meanwhile, we propose a thread utilization and data loading strategy to optimize the efficiency. Furthermore, to support graph computing applications, we integrate \textsf{HC-SpMM} into \textsf{PyTorch} and employ it to accelerate GNN training. To better improve the efficiency of GNN training, we devise a kernel fusion strategy and an algorithm \textsf{LOA} to reuse data, improve the graph layout, and better fit the computational models of hybrid cores. Experimental results consistently demonstrate the superiority of {\sf HC-SpMM} over existing SpMM kernels and GNN training frameworks. 

% \newpage
% \balance
% \clearpage

\bibliographystyle{abbrv}
\bibliography{ref}

% \newpage
\appendix
\label{supexp}
\subsection{Evaluation on various GPU architectures}
\label{absoluteSpMM}
We report the absolute numbers related to the execution times of SpMM kernels and compare the SpMM kernels on {Nvidia A100 and RTX 4090 GPUs} to evaluate the generality of the regression model, which are depicted in Table \ref{spmm_abs_numbers}. 

Our proposed SpMM kernel is superior to other SpMM kernels in most cases, and the performance of the logistic regression model is stable on different types of GPUs. 

\subsection{Evaluation on various FP types}
\label{evaFPtype}
{We evaluate 3 commonly used FP types: {\sf TF32}, {\sf half}, and {\sf bfloat16}. 
In this evaluation, we exclude {\sf GE-SpMM} and {\sf DTC-SpMM} from the comparison, as they do not support other FP types. {\sf Sputnik} supports both {\sf float} and {\sf half} precision. We modify {\sf TC-GNN} to enable {\sf half} precision. The results are reported in Table \ref{tab:fptype}. 
For {\sf HC-SpMM}, the performance of {\sf half} and {\sf bfloat16} is similar. {\sf Sputnik}, having been specifically optimized for {\sf half} precision, achieves up to more than 2$\times$ speedup over its own full precision implementation. Although {\sf HC-SpMM} doesn't include optimizations for {\sf half} precision, it outperforms {\sf Sputnik} by up to 1.26$\times$ and exhibits superior performance in most datasets. This demonstrates the adaptability and scalability of our proposed method across different FP types. Due to the 16$\times$16$\times$16 matrix input requirement of the {\sf WMMA} API, {\sf TC-GNN} exhibits worse performance than itself with {\sf TF32}. This is because {\sf TF32} has an input requirement of 16$\times$8$\times$16, which has a smaller granularity than 16$\times$16$\times$16, resulting in fewer unnecessary calculations involving zeros.}  

\begin{table}[tbp]
    \centering
    \caption{Overhead of SpMM on different FP types.($\times 10^{-6}$s)}
    \vspace{-3mm}
    \begin{tabular}{|c|c|c|c|c|}
    \hline
            & {\sf Sputnik} & {\sf TC-GNN} & {\sf HC-SpMM}(half) & {\sf HC-SpMM}(bfloat) \\
    \hline
        \textit{CS} & 4.99 & 23.29 & \textbf{4.06} & 4.06 \\
    \hline
    \textit{CR} & 6.26 & 41.95 & 5.54 & \textbf{5.41}  \\
    \hline
    \textit{PM} & 9.09 & 94.2 & \textbf{10.43} & 10.56  \\
    \hline 
    \textit{PT} & \textbf{12.77} & 37.63 & 15.42 & 15.39  \\
    \hline
    \textit{DD} & 117.82 & 248.64 & \textbf{105.28} & 105.47  \\
    \hline
    \textit{AZ} & \textbf{159.87} & 618.24 & 223.14 & 223.10 \\
    \hline
    \textit{YS} & 575.90 & 683.55 & \textbf{477.31} & 477.34  \\
    \hline
    \textit{OC} & 585.79 & 813.49 & 515.74 & \textbf{515.10} \\
    \hline
    \textit{GH} & 591.04 & 1226.78  & 476.93 & \textbf{476.51}  \\
    \hline
    \textit{YH} & 975.03 & 1319.19 & 851.84 & \textbf{851.16}  \\
    \hline
    \textit{RD} & 1652.57 & 1916.91 & 1304.47 & \textbf{1301.69} \\
    \hline
    \textit{TT} & 1484.54 & 2207.99 & \textbf{1200.03} & 1201.24  \\
    \hline
    \textit{DP} & \textbf{9096.03} & 149166.60 & 15138.63 & 15119.01  \\
    \hline
    \end{tabular}
    \vspace{-2mm}
    \label{tab:fptype}
\end{table}

\subsection{Absolute Numbers of End-to-end Training}
\label{absoluteGNN}
We report the absolute numbers of the training overhead of GCN in Table \ref{gnn_abs_numbers} (corresponding to Figure \ref{backward}) and GIN in Table \ref{gin_abs_numbers} (corresponding to Figure \ref{gineva}). 

\begin{table}[tbp]
    \centering
        \caption{Average epoch time of GCN training. ($\times 10^{-3}$ s)}  
        \vspace{-3mm}
        \label{gnn_abs_numbers} 
        \begin{tabularx}{0.99\linewidth}{|c|X|X|X|X|}
            \hline 
            \multicolumn{2}{|c|}{} & \textsf{GE-SpMM} & \textsf{TC-GNN} & \textsf{HC-SpMM}  \\
            \hline
            \multirow{2}{*}{\textit{CS}} & Forward& 0.33 & 0.37  & \textbf{0.31}  \\
            \cline{2-5} & Backward & 0.53 & 0.51 & \textbf{0.45} \\
            \hline
            \multirow{2}{*}{\textit{CR}} & Forward & 0.30 & 0.45 & \textbf{0.26} \\
            \cline{2-5} & Backward & 0.45 & 0.45 & \textbf{0.36}  \\
            \hline
            \multirow{2}{*}{\textit{PM}} & Forward & 0.32 & 0.72 & \textbf{0.28}  \\
            \cline{2-5} & Backward & 0.43 & 0.78 & \textbf{0.43} \\
            \hline
            \multirow{2}{*}{\textit{PT}} & Forward & 0.35 & 0.42  & \textbf{0.32} \\
            \cline{2-5} & Backward & 0.49 & 0.51 & \textbf{0.42}  \\
            \hline
            \multirow{2}{*}{\textit{DD}} & Forward & 2.45 & 2.81 & \textbf{2.17}  \\
            \cline{2-5} & Backward  & 2.85 & 3.22 & \textbf{2.09} \\
            \hline
            \multirow{2}{*}{\textit{AZ}} & Forward  & 3.94 & 5.63 & \textbf{3.41} \\
            \cline{2-5} & Backward  & 4.36 & 6.59 & \textbf{3.82} \\
            \hline
            \multirow{2}{*}{\textit{YS}} & Forward & 11.46 & 11.01  & \textbf{10.12} \\
            \cline{2-5} & Backward & 13.44 & 13.02 & \textbf{9.24}  \\
            \hline
            \multirow{2}{*}{\textit{OC}} & Forward  & 12.19 & 12.32 & \textbf{10.98} \\
            \cline{2-5} & Backward & 14.56 & 14.75 & \textbf{10.12} \\
            \hline
            \multirow{2}{*}{\textit{GH}} & Forward & 9.15 & 12.10 & \textbf{7.88} \\
            \cline{2-5} & Backward  & 11.76 & 14.67 & \textbf{8.30} \\
            \hline
            \multirow{2}{*}{\textit{YH}} & Forward  & 20.73 & 20.98 & \textbf{18.74} \\
            \cline{2-5} & Backward  & 23.90 & 24.20 & \textbf{16.82} \\
            \hline
            \multirow{2}{*}{\textit{RD}} & Forward & 28.67 & 28.48 & \textbf{25.30}  \\
            \cline{2-5} & Backward & 38.03 & 37.77 & \textbf{26.46}  \\
            \hline
            \multirow{2}{*}{\textit{TT}} & Forward  & 23.86 & 26.49 & \textbf{20.46} \\
            \cline{2-5} & Backward  & 31.06 & 33.40 & \textbf{21.94} \\
            \hline
        \end{tabularx}
        \vspace{-2mm}
\end{table}

\begin{table}[tbp]
    \centering
    \vspace{-2mm}
        \caption{Average epoch time of GIN training. ($\times 10^{-3}$ s)}  
        \vspace{-3mm}
        \label{gin_abs_numbers} 
        \begin{tabularx}{\linewidth}{|c|X|X|X|X|}
            \hline 
            \multicolumn{2}{|c|}{} & \textsf{GE-SpMM} & \textsf{TC-GNN} & \textsf{HC-SpMM} \\
            \hline
            \multirow{2}{*}{\textit{YS}} & Forward & 12.70 & 11.75 & \textbf{8.16}  \\
            \cline{2-5} & Backward & 14.89 & 13.82 & \textbf{13.26}  \\
            \hline
            \multirow{2}{*}{\textit{OC}} & Forward & 13.55 & 13.10 & \textbf{8.92}  \\
            \cline{2-5} & Backward  & 15.81 & 15.44 & \textbf{14.65} \\
            \hline
            \multirow{2}{*}{\textit{YH}} & Forward  & 23.09 & 23.32 & \textbf{15.11} \\
            \cline{2-5} & Backward  & 26.38 & 25.46 & \textbf{24.14} \\
            \hline
            \multirow{2}{*}{\textit{RD}} & Forward & 32.55 & 31.36 & \textbf{21.49}  \\
            \cline{2-5} & Backward  & 41.92 & 40.65 & \textbf{39.27} \\
            \hline
            \multirow{2}{*}{\textit{TT}} & Forward  & 26.88 & 29.19 & \textbf{20.15} \\
            \cline{2-5} & Backward  & 34.09 & 36.26 & \textbf{32.92} \\
            \hline
        \end{tabularx}
        \vspace{-2mm}
\end{table}

\subsection{Evaluations on various sparsity}
\label{additionexp}
To verify the adaptability of \textsf{HC-SpMM} to different densities and demonstrate the necessity of using hybrid GPU cores, we conduct experiments on synthetic matrices with different sparsity. 
We vary the number of non-zero elements in $16 \times 8$ non-zero blocks to generate various synthetic matrices. 

\begin{table}[tbp]
    \centering
    \vspace{-2mm}
    \caption{Runtime of SpMM kernels with various sparsity. ($\times 10^{-6}$ s)}
    \vspace{-3mm}
    \label{tab:density}
    \begin{tabularx}{\linewidth}{|c|X|X|X|X|}
    \hline
    \diagbox{Methods}{Sparsity}  & 80\% & 85\% & 90\% & 95\% \\
        \hline
    \textsf{Sputnik} & 9.28 & 8.58 & 6.67 & 6.10 \\
    \hline
    \textsf{GE-SpMM} & 9.34 & 8.93 & 8.77 & 7.90 \\
    \hline
    \textsf{TC-GNN} & 14.85 & 14.56 & 13.41 & 10.75 \\
    \hline
    \textsf{DTC-SpMM} & 8.21 & 8.35 & 7.94 & 6.45 \\
    \hline
    \textsf{HC-SpMM} & \textbf{7.49} & \textbf{6.62} & \textbf{5.73} & \textbf{5.31} \\
    \hline
    \end{tabularx}
    \vspace{-2mm}
\end{table}

\begin{figure*}[tbp]
    \centering    
    \hspace{-2mm}
    \includegraphics[width=0.16\textwidth]{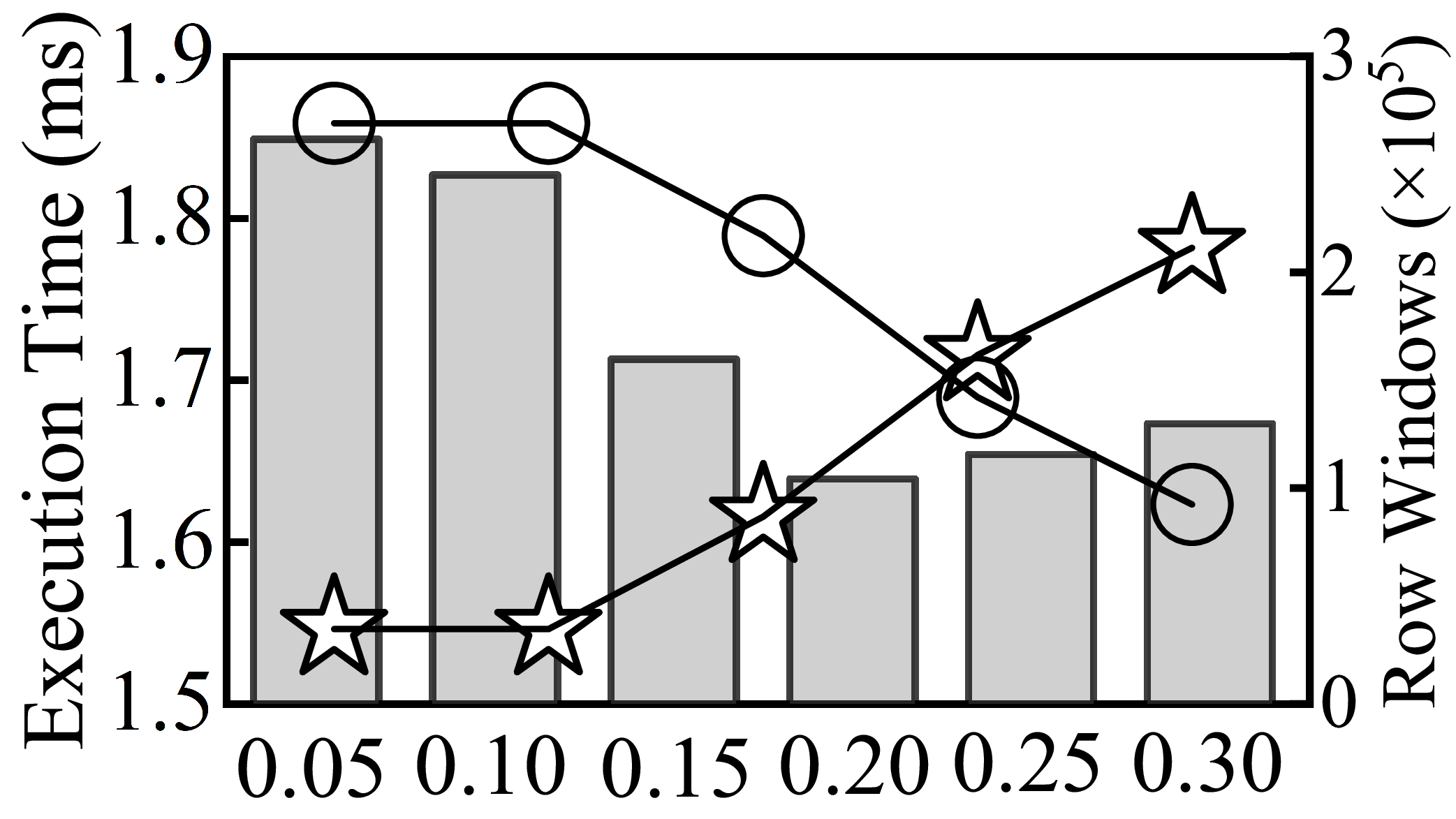}
    % \hspace{-6mm}
    \includegraphics[width=0.17\textwidth]{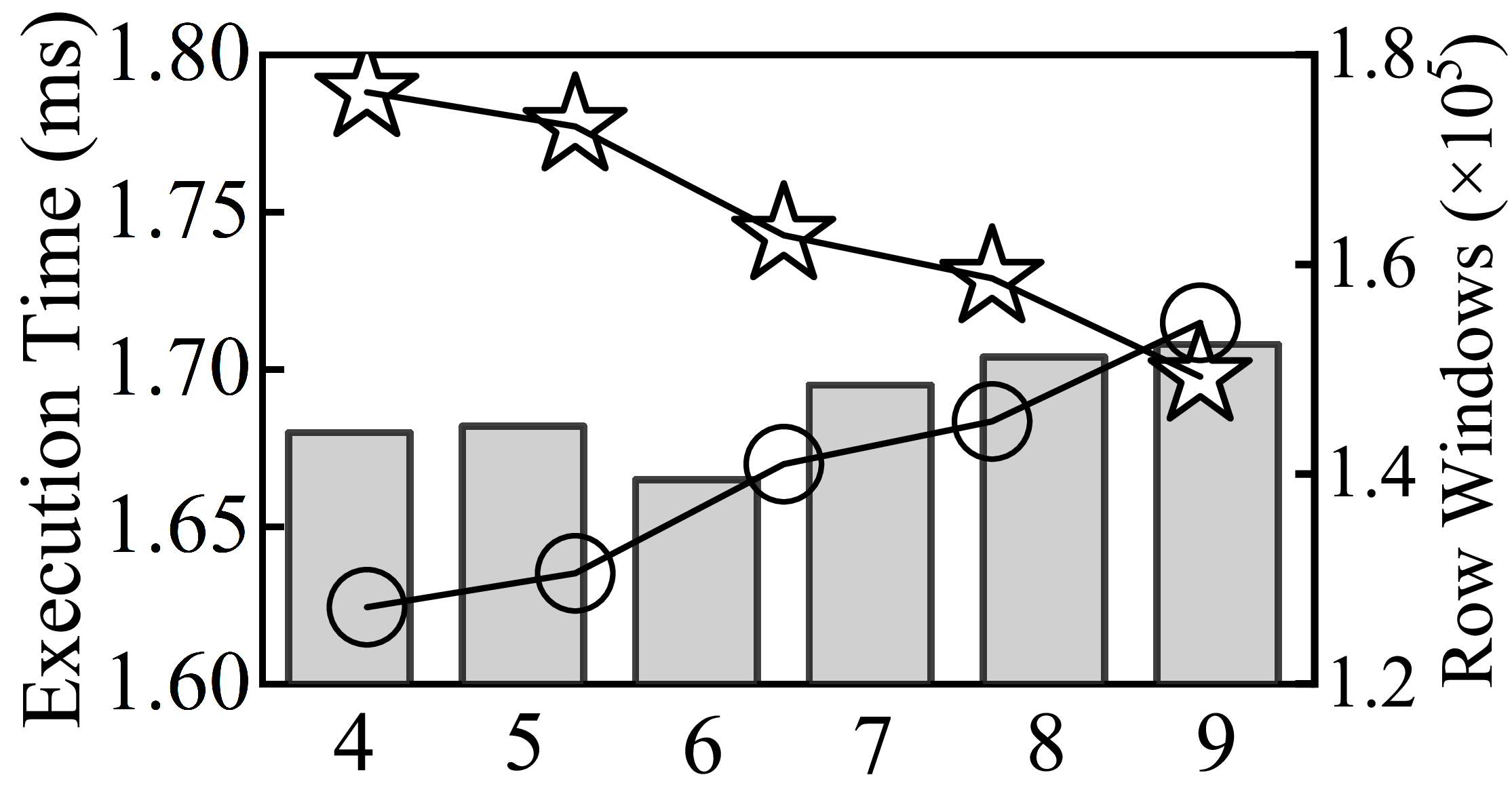}
    % \hspace{6mm}
    \includegraphics[width=0.16\textwidth]{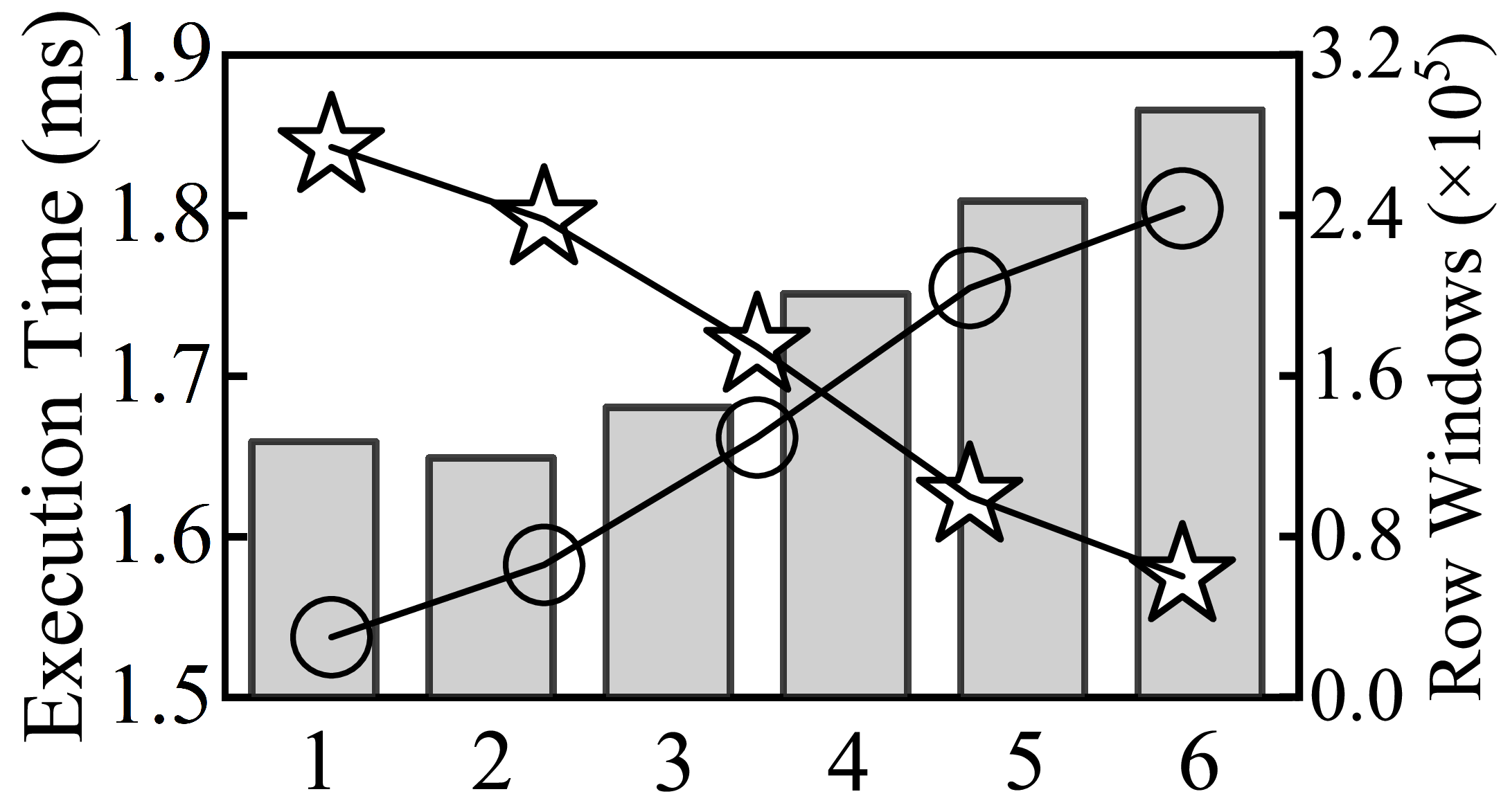}
    \includegraphics[width=0.16\textwidth]{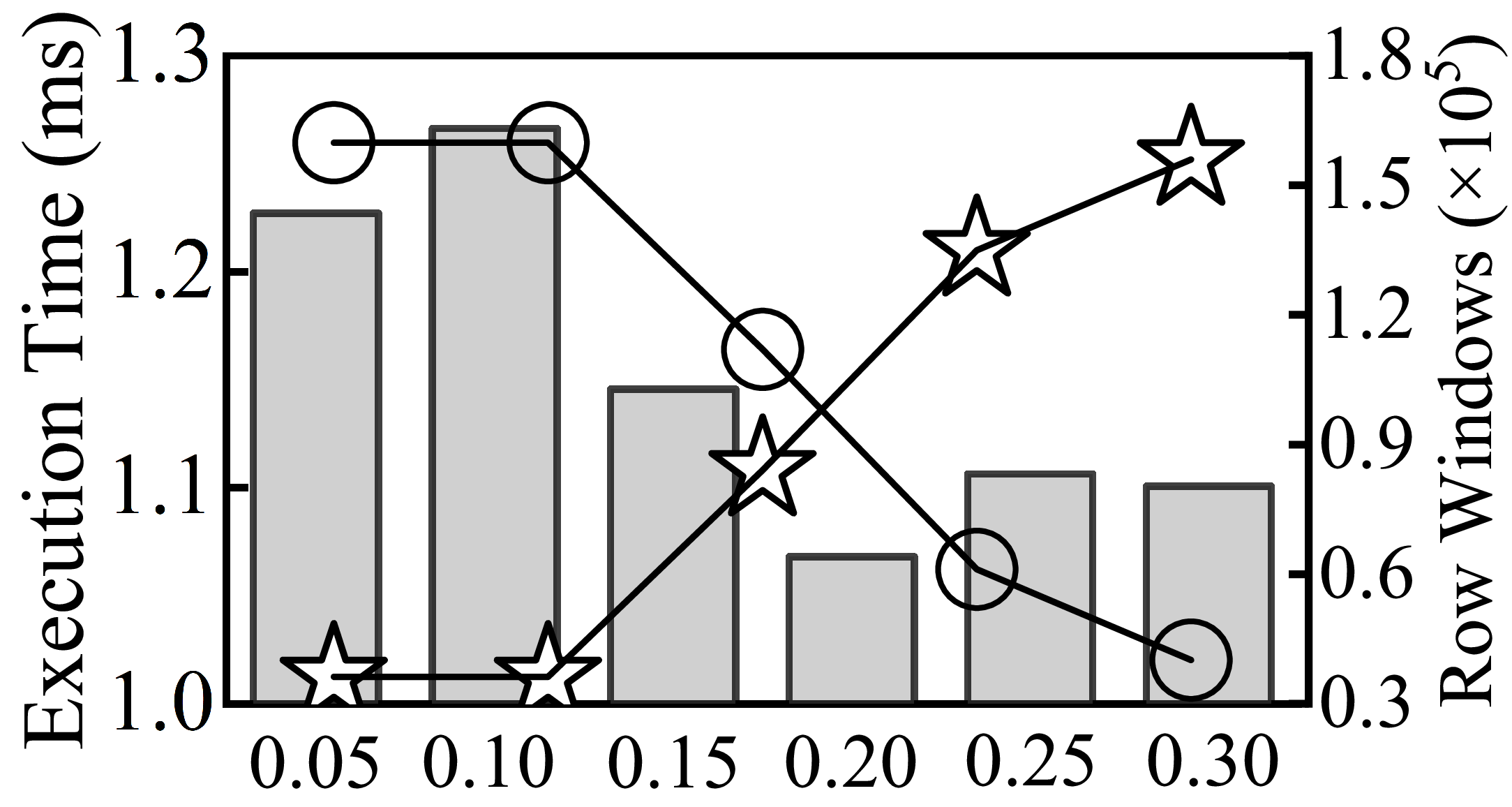}
    % \hspace{-6mm}
    \includegraphics[width=0.165\textwidth]{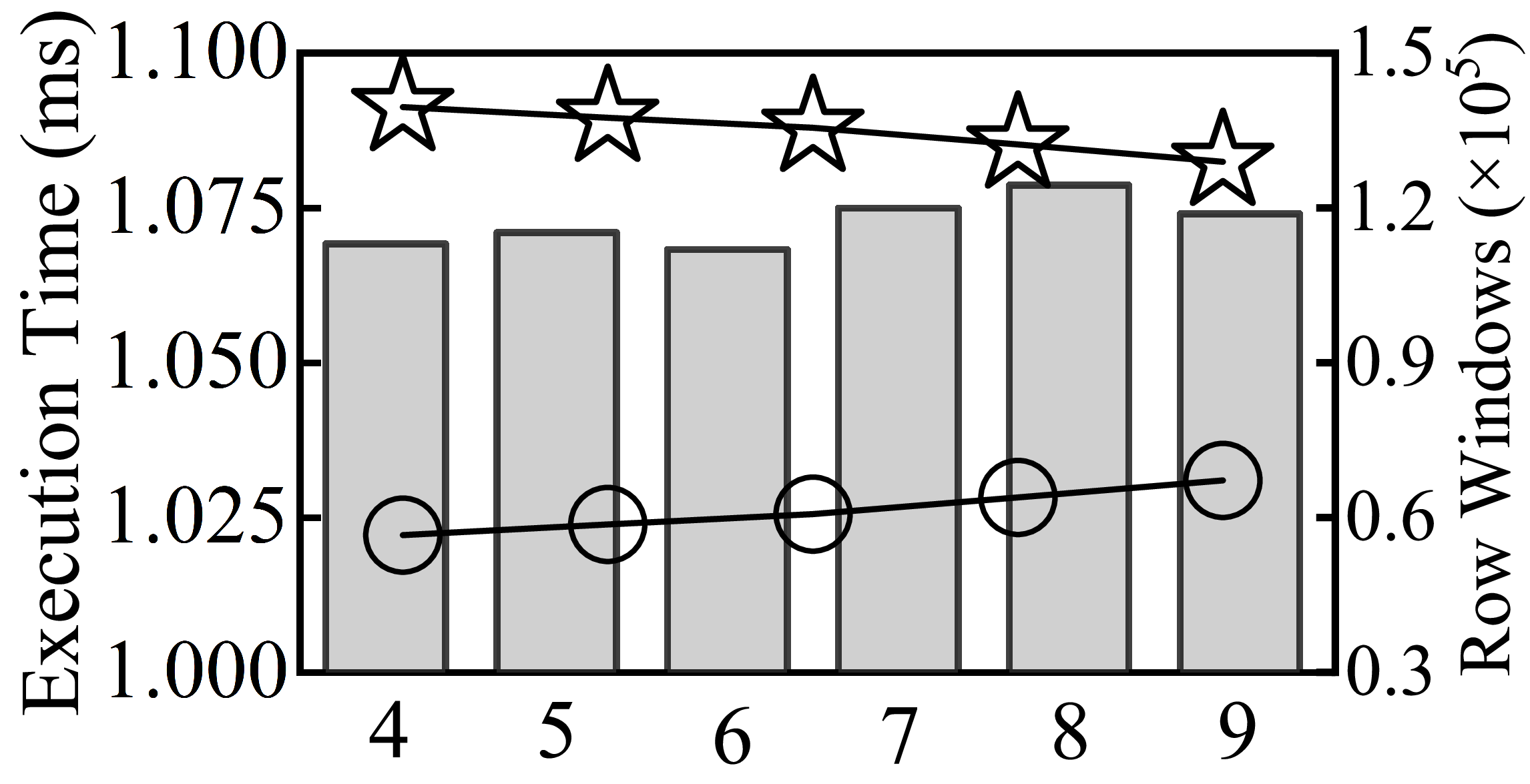}
    % \hspace{6mm}
    \includegraphics[width=0.16\textwidth]{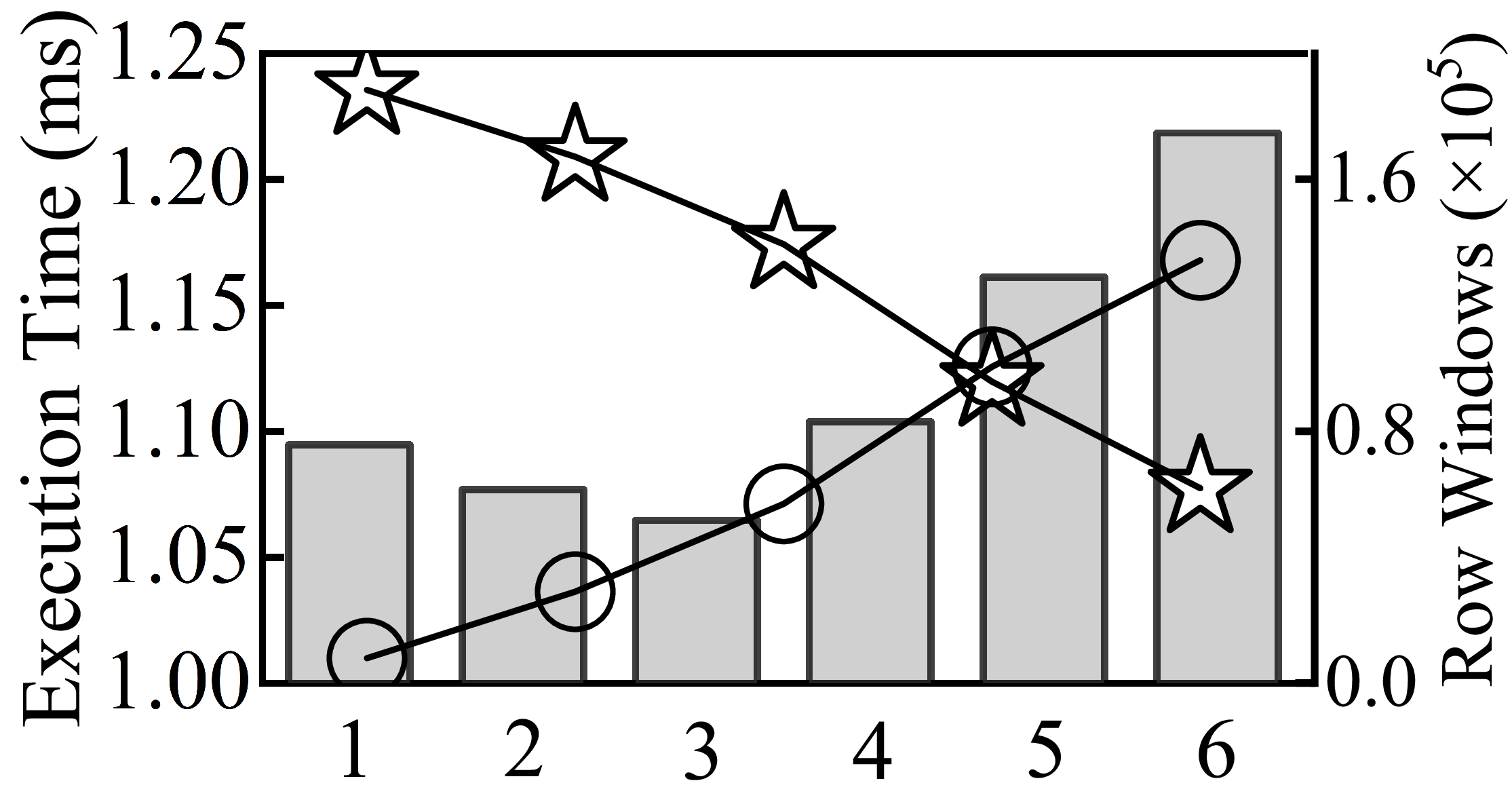}
    
    \vspace{-4mm}
    \caption{Sensity of regression model parameters to SpMM performance.}
    \label{fig:sensity}
\vspace{-6mm}
\end{figure*}

Execution times of different SpMM kernels are reported in Table \ref{tab:density}. \textsf{HC-SpMM} achieves the best performance on all synthetic matrices. When the sparsity is less than 85\%, \textsf{DTC-SpMM} exhibits less execution time than \textsf{Sputnik}, while \textsf{Sputnik} performs better than \textsf{DTC-SpMM} when the sparsity is greater than 90\%. This phenomenon is consistent with our experimental results in Figure \ref{executiontime}, which demonstrate that when there are more dense parts in a matrix, Tensor cores (represented by \textsf{DTC-SpMM}) have higher efficiency than CUDA cores (represented by \textsf{Sputnik}). 

\subsection{Sensity Evaluation}
\label{exp:sensity}
{We evaluate the sensitivity of performance to the specific logistic regression model parameters on two datasets \textit{YH} and \textit{RD}. The model includes three parameters: $w_1$, corresponding to the number of non-zero columns; $w_2$ representing the coefficient of the sparsity; and $b$, the intercept. Figure \ref{fig:sensity} depicts the results.
The performance sensitivity to $w_1$ and $b$ is similar, where a 50\% change in parameter values leads to a performance variation of around 14\%. The changes in parameters $w_1$ and $b$ significantly affect the number of row windows classified into CUDA cores and Tensor cores. In contrast, a 50\% change in the value of $w_2$ results in a performance difference of up to 3\%. Overall, the performance is more sensitive to variations in $w_1$ and $b$ than in $w_2$. }

\subsection{Evaluation of the Preprocessing Overhead}
\label{exp:preprocess}
{
We adopt the preprocessing method from {\sf DTC-SpMM}, which performs preprocessing directly on the GPU, eliminating additional PCIe transfers. The preprocessing overhead consists of two parts: preparing data for Tensor cores and selecting appropriate GPU cores for row windows. }

{The preprocessing overhead related to Tensor cores is unavoidable due to the sparsity of input matrices. This step involves densifying each row window by moving non-zero columns to the front of the row windows. Existing SpMM algorithms on Tensor cores such as {\sf TC-GNN} and {\sf DTC-SpMM} have the same preprocessing process. We adopt the preprocessing kernel from {\sf DTC-SpMM} while removing the unnecessary process in {\sf HC-SpMM}. The comparison of preprocessing overhead is presented in Table \ref{tab:preprocess}. Specifically, the preprocessing of {\sf HC-SpMM} outperforms that of {\sf DTC-SpMM} and {\sf TC-GNN} by factors of 1.3$\times$ and 36.0$\times$, respectively. On average, the preprocessing overhead in {\sf HC-SpMM} is about 13.0$\times$ that of a single SpMM execution. }

{The selection overhead is negligible because the inference of the logistic regression model simply involves calculating $w_1*x_1+w_2*x_2+b$, where $x_1$ and $x_2$ are computed during the preprocessing of Tensor cores, and $w_1$, $w_2$, and $b$ are parameters retrieved from the regression model trained offline. Consequently, this selection step takes only a few nanoseconds to complete. There is no preprocessing overhead for SpMM on CUDA cores. The transferred data through PCIe consists only of the CSR format of graphs.} 

{We didn't include the preprocessing overhead in Figure 10 because, in many real-world applications such as GNNs, thousands of SpMM operations are performed on unchanged sparse matrices~\cite{fan2024dtc,wen2018thundersvm}. The preprocessing overhead is negligible in these cases. For scenarios where sparse matrices are constantly changing, SpMM methods optimized for CUDA cores such as {\sf Sputnik} are more suitable.} 

\begin{table}[tbp]
    \centering
    \renewcommand{\arraystretch}{1.0}
    \vspace{-2mm}
    \caption{Comparison of preprocessing overhead (ms). }
    \vspace{-3mm}
    \begin{tabularx}{\linewidth}{|c|X|X|X|}
    \hline
        Datasets & {\sf DTC-SpMM} & {\sf TC-GNN} & {\sf HC-SpMM} \\
         \hline
        \textit{YS} & 11.48 & 241.50 & \textbf{8.72}\\
         \hline
        \textit{OC} & 11.56 & 284.81 & \textbf{9.38} \\
         \hline
        \textit{YH} & 15.03 & 457.70 & \textbf{11.82}  \\
         \hline
        \textit{RD} & 20.44 & 671.76 & \textbf{15.72}  \\
         \hline
        \textit{TT} & 33.94 & 966.86 & \textbf{24.02}  \\
         \hline
    \end{tabularx}
    \vspace{-2mm}
    \label{tab:preprocess}
\end{table}

\subsection{Evaluation of Memory Usage}
\label{exp:memory}
{Note that an out-of-memory error occurs during training GNNs on \textit{DP} for {\sf HC-SpMM}, {\sf GE-SpMM}, and {\sf TC-GNN}. To demonstrate the comparable memory usage of {\sf HC-SpMM}, we evaluate the memory usage of the three methods on 5 datasets. Table \ref{tab:memoryusage} depicts the results. The memory usage of {\sf HC-SpMM} is only up to 2\% more than {\sf GE-SpMM} and up to 6\% more than {\sf TC-GNN} due to the additional data structure for efficient SpMM with hybrid GPU cores.} 

\begin{table}[tbp]
\renewcommand{\arraystretch}{1.0}
    \centering
    \vspace{-2mm}
    \caption{Comparison of memory usage (MB). }
    \vspace{-3mm}
    \begin{tabularx}{\linewidth}{|c|X|X|X|X|}
    \hline
        Datasets & {\sf GE-SpMM} & {\sf TC-GNN} & {\sf HC-SpMM} \\
        \hline
        \textit{YS} & 4736 & \textbf{4542} & 4752 \\
         \hline
        \textit{OC} & 5096 & \textbf{4870} & 5122  \\
         \hline
        \textit{YH} & 7558 & \textbf{7210} & 7608  \\
         \hline
        \textit{RD} & 10976 & \textbf{10432} & 11046 \\
         \hline
        \textit{TT} & 8938 & \textbf{8570} & 9050  \\
         \hline
    \end{tabularx}
    \vspace{-2mm}
    \label{tab:memoryusage}
\end{table}

\subsection{Utilization Metrics of CUDA and Tensor Cores}
\label{exp:utilization}
{We evaluate several utilization metrics of the GPU cores. Accurately measuring the CUDA cores' utilization during SpMM is challenging, as they not only handle SpMM calculations but also manage data loading and preparation before the Tensor cores' computations. Therefore, we report only the utilization of Tensor cores in Table \ref{tab:utilization}. In {\sf HC-SpMM}, CUDA and Tensor cores don't execute concurrently, resulting in lower overall utilization because one type of GPU cores remains idle while the other is active.}

{We also evaluate the execution time for CUDA and Tensor cores, as shown in Table \ref{tab:tctime}. The execution time is proportional to the number of row windows processed by CUDA and Tensor cores in Figure \ref{fig:effect}(b). }

{Additionally, we measure and compare the computing and memory throughput using {\sf Nsight Compute}. The results, presented in Table \ref{tab:throughput}, demonstrate that {\sf HC-SpMM} achieves the highest computing and memory throughput compared to the other methods evaluated. }

\begin{table}[tbp]
\renewcommand{\arraystretch}{1.0}
    \centering
    \vspace{-2mm}
    \caption{Comparison of Tensor cores' utilization (\%). }
    \vspace{-3mm}
    \begin{tabularx}{\linewidth}{|c|X|X|X|}
    \hline
       Datasets & {\sf DTC-SpMM} & {\sf TC-GNN} & {\sf HC-SpMM}  \\
       \hline
        \textit{YS} & 3.69 & 2.77 & \textbf{3.85} \\
         \hline
        \textit{OC} & \textbf{3.82} & 2.84 & 2.82  \\
         \hline
        \textit{YH} & \textbf{3.79} & 2.82 & 2.94  \\
         \hline
        \textit{RD} & \textbf{3.51} & 2.69 & 2.72  \\
         \hline
        \textit{TT} & \textbf{4.07} & 2.96 & 2.36  \\
         \hline
    \end{tabularx}
    \vspace{-2mm}
    \label{tab:utilization}
\end{table}

\begin{table}[tbp]
\renewcommand{\arraystretch}{1.0}
    \centering
    \caption{Comparison of GPU cores' execution time (ms). }
    \vspace{-3mm}
    \begin{tabularx}{\linewidth}{|c|X|X|X|X|X|}
    \hline
       GPU cores & \textit{YS} & \textit{OC} & \textit{YH} & \textit{RD} & \textit{TT} \\
       \hline
        CUDA cores & 563.00 & 512.87 & 803.04 & 990.30 & 992.99 \\
         \hline
        Tensor cores & 110.68 & 202.73 & 327.17 & 786.15 & 245.35 \\
         \hline
    \end{tabularx}
    \vspace{-2mm}
    \label{tab:tctime}
\end{table}

\begin{table}[tbp]
\renewcommand{\arraystretch}{1.0}
    \centering
    \caption{Comparison of computing and memory throughput (\%). }
    \vspace{-3mm}
    \begin{tabular}{|c|c|c|c|c|c|c|}
        \hline 
        Types & Methods & \textit{YS} & \textit{OC} & \textit{YH} & \textit{RD} & \textit{TT} \\
        \hline
        \multirow{5}{*}{Computing} & {\sf TC-GNN} & 31.35 & 31.66 & 31.63 & 30.68 & 34.35 \\
        \cline{2-7}
          & {\sf Sputnik} & 18.36 & 18.51 & 18.53 & 20.04 & 26.31 \\
        \cline{2-7}
          & {\sf GE-SpMM} & 40.61 & 39.74 & 39.52 & 38.28 & 57.05 \\
          \cline{2-7}
          & {\sf DTC-SpMM} & 44.08 & 45.24 & 45.09 & 42.17 & 47.61 \\
          \cline{2-7}
          & {\sf HC-SpMM} & \textbf{53.77} & \textbf{52.22} & \textbf{51.94} & \textbf{51.62} & \textbf{75.97} \\
          \hline
          \multirow{5}{*}{Memory} & {\sf TC-GNN} & 31.24 & 31.46 & 31.45 & 31.38 & 33.11 \\
        \cline{2-7}
          & {\sf Sputnik} & 64.22 & 64.71 & 64.90 & 56.84 & 54.04 \\
        \cline{2-7}
          & {\sf GE-SpMM} & 55.49 & 58.72 & 58.84 & 55.51 & 57.05 \\
          \cline{2-7}
          & {\sf DTC-SpMM} & 78.70 & 75.06 & 76.58 & 79.48 & 67.16 \\
          \cline{2-7}
          & {\sf HC-SpMM} & \textbf{84.58} & \textbf{87.11} & \textbf{87.51} & \textbf{89.76} & \textbf{82.77} \\
          \hline
    \end{tabular}
    \vspace{-2mm}
    \label{tab:throughput}
\end{table}

\begin{table*}[tbp]
\centering
    % \begin{center}   
        \caption{Overhead of SpMM on different GPUs. ($\times 10^{-6}$ s)}  
        \vspace{-2mm}
        \label{spmm_abs_numbers} 
        % \resizebox{1\linewidth}{!}{
        \begin{tabular}{|c|c|c|c|c|c|c|c|c|}
            \hline   \multicolumn{2}{|c|}{} & \textsf{Sputnik} & \textsf{GE-SpMM} & \textsf{TC-GNN} & \textsf{DTC-SpMM} & \textsf{cuSPAESE} & \textsf{HC-SpMM} \\
            \hline
            \multirow{3}{*}{\textit{CS}} & 3090 & 5.63 & 6.62 & 22.40 & 10.73 & 12.77 & \textbf{5.25}  \\
            \cline{2-8} & 4090 & 4.06 & 4.19 & 18.08 & 8.45 & 9.89 & \textbf{3.93}  \\
            \cline{2-8} & A100 & 14.30 & 15.43 & 56.07 & 27.07 & 23.07 & \textbf{10.66} \\
            \hline
            \multirow{3}{*}{\textit{CR}} & 3090 & 6.98 & 8.64 & 6.98 & 18.33 & 13.28 & \textbf{6.05}  \\
            \cline{2-8} & 4090 & 5.44 & 5.98 & 29.92 & 12.80 & 9.76 & \textbf{4.83}  \\
            \cline{2-8} & A100 & 17.44 & 20.83 & 76.39 & 42.11 & 25.02 & \textbf{13.76}  \\
            \hline
            \multirow{3}{*}{\textit{PM}} & 3090 & 13.41 & 13.22 & 78.53 & 33.79 & 59.23 & \textbf{11.62}  \\
            \cline{2-8} & 4090 & 7.46 & \textbf{7.20} & 72.48 & 26.72 & 47.52 & 7.87  \\
            \cline{2-8} & A100 & 25.02 & 21.54 & 211.27 & 84.26 & 90.91 & \textbf{15.43}  \\
            \hline
            \multirow{3}{*}{\textit{PT}} & 3090 & 23.90 & 24.70 & 38.21 & 25.40 & 51.49 & \textbf{17.76}  \\
            \cline{2-8} & 4090 & \textbf{7.84} & 8.90 & 25.22 & 12.00 & 27.71 & 11.23  \\
            \cline{2-8} & A100 & 23.46 & 27.87 & 59.11 & 32.48 & 61.25 & \textbf{17.82}  \\
            \hline
            \multirow{3}{*}{\textit{DD}} & 3090 & 171.42 & 185.98 & 249.98 & 164.76 & 354.85 & \textbf{121.57}  \\
            \cline{2-8} & 4090 & \textbf{67.68} & 106.50 & 138.27 & 112.90 & 249.54 & 73.54  \\
            \cline{2-8} & A100 & \textbf{155.78} & 267.30 & 433.26 & 202.24 & 509.10 & 178.79  \\
            \hline
            \multirow{3}{*}{\textit{AZ}} & 3090 & 353.57 & 358.53 & 743.17 & 414.25 & 3833.33 & \textbf{240.67}  \\
            \cline{2-8} & 4090 & 118.37 & 584.84 & 293.83 & 171.62 & 856.68 & \textbf{112.61}  \\
            \cline{2-8} & A100 & 292.26 & 384.36 & 1051.22 & 471.43 & 1081.33 & \textbf{271.97}  \\
            \hline
            \multirow{3}{*}{\textit{YS}} & 3090 & 769.09 & 866.75 & 756.80 & \textbf{581.36} & 1226.88 & 581.41  \\
            \cline{2-8} & 4090 & 479.24 & 585.10 & 548.04 & 498.15 & 1048.91 & \textbf{461.51}  \\
            \cline{2-8} & A100 & 790.67 & 1181.46 & 1211.83 & \textbf{669.47} & 1286.10 & 743.82  \\
            \hline
            \multirow{3}{*}{\textit{OC}} & 3090 & 841.21 & 909.57 & 899.20 & 660.56 & 1313.76 & \textbf{624.58}  \\
            \cline{2-8} & 4090 & 533.29 & 634.89 & 627.75 & 577.55 & 1178.73 & \textbf{505.99}  \\
            \cline{2-8} & A100 & 851.28 & 1000.53 & 1451.10 & \textbf{786.56} & 1471.58 & 825.17 \\
            \hline
            \multirow{3}{*}{\textit{GH}} & 3090 & 850.59 & 836.10 & 1330.56 & 764.60 & 1296.15 & \textbf{568.41} \\
            \cline{2-8} & 4090 & 475.97 & 549.74 & 806.22 & 534.95 & 1104.04 & \textbf{441.32} \\
            \cline{2-8} & A100 & 1046.29 & 1102.03 & 2221.16 & 811.76 & 1863.42 & \textbf{761.20} \\
            \hline
            \multirow{3}{*}{\textit{YH}} & 3090 & 1395.39 & 1510.75 & 1461.47 & 1085.29 & 2145.85 & \textbf{1045.92} \\
            \cline{2-8} & 4090 & 917.48 & 1038.70 & 1045.17 & 954.03 & 1901.68 & \textbf{869.32} \\
            \cline{2-8} & A100 & 1431.06 & 2011.01 & 2073.99 & 1129.12 & 2502.35 & \textbf{1032.05} \\
            \hline
            \multirow{3}{*}{\textit{RD}} & 3090 & 2441.08 & 2469.04 & 2182.26 & 1651.71 & 2917.27 & \textbf{1574.69} \\
            \cline{2-8} & 4090 & 1484.53 & 1639.77 & 1597.62 & 1474.62 & 2945.31 & \textbf{1436.98} \\
            \cline{2-8} & A100 & 2617.16 & 2870.19 & 3167.70 & \textbf{1846.44} & 2856.08 & 2013.03 \\
            \hline
            \multirow{3}{*}{\textit{TT}} & 3090 & 2164.82 & 2153.66 & 2426.94 & 1601.94 & 4003.09 & \textbf{1382.53} \\
            \cline{2-8} & 4090 & 1276.56 & 1347.54 & 1541.43 & 1231.86 & 3109.70 & \textbf{1126.61} \\
            \cline{2-8} & A100 & 2245.99 & 2918.77 & 2073.99 & 1963.72 & 5752.35 & \textbf{1880.32} \\
            \hline
            \multirow{3}{*}{\textit{DP}} & 3090 & 19696.12 & 18312.22 & 87216.13 & 21148.42 & 327079.96 & \textbf{16718.30} \\
            \cline{2-8} & 4090 & 12061.82 & \textbf{11206.81} & 95456.71 & 14240.56 & 169181.09 & 11350.72 \\
            \cline{2-8} & A100 & 16926.84 & 15532.93 & 178816.99 & 15808.13 & 145479.16 & \textbf{13778.36} \\
            \hline
        \end{tabular}
        % }
    % \end{center}
\end{table*}

% \balance

\end{document}